\documentclass[aps,prb,twocolumn]{revtex4-2}
\usepackage[latin9]{inputenc}
\usepackage{graphicx,epsf}
\usepackage{amssymb}
\usepackage{amsmath}
\usepackage{color,soul}
\usepackage{bm,bbm}
\usepackage{placeins}
\usepackage{verbatim}
\usepackage{textcomp}
\usepackage{esint}
\usepackage{array}
\usepackage{braket}
\usepackage{dsfont}
\usepackage{enumitem}
\usepackage[unicode=true,pdfusetitle,
 bookmarks=true,bookmarksnumbered=false,bookmarksopen=false,
 breaklinks=true,pdfborder={0 0 0},pdfborderstyle={},backref=false,colorlinks=true]{hyperref}
\hypersetup{linkcolor=blue,citecolor=blue,urlcolor=blue}

\newcommand{\llangle}{\langle\!\langle}
\newcommand{\rrangle}{\rangle\!\rangle}

\newcommand{\MSS}{MnSc$_2$S$_4$}
\newcommand{\LYO}{LiYbO$_2$}
\newcommand{\Jrat}{\alpha}
\newcommand{\ndis}{N_\mathrm{dis}}
\newcommand{\ninit}{N_\mathrm{init}}
\newcommand{\QQ}{\mathbf{Q}}
\newcommand{\Egs}{E_{\mathrm{gs}}}
\newcommand{\alphaQ}{\theta_\mathbf{Q}}
\newcommand{\phiQ}{\phi_\mathbf{Q}}
\newcommand{\ii}{\mathrm{i}}

\newcommand{\Htex}{\mathcal{H}_{\mathrm{tex}}}

\newcommand{\absval}[1]{\lvert #1 \rvert}

\definecolor{darkgreen}{rgb}{0,0.5,0}
\definecolor{darkblue}{rgb}{0,0,0.5}
\definecolor{purple}{rgb}{0.35,0,0.35}
\definecolor{orange}{rgb}{0.9,0.4,0}

\graphicspath{{./}{./figs/}}

\begin{document}

%%%%%%%%%%%%%%%%%%%%%%%%%%%%%%%%%%%%%%%%%%%%%%%%%%%%%%%%%%%%%%%%%%%%%%%
%%%%%%%%%%%%%%%%%%%%%%%%%%%%%%%%%%%%%%%%%%%%%%%%%%%%%%%%%%%%%%%%%%%%%%%
%%%%%%%%%%%%%%%%%%%%%%%%%%%%%%%%%%%%%%%%%%%%%%%%%%%%%%%%%%%%%%%%%%%%%%%

\title{
Disorder effects in spiral spin liquids: \\ Long-range spin textures, Friedel-like oscillations, and spiral spin glasses
%From spiral spin liquid to spiral spin glass via quenched disorder
}

\author{Pedro M. C\^onsoli}
\author{Matthias Vojta}
\affiliation{Institut f\"ur Theoretische Physik and W\"urzburg-Dresden Cluster of Excellence ct.qmat, Technische Universit\"at Dresden,
01062 Dresden, Germany}

\date{\today}

\begin{abstract}
Spiral spin liquids are correlated states of matter in which a frustrated magnetic system evades order by fluctuating between a set of (nearly) degenerate spin spirals. Here, we investigate the response of spiral spin liquids to quenched disorder in a $J_1$-$J_2$ honeycomb-lattice Heisenberg model. At the single-impurity level, we identify different order-by-quenched-disorder mechanisms and analyze the ensuing spin textures. In particular, we show that the latter generally display Friedel-like oscillations, which encode direct information about the spiral contour, i.e., the classical ground-state manifold. At finite defect concentrations, we perform extensive numerical simulations and characterize the resulting phases at zero temperature. As a result, we find that the competition between incompatible order-by-quenched-disorder mechanisms can lead to spiral spin glass states already at low to moderate disorder. Finally, we discuss extensions of our conclusions to nonzero temperatures and higher-dimensional systems, as well as their applications to experiments.
\end{abstract}

\maketitle

%%%%%%%%%%%%%%%%%%%%%%%%%%%%%%%%%%%%%%%%%%%%%%%%%%%%%%%%%%%%%%%%%%%%%%%
%%%%%%%%%%%%%%%%%%%%%%%%%%%%%%%%%%%%%%%%%%%%%%%%%%%%%%%%%%%%%%%%%%%%%%%
%%%%%%%%%%%%%%%%%%%%%%%%%%%%%%%%%%%%%%%%%%%%%%%%%%%%%%%%%%%%%%%%%%%%%%%

\section{Introduction}

Within the realm of frustrated magnetism, one of many possible outcomes of frustration is to generate a correlated paramagnetic state known as a spiral spin liquid (SSL). In such a state, the system fluctuates between (nearly) degenerate coplanar spin spirals whose ordering wave vectors $\QQ$ span a contour or surface in reciprocal space.

Following an early theoretical proposal \cite{Bergman2007}, neutron scattering experiments were able to verify the existence of a spiral surface in the diamond-lattice antiferromagnet {\MSS} and confirm it as a realization of a SSL \cite{Gao2017}.
Similar behavior was recently observed in two other compounds: the spin-$5/2$ honeycomb material FeCl$_{3}$ \cite{Gao2022a} and LiYbO$_2$, where spin-$1/2$ Yb$^{3+}$ ions form an elongated diamond lattice \cite{Graham2023}.
Meanwhile, the proximity to a SSL regime has also been associated with the absence of long-range order (LRO) in CoAl$_2$O$_4$ \cite{Zaharko2011,MacDougall2011,MacDougall2016} and the cluster-like scattering in the B-site spinel MgCr$_2$O$_4$ at intermediate temperatures \cite{Bai2019}.
Motivated by these developments, current theoretical work continues to identify spin models with similar classical ground-state degeneracies \cite{Rastelli1979, Niggemann2020, Yao2021, Gao2022b}, while also investigating more fundamental issues related to the nature of the low-energy excitations of SSLs \cite{Yan2022}.

Arguably, one of the most appealing traits of a SSL lies in the wealth of phases and phenomena it can lead to when subjected to different perturbations. While small thermal fluctuations generally favor magnetic (or, in two dimensions, nematic) order via entropic selection \cite{Bergman2007,Mulder2010,Okumura2010,Yao2021}, magnetic anisotropies are known to give rise to commensurate-incommensurate transitions which can render the thermal ordering process highly nontrivial \cite{Lee2008,Gao2017}.
In systems with small spin $S$, enhanced quantum fluctuations can circumvent order by disorder at sufficiently low $T$ and stabilize a quantum paramagnetic phase, such as a valence bond solid \cite{Mulder2010,Ferrari2020} or a quantum spin liquid \cite{Buessen2018,Niggemann2020}.

An additional route toward unconventional behavior in proximate SSLs comes from the application of a magnetic field. Experimentally, this is seen in {\MSS} through the emergence of an antiferromagnetic skyrmion lattice in a field-induced triple-$\QQ$ phase \cite{Gao2017,Gao2020}. In parallel, large-scale classical Monte Carlo simulations have also uncovered different multi-$\QQ$ phases and nontrivial topological textures, including a fascinating ``ripple state", by studying minimal two-dimensional SSL models in magnetic fields \cite{Shimokawa2019a,Shimokawa2019b}.

\begin{figure}
\centering
\includegraphics[width=\columnwidth]{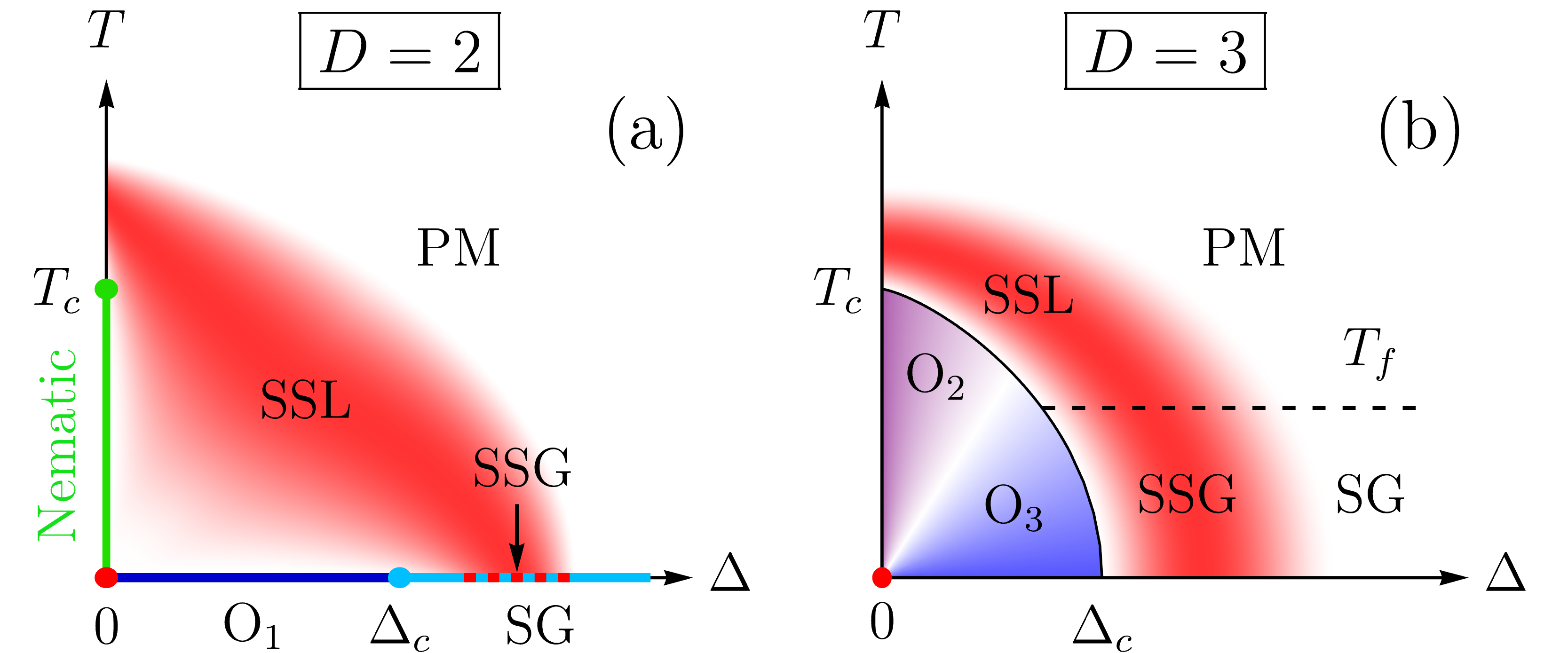}
\caption{
	Schematic phase diagrams illustrating the response of spiral spin liquids to quenched bond disorder in the classical limit in (a) $D=2$ and (b) $D=3$ space dimensions. The axes represent temperature $T$ and disorder strength $\Delta$; $T_c$ and $\Delta_{c}$ correspond to the critical values above which the system enters a paramagnetic (PM) and a spin glass (SG) phase, respectively. The latter has a freezing temperature $T_f$ and, at intermediate $\Delta$, exhibits a ``spiral spin-glass'' (SSG) regime. $\mathrm{O}_{1,2,3}$ represent possibly different phases with spiral LRO, selected either by quenched disorder (O$_1$ and O$_3$) or thermal fluctuations (O$_2$). The red shading encodes the extent to which the static spin structure factors delineates the spiral contour or surface.
}
	\label{fig:pds}
\end{figure}

In contrast, the influence of quenched disorder on SSLs remained relatively unexplored to date \cite{Savary2011}. The aim of this paper is to deepen our understanding of such effects by providing a comprehensive account of impurity-induced phenomena in the simplest model to realize a SSL, the classical $J_1$-$J_{2}$ Heisenberg model on the honeycomb lattice.
Our primary results are for zero temperature, but we shall also discuss their implications for $T>0$ and extensions to higher spatial dimensions. This includes, for instance, scenarios where a nontrivial competition between different ordered phases arises due to the incompatibility between order by quenched disorder (ObQD) and other selection mechanisms, as observed in pyrochlore XY antiferromagnets \cite{Maryasin2014,Ross2016,Sarkar2017,Andrade2018}.

\subsection{Summary of results}

The main results of the paper can be divided into two categories: (a) single-impurity effects, which are relevant to the limit of dilute disorder, and (b) physics of a finite concentration of defects.

Within the first category, we employed perturbative analytic techniques to study the outcome of ObQD for different types of impurities. The results show a dependence not only on the type, but, when applicable, also on the sign and spatial orientation of the defect. In the specific case of a single spin vacancy, we even found a discontinuity in evolution of the ordering wave vectors of the selected states as a function of  $J_{2}/J_{1}$. At the same time, we showed that not in all, but in most of those cases, the impurity distorts the selected spiral state.

By using linear response theory, we determined the spin texture induced by a weak defect at $T=0$. This revealed that, whenever a texture emerges, the zero mode along the spiral contour drives the texture to become non-coplanar as a means to relieve frustration. Moreover, the resulting out-of-plane component of the texture oscillates while decaying in space, in close similarity to Friedel oscillations in a Fermi liquid.
We argue that this feature is not exclusive to the system considered here and should be realized for any model with a spin-wave mode which is gapless along and linearly-dispersing around a spiral manifold of dimension $d\geq1$ (in space dimension $D\geq 2$). In such cases, the out-of-plane component of the texture decays as $r^{-d/2}$ at large distances $r$ from the impurity and at $T = 0$, while the decay becomes exponential at $T>0$. Furthermore, the period and angular dependence of the oscillations are directly related to the geometry of the spiral manifold.
We propose imaging experiments for these Friedel-like oscillations to turn them into a useful experimental tool to reconstruct spiral manifolds in SSL candidates.

By combining the same linear-response method with careful finite-size extrapolations, we also characterized the long-distance behavior of the in-plane components of the textures generated by different bond defects (nearest- and next-nearest-neighbor) as well as by a vacancy.
In the former cases, the textures display $p$-wave-like symmetry and decay as $1/\sqrt{r}$ over an extended range of $J_2/\absval{J_1}$. By analogy with previous results on the triangular-lattice antiferromagnet \cite{Utesov2015,Dey2020}, this suggests that any finite concentration of bond defects destroys the long-range magnetic order which ObQD would otherwise stabilize.
On the other hand, the texture induced by a vacancy changes according to the outcome of the ObQD selection. Near the first Lifshitz point in parameter space, defined by the ratio $J_2/\absval{J_1}$ at which the SSL first emerges, the texture is $s$-wave-like and decays more slowly than any power law. However, at slightly larger $J_2/\absval{J_1}$, a different spiral state is selected, and it develops a texture which is $d$-wave-like and decays as $r^{-3/2}$.

Within the second category, we performed numerical simulations to determine the classical ground states of finite-size systems with a reference ratio $J_2/\absval{J_1} = 0.2$ and different types of disorder. Thus, besides obtaining an independent confirmation of some of our ObQD results, we also identified two distinct mechanisms for the destruction of long-range magnetic order. The first stems from the interference between different textures and tends to lead to short-range order. The second mechanism, in contrast, results from the competition between incompatible ObQD mechanisms and has far more drastic consequences. For a low to moderate concentration $\Delta$ of defects, the system enters a spin glass phase with a ``spiral spin glass'' (SSG) regime whose rugged energy landscape produces a diffuse $T=0$ structure factor which nonetheless delineates the spiral contour.

At low temperatures $T>0$, where entropic order by disorder becomes active and the system retains only a sixfold ground-state degeneracy, the second mechanism above gives rise to effective random-field disorder \cite{Vojta2019}. This imposes even stronger restrictions on the occurrence of spontaneous symmetry breaking than the Hohenberg-Mermin-Wagner theorem \cite{Hohenberg1967,Mermin1966}.  Indeed, the Imry-Ma argument \cite{Imry1975,Aizenman1989} implies that, in $D=2$ space dimensions, phases with discrete order-parameter symmetry also become unstable at $T>0$ if the system contains \emph{any} concentration $\Delta>0$ of bond defects (see Sec. V for subtleties related to vacancies). When combined with the fact that the lower critical dimension of the glass transition is larger than two \cite{Maiorano2018,Lee2007,Kawamura2011,Ogawa2020}, these considerations yield the qualitative phase diagrams depicted in Fig.~\ref{fig:pds}. This shows that disorder can extend the intermediate-temperature SSL regime, an effect which is most pronounced for $D=2$.

\subsection{Outline}

The remainder of the paper is organized as follows.
In Section~\ref{sec:model}, we introduce the $J_1$-$J_2$ Heisenberg model, review the properties of its classical ground-state manifold in the absence of disorder, and discuss caveats arising in finite-size numerics.
Section~\ref{sec:sip} proceeds by providing analytic insights into the effects of different types of single impurities, paying close attention to the possible ObQD mechanisms and characterizing the long-distance behavior of the impurity-induced spin textures.
Armed with this background, we investigate various scenarios with finite impurity concentrations in Section~\ref{sec:mip} and demonstrate the existence of two independent mechanisms promoting glassiness.
In Section~\ref{sec:disclaimers}, we address the influence of thermal fluctuations and the presence of additional interactions. We also discuss important differences between two- and three-dimensional SSL systems.
Finally, Section~\ref{sec:experiments} covers the connection between our work and experiment, while Section~\ref{sec:concl} closes the paper with a summary and an outlook.

%%%%%%%%%%%%%%%%%%%%%%%%%%%%%%%%%%%%%%%%%%%%%%%%%%%%%%%%%%%%%%%%%%%%%%%
%%%%%%%%%%%%%%%%%%%%%%%%%%%%%%%%%%%%%%%%%%%%%%%%%%%%%%%%%%%%%%%%%%%%%%%
%%%%%%%%%%%%%%%%%%%%%%%%%%%%%%%%%%%%%%%%%%%%%%%%%%%%%%%%%%%%%%%%%%%%%%%

\section{Model and results for the clean system}
\label{sec:model}

We consider a classical Heisenberg model on the honeycomb lattice with nearest-neighbor (NN) and next-nearest-neighbor (NNN) interactions. The Hamiltonian for this system can be written as
\begin{equation}
\mathcal{H} =
\sum_{\langle i0, j1 \rangle} J_{1,ij} \mathbf{S}_{i0} \cdot \mathbf{S}_{j1} +
\sum_{\mu} \sum_{\llangle i\mu , j\mu \rrangle} J_{2,ij\mu} \mathbf{S}_{i \mu} \cdot \mathbf{S}_{j \mu}.
\label{eq:H}
\end{equation}
Here, we identify sites on the lattice by a pair of indices $i\mu$, which refer to one of the two basis sites $\mu \in \left\{0,1\right\}$ in the unit cell located at position $\mathbf{R}_i$. As we are interested in the classical limit of the model, the spins $\mathbf{S}_{i\mu}$ will be treated as three-component unit vectors; for the semiclassical linear-response calculations we formally work with spins of size $S$ and perform a $1/S$ expansion.

\subsection{Ground state of the clean infinite system}
\label{subsec:clean}

\begin{figure}
\centering
\includegraphics[width=\columnwidth]{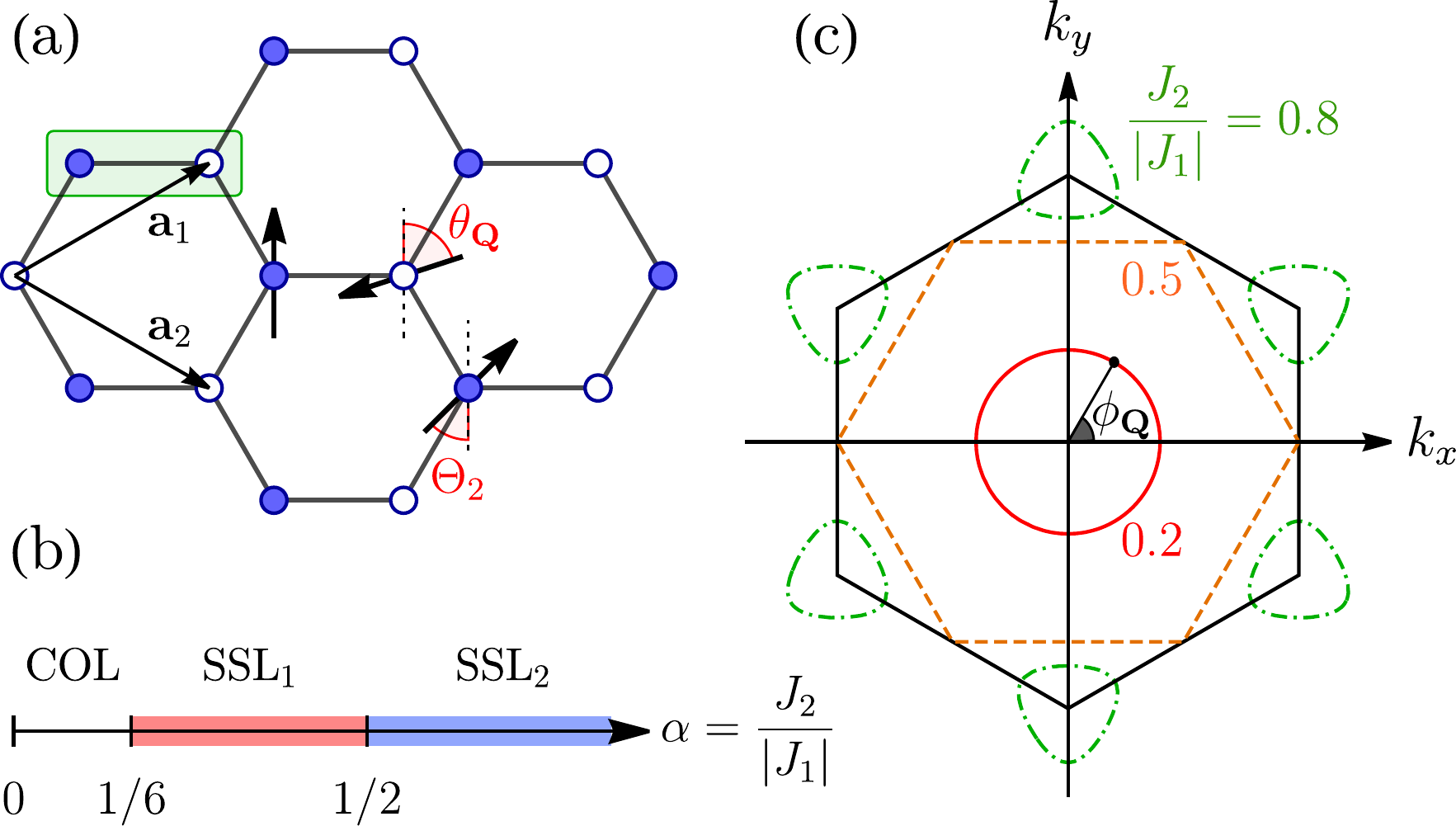}
\caption{
(a) Honeycomb lattice with unit cell (green box) and the primitive vectors $\mathbf{a}_1$ and $\mathbf{a}_2$. For $J_1>0$, the angle between two spins inside the same unit cell is given by $\pi + \alphaQ$.
(b) Ground-state phase diagram of the clean classical $J_1$-$J_2$ Heisenberg model as function of $\alpha=J_2/|J_1|$, valid for both ferromagnetic and antiferromagnetic $J_1$. The label COL stands for ``collinear" and can refer either to a ferromagnetic or N\'eel state, while SSL$_1$ and SSL$_2$ are different SSL regimes.
(c) Spiral contours representing the ground-state manifold of the clean Hamiltonian for three different ratios $\Jrat = J_2 / \absval{J_1}$. In each case, the spiral contour can be parameterized by a polar angle $\phi_{\QQ}$, which is measured w.r.t. the origin and $\mathbf{K} = \left(0, 4\pi/ \left(3\sqrt{3}a \right)\right)$ in the SSL$_1$ and SSL$_2$ regimes, respectively.
}
\label{fig:schems}
\end{figure}

Without disorder, the couplings in Eq.~\eqref{eq:H} simplify to $J_{1,ij} = J_1$ and $J_{2,ij\mu} = J_2$. Given that the honeycomb lattice is bipartite, antiferromagnetic $J_2>0$ leads to frustration for both signs of $J_1$. Our main interest will be on the antiferromagnetic case when both $J_{1,2}>0$, but many results can also be adapted to the case $J_1<0$, $J_2>0$ by inverting the spins on one of the sublattices.

The energy of the system can be minimized via the Luttinger-Tisza method \cite{Luttinger1946,Lyons1960,Attig2017}, which is exact not only for Bravais lattices, but also for any bipartite lattice with equivalent sublattices \cite{Niggemann2020}. It predicts that the lowest-energy states are coplanar spin spirals, characterized by an ordering wave vector $\QQ$ and parameterized as
\begin{equation}
\mathbf{S}_{i \mu}
= s^{\mu} \left[
\cos \varphi_{i \mu} \left(\QQ\right) \hat{\mathbf{z}} +
\sin \varphi_{i \mu} \left(\QQ\right) \hat{\mathbf{x}}
\right].
\label{eq:Spar}
\end{equation}
The prefactor $s = -\mathrm{sign} \left(J_1\right)$ builds a staggered structure into the configuration when $J_1 > 0$, while the function $\varphi_{i \mu} \left( \QQ \right) = \QQ \cdot \mathbf{R}_i + \delta_{\mu1} \alphaQ$ encodes the spatial dependence of the spiral. The angle $\alphaQ$ determines the phase shift between two spins in the same unit cell and follows from the set of equations \cite{Katsura1986,Okumura2010}
\begin{align}
\sin \alphaQ &= \left( \sin \Theta_1 + \sin \Theta_2 \right) / \Lambda_{\QQ},
\notag \\
\cos \alphaQ &= \left( 1 +  \cos \Theta_1 + \cos \Theta_2 \right) / \Lambda_{\QQ},
\notag \\
\Lambda_{\QQ} &= \sqrt{ 3 + 2\left( \cos \Theta_1 + \cos \Theta_2 + \cos \Theta_3 \right) },
\label{eq:alphaQ}
\end{align}
where $\Theta_n = \QQ \cdot \mathbf{a}_n$, $n=1,2,3$, are the projections of $\QQ$ onto the primitive lattice vectors $\mathbf{a}_1$, $\mathbf{a}_2$, and $\mathbf{a}_3 = \mathbf{a}_1 - \mathbf{a}_2$ [see Fig.~\ref{fig:schems}(a)].

At a fixed value of $\Jrat = J_2 / \absval{J_1}$, the set of possible ordering wave vectors $\QQ$ is determined by minimizing the Fourier transform of $J_{ij}^{\mu \nu}$, a $2\times2$ matrix containing the couplings between sites that belong to sublattices $\mu$ and $\nu$ and whose unit cells are separated by $\mathbf{R} = \mathbf{R}_{j} - \mathbf{R}_{i}$ \cite{Katsura1986,Niggemann2020}. When $\Jrat \le 1/6$, this yields a unique solution, $\QQ=\mathbf{0}$, corresponding to the ferromagnetic ($J_1<0$) or N\'eel ($J_1>0$) state one expects for the nearest-neighbor model. On the other hand, once $\Jrat>1/6$, the increased degree of frustration gives rise to a macroscopic ground-state degeneracy as the system enters a SSL regime, see Fig.~\ref{fig:schems}(b). There, the energy is minimized by satisfying
\begin{equation}
	2\absval{\Jrat} \Lambda_{\QQ} = 1.
	\label{eq:spiralcontour}
\end{equation}
As depicted in Fig.~\ref{fig:schems}(c), the solutions of Eq.~\eqref{eq:spiralcontour} form a contour in reciprocal space, parameterized by an angle $\phi_{\QQ}$, which starts as a ring centered at zero momentum for small $\left( \Jrat - 1/6 \right)$ and expands into a hexagon as $\Jrat$ approaches $1/2$. Once $\Jrat>1/2$, the contour changes topology as it splits into pockets that enclose the corners of the Brillouin zone. These pockets gradually shrink to points as $\Jrat \to \infty$, where the ground states become two independent $120^\circ$ orders on each triangular sublattice.

The observed topological transition is enabled by the fact that the second Lifshitz point $\Jrat = 1/2$ hosts an augmented ground-state degeneracy \cite{Fouet2001}, which can be parameterized by a subextensive number of continuous degrees of freedom (as opposed to a single such degree of freedom for $\Jrat>1/6, \neq 1/2$) \cite{footnoteJrat0p5}.
The nontrivial evolution of the ``spiral contour'' with $\Jrat$ prompts interesting questions: How do the ground-state configurations differ between the regimes below (SSL$_1$) and above (SSL$_2$) $\Jrat = 1/2$? What are the consequences when perturbed by defects? We will address these issues in Sec.~\ref{sec:sip} by showing that each regime imposes a distinct set of constraints on ground states and thereby has its own characteristic response to different types of impurities.

Finally, the Luttinger-Tisza method also allows us to compute the ground-state energy, $\Egs$, as a function of $\Jrat$. If $N$ denotes the total number of spins, then we find that, in the entire SSL regime,
\begin{equation}
	\Egs = -\frac{N J_2}{8} \left( 12 + \frac{1}{\Jrat^2} \right).
	\label{eq:Egs}
\end{equation}

%%%%%%%%%%%%%%%%%%%%%%%%%%%%%%%%%%%%%%%%%%%%%%%%%%%%%%%%%%%%%%%%%%%%%%%

\subsection{Thermal order by disorder}
\label{subsec:ObTD}

Given that the subextensive ground-state degeneracy is accidental, i.e., not enforced by symmetries, the states along the spiral contour differ in entropy and acquire different free energies at temperatures $T>0$. This typically results in the selection of a discrete set of ground states and the stabilization of an ordered phase at sufficiently low $T$ through a mechanism known as order by thermal disorder \cite{Villain1980,Henley1987}.

The outcome of such an effect for the model at hand was first studied in Ref.~\onlinecite{Okumura2010}. By considering Gaussian fluctuations around different classical ground states and computing their contributions to the free energy at leading order in $T$, the authors showed that the selected states are $\phiQ = \left(2n + 1\right) \pi/6$ when $1/6 < \Jrat \lesssim 0.232$ and $\phiQ = n\pi/3$ when $0.232 \lesssim \Jrat < 1/2$. These results are consistent with classical Monte Carlo simulations performed for $\Jrat = 0.18$ and $\Jrat = 0.3$ \cite{Okumura2010,Shimokawa2019a,Shimokawa2019b}.
To the best of our knowledge, no explicit calculations have been reported for $\Jrat>1/2$. By employing the same theory of Gaussian fluctuations, we find that, in this SSL$_2$ regime, the selected wave vectors are given by $\phiQ = \left(4n+3\right)\pi/6$, which correspond to the corners of the pockets surrounding the $\mathbf{K}$ and $-\mathbf{K}$ points, as one might have suspected on the grounds of continuity.

%%%%%%%%%%%%%%%%%%%%%%%%%%%%%%%%%%%%%%%%%%%%%%%%%%%%%%%%%%%%%%%%%%%%%%%

\subsection{Finite systems and the influence of boundaries}
\label{subsec:cleanfs}

\begin{figure*}
\includegraphics[width=\textwidth]{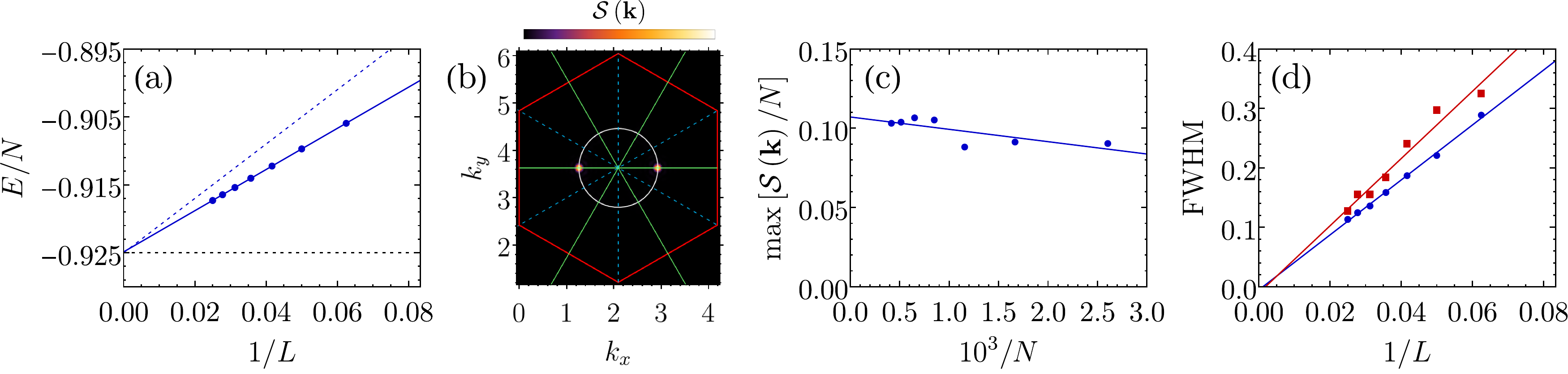}
\caption{
Illustration of finite-size effects in the clean limit. Results are shown for $\Jrat\equiv J_2/J_1= 0.2$ and hexagonal samples with open zigzag boundary conditions.
(a) Finite-size scaling of the ground-state energy per site (circles) fitted to a linear function (solid line). The dashed blue line indicates the energy of the infinite-system solution [Eqs.~\eqref{eq:Spar}-\eqref{eq:spiralcontour}] when placed into the cluster. In the limit $L\to\infty$, the results extrapolate to the value predicted by Eq.~\eqref{eq:Egs} (horizontal dashed line).
(b) Static structure factor $\mathcal{S}(\mathbf{k})$ of a single lowest-energy state for $L=40$. The red hexagon delimits a second Brillouin zone centered at $2\pi \left( 1/3,1/\sqrt{3} \right)$, while the white ring represents the spiral contour of the infinite system (shown as guide to the eye). The solid green and dashed cyan lines correspond to different high-symmetry directions.
(c) Finite-size scaling of ${\rm max}_k\mathcal{S}(\mathbf{k})$ normalized by the total number of sites $N$.
(d) Full width at half maximum (FWHM) of the peak in $\mathcal{S}(\mathbf{k})$ along a radial (blue circles) and tangential (red squares) cut through the second Brillouin zone.
}
\label{fig:cleanzigzag}
\end{figure*}

In the presence of quenched disorder, translational invariance is spoiled and the Luttinger-Tisza method, which was used to derive the results summarized in Sec.~\ref{subsec:clean}, can no longer be applied. Consequently, one has to resort either to perturbation theory or to numerical methods that minimize the energy on finite clusters and perform extrapolations to extract valid conclusions for the thermodynamic limit.

However, doing this in a system that hosts a SSL requires caution. Every spin spiral in the ground-state manifold has a unique combination of a wavelength and a propagation direction, and therefore responds differently when placed in a finite cluster with specific boundary conditions.
For instance, under periodic boundary conditions, spin spirals that are (nearly) commensurate with the cluster are favored by an energy difference proportional to the perimeter (or area, in three dimensions) of the boundary. Imbalances such as this generally lift the ground-state degeneracy and can induce strong selection mechanisms that occlude bulk effects caused by other physical sources in the range of system sizes that are accessible to numerics.
Hence, before adding disorder to system and investigating its consequences, it is important to select boundary conditions that permit a meaningful extrapolation to the thermodynamic limit and carefully characterize their impact on the finite-size behavior of the clean system.

Here, we implemented an energy minimization algorithm (see Appendix \ref{app:algorithm} for details) on finite hexagonal clusters and, in light of the previous considerations, imposed open boundary conditions. For concreteness, we considered an antiferromagnetic $J_1 > 0$, but our conclusions in this and following sections can be extended to the case $J_1 < 0$.
Moreover, all results shown in the main text were obtained for zigzag-type boundaries, such that $N$ is related to the linear system size, $L$, via $N = \left(3/2\right)\left(L^2 - L \, \mathrm{mod} \, 2\right)$. For such systems, we observe that all coplanar spin spirals described by Eq.~\eqref{eq:Spar} remain degenerate for any $L$. However, as the numerical results depicted in Fig.~\ref{fig:cleanzigzag}(a) indicate, they are not minimal-energy states away from the thermodynamic limit. To characterize the actual finite-size ground states, we computed the static structure factor
\begin{equation}
\mathcal{S}\left( \mathbf{k} \right) = \frac{1}{N} \sum_{ij} \sum_{\mu\nu} e^{\ii \mathbf{k} \cdot \left( \mathbf{r}_{i \mu} - \mathbf{r}_{j \nu} \right)} \mathbf{S}_{i \mu} \cdot \mathbf{S}_{j \nu},
\end{equation}
where $\mathbf{r}_{i \mu} = \mathbf{R}_i + \delta_{\mu1} \hat{\mathbf{x}}$ denotes the position of site $i \mu$, for clusters of sizes ranging from $L=16$ to $40$.
The result for the largest of these systems is shown in Fig.~\ref{fig:cleanzigzag}(b). There, one sees a pair of Bragg peaks lying on the intersection between the spiral contour (white ring) and a high-symmetry line of the second Brillouin zone. By performing a finite-size analysis, we find that the height of the Bragg peaks diverges with $N$ [Fig.~\ref{fig:cleanzigzag}(c)], whereas their full width at half maximum (FWHM) vanishes as $1/L$ [Fig.~\ref{fig:cleanzigzag}(d)].
Put together, this evidence signals that our choice of open boundary conditions still induces long-range magnetic order despite being less biased than periodic boundary conditions. The degeneracy of the spiral surface is lifted such that the ground-state manifold reduces to a discrete set of six states, which are
related (up to global spin rotations) by the sixfold rotational symmetry around the center of a hexagonal plaquette.
In Appendix \ref{app:clean}, we explain the origin of the underlying selection mechanism and show that the finite-size ground states acquire significant out-of-plane components near four of the six edges of the cluster.

To cope with these artifacts, we performed numerical simulations with impurity concentrations that were large enough to overcome the boundary selection and, at the same time, within a reasonable range for realistic systems. When suitable, we validated our results by testing their reproducibility in clusters with armchair edges, which favor a different set of wave vectors. Aside from that, we studied single-impurity effects using analytic techniques which are immune to finite-size effects. This will be the topic of the next section.

%%%%%%%%%%%%%%%%%%%%%%%%%%%%%%%%%%%%%%%%%%%%%%%%%%%%%%%%%%%%%%%%%%%%%%%
%%%%%%%%%%%%%%%%%%%%%%%%%%%%%%%%%%%%%%%%%%%%%%%%%%%%%%%%%%%%%%%%%%%%%%%
%%%%%%%%%%%%%%%%%%%%%%%%%%%%%%%%%%%%%%%%%%%%%%%%%%%%%%%%%%%%%%%%%%%%%%%

\section{Isolated impurities: Order by quenched disorder}
\label{sec:sip}

In this section, we discuss how different types of isolated impurities lift the continuous degeneracy of the ground-state manifold in the thermodynamic limit -- a necessary ingredient for ObQD -- and draw distinctions between the SSL$_1$ and SSL$_2$ regimes. We also analyze spin textures surrounding isolated impurities. The physics of a finite concentration of impurities will be discussed in Sec.~\ref{sec:mip} below.

It is convenient to rewrite the clean version of the Hamiltonian \eqref{eq:H} in the form $\mathcal{H} = -\sum_{i\mu} \textbf{h}_{i\mu} \cdot \mathbf{S}_{i\mu}$, where
\begin{equation}
\mathbf{h}_{i\mu}
= -\frac{J_1}{2} \sum_{\mathrm{NN}} \mathbf{S}_{j\nu} - \frac{J_2}{2} \sum_{\mathrm{NNN}} \mathbf{S}_{j\mu}
\label{eq:hloc}
\end{equation}
is the local field generated at site $i\mu$ by exchange interactions. The first and second sums above run over the NN and NNN sites of $i\mu$, respectively. In these terms, every spin configuration belonging to the ground-state manifold yields a different set of local fields satisfying $\mathbf{h}_{i\mu} \left( \Jrat,\phiQ \right) \parallel \mathbf{S}_{i\mu} \left( \Jrat,\phiQ \right)$ for all $i\mu$. Due to translation invariance and the equivalence between the two sublattices, this implies that $\mathbf{h}_{i\mu} = -h \, \mathbf{S}_{i\mu}$ for such states, with $h = \absval{\Egs}/N$.

When a single impurity is added to the system, its leading effect is to change the local fields at the sites $i\mu$ to which it couples, i.e. $\mathbf{h}_{i\mu} \to \mathbf{h}_{i\mu} + \mathbf{\delta h}_{i\mu}$. Generally, the deviation field $\mathbf{\delta h}_{i\mu}$ lifts the degeneracy of the ground-state manifold and selects a discrete set of ground states via ObQD \cite{Villain1980,Henley1987,Henley1989}. As we describe in detail below, the selection mechanism is in some cases simple enough to allow one to directly predict the outcome for an infinite system by analytic means.

\begin{figure}
	\centering
	\includegraphics[width=0.95\columnwidth]{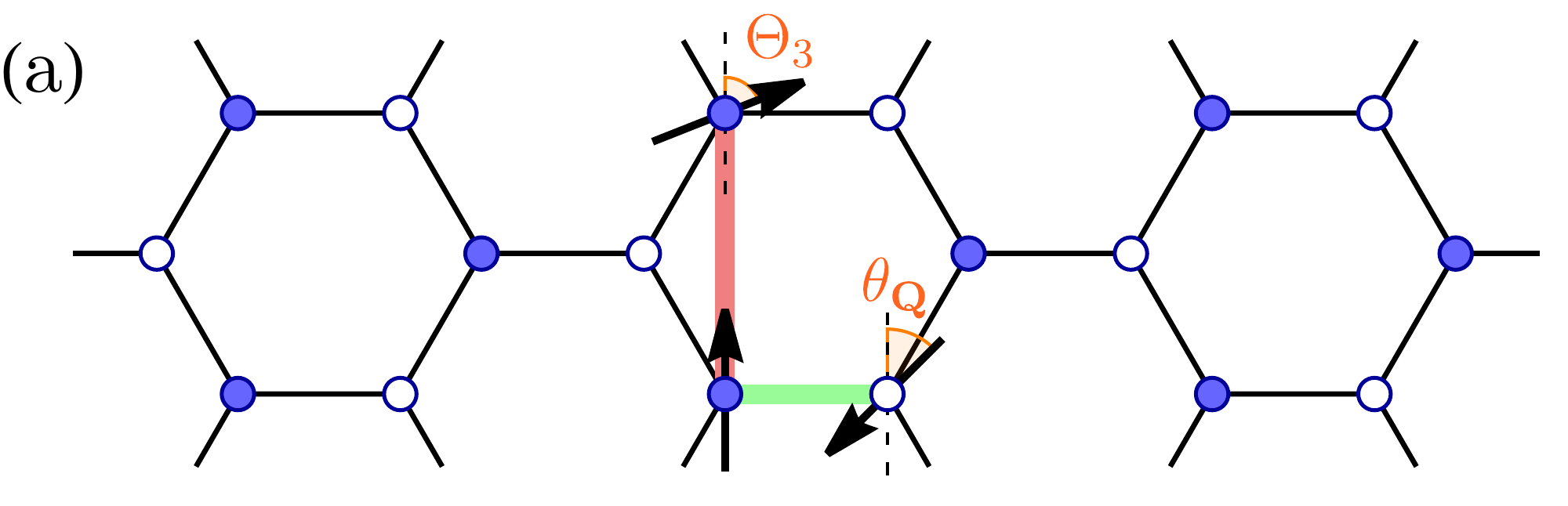}
	\\
	\vspace{0.3cm}
	\includegraphics[width=\columnwidth]{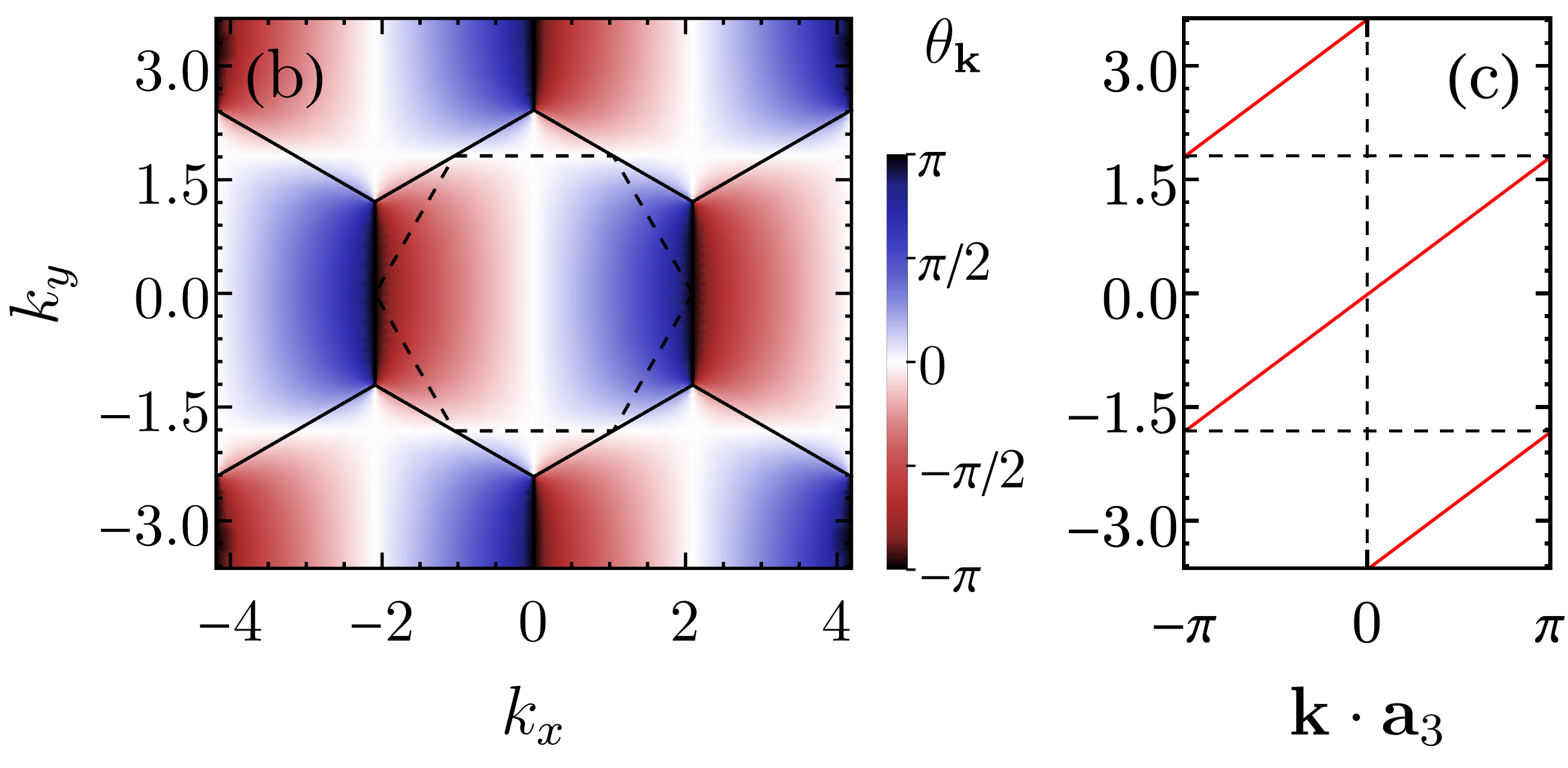}
	\caption{(a) Schematic showing the angles $\alphaQ$ between NN spins in the same unit cell and $\Theta_3$ between NNN spins separated by a vector $\mathbf{a}_3$. The green and red lines highlight the directions of the NN and NNN bond defects we considered for the ObQD arguments.
	(b) Momentum-dependence of the function $\theta_{\mathbf{k}}$ \eqref{eq:alphaQ}. Solid hexagons delimit Brillouin zones, and the inner dashed hexagon corresponds to the spiral contour at the Lifshitz point $\Jrat = 1/2$, which separates the SSL$_1$ and SSL$_2$ regimes.
	(c) Variation of $\mathbf{k}\cdot\mathbf{a}_3$ with $k_y$. For wave vectors $\mathbf{k}$ in the spiral contour, this sets the angle $\Theta_3$ between NNN spins, which is why we limited the results to the compact interval $\left]-\pi,\pi\right]$. We also took $k_y$ as the $y$ axis to match the visualization in (b). The horizontal dashed lines indicate the locations of the upper and lower edges of the $\Jrat = 1/2$ spiral contour. Panels (b) and (c) can be used to determine under which circumstances bond defects do \textit{not} produce a texture, for details see text.
	}
	\label{fig:analytics_sip}
\end{figure}

%%%%%%%%%%%%%%%%%%%%%%%%%%%%%%%%%%%%%%%%%%%%%%%%%%%%%%%%%%%%%%%%%%%%%%%

\subsection{Bond defect}
\label{subsec:sbd}

We begin by considering a single bond defect between a pair of NN sites $m0$ and $n1$, such that the couplings in Eq.~\eqref{eq:H} take the form $J_{1,ij} = J_1 + \delta_{im}\delta_{jn}\delta J_1$ and $J_{2,ij\mu} = J_{2}$. For concreteness, we assume that this bond is horizontal, as depicted in Fig.~\ref{fig:analytics_sip}(a).

When $\delta J_1 > 0$ and $J_1>0$, the defective bond is strengthened and therefore favors states in which $\mathbf{S}_{m0}$ and $\mathbf{S}_{n1}$ are antiparallel, i.e., $\alphaQ = 0$. According to Figs.~\ref{fig:schems}(c) and \ref{fig:analytics_sip}(b), this condition is, for any $\Jrat$, fulfilled by the pair of wave vectors $\QQ$ in the spiral contour that satisfy $Q_x = 0$ in the first Brillouin zone. Formally, we can verify this conclusion by tracking how $\delta \mathbf{h}_{m0} = -\left( \delta J_1/2 \right) \mathbf{S}_{n1}$ changes relative to $\mathbf{S}_{m0}$ as a function of $\phi_{\QQ}$. It follows from Eq.~\eqref{eq:Spar} that
\begin{align}
	\frac{\delta \mathbf{h}_{m0}}{\delta J_1/2} =
	 \cos \! \left( \QQ \cdot \mathbf{R}_{nm} \!+\! \alphaQ \right) \hat{\mathbf{e}}_{\parallel}
	 + \sin \! \left( \QQ \cdot \mathbf{R}_{nm} \!+\! \alphaQ \right) \hat{\mathbf{e}}_{\perp} ,
	 \label{eq:dhJ1bd}
\end{align}
where $\mathbf{R}_{nm} = \mathbf{R}_n - \mathbf{R}_m$ and $\hat{\mathbf{e}}_{\parallel}$ $\left(\hat{\mathbf{e}}_{\perp}\right)$ is a unit vector that is parallel (perpendicular) to $\mathbf{S}_{m0}$. Since $\mathbf{R}_{nm} = 0$ for a horizontal bond defect, we confirm that the parallel component $\delta h_{m0}^{\parallel}$ is indeed maximized when $\alphaQ = 0$. At the same time, Eq.~\eqref{eq:dhJ1bd} reveals that the transverse component $\delta h_{m0}^{\perp}=0$ for these optimal states, such that they remain intact regardless of the magnitude of $\delta J_1$. In fact, this analysis implies that any distribution of equally-oriented strong NN bond defects selects a pair of spiral states without inducing any additional spin textures.

An entirely different result arises when $\delta J_1<0$. In this case, Eq.~\eqref{eq:dhJ1bd} indicates that the impurity favors the states in the spiral contour with minimal $\cos \alphaQ$, for which $\mathbf{S}_{m0}$ and $\mathbf{S}_{n1}$ are as aligned as possible. The system thereby transfers the maximum amount of frustration to the defect in order to optimize the exchange interactions on the unweakened NN bonds connected to $m0$ and $n1$.
However, as depicted in Fig.~\ref{fig:analytics_sip}(b), this condition cannot be fully satisfied through states with $\alphaQ = \pi$ unless the spiral contour intersects the boundaries of the Brillouin zone, i.e., $\Jrat \ge 1/2$. Remarkably, this shows that the effect of a weak NN bond defect depends sensitively on the topology of the spiral contour.
In the SSL$_1$ regime, the preferred states $\phiQ = 0$ or $\pi$ always have a transverse deviation field $\delta h_{m0}^{\perp} \ne 0$, since $\mathbf{S}_{m0}$ and $\mathbf{S}_{n1}$ are not parallel for any ground state. As a result, the selected spin spiral acquires a texture due to the defect.
In contrast, there is no such constraint in SSL$_2$; the impurity simply selects the states with wave vectors $\QQ$ on the vertical edges of the Brillouin zone without inducing any textures.

Next, we turn to the case of a bond defect between two NNN sites $m\mu$ and $n\mu$. Without any loss of generality, we can assume that both sites belong to the sublattice $\mu = 0$, such that $J_{1,ij} = J_{1}$ and $J_{2,ij\mu} = J_2 + \delta_{im}\delta_{jn}\delta_{\mu 0} \delta J_2$. The deviation field $\delta \mathbf{h}_{m0} = -\left( \delta J_2/2 \right) \mathbf{S}_{n0}$ at site $m0$ is then given by
\begin{equation}
	\frac{\delta \mathbf{h}_{m0}}{\delta J_2/2} = -\cos \left(\QQ \cdot \mathbf{R}_{nm} \right) \hat{\mathbf{e}}_{\parallel} + \sin \left(\QQ \cdot \mathbf{R}_{nm} \right) \hat{\mathbf{e}}_{\perp}.
	\label{eq:dhJ2bd}
\end{equation}
For now, let us focus on the case of a vertical bond defect with $\mathbf{R}_{nm} = \mathbf{a}_3$ [see Fig.~\ref{fig:analytics_sip}(a)]. As before, we consider the cases of positive and negative $\delta J_2$ separately.

When $\delta J_2 < 0$, Eq.~\eqref{eq:dhJ2bd} shows that the system benefits from minimizing the angle $\Theta_3$ between $\mathbf{S}_{m0}$ and $\mathbf{S}_{n0}$. Similarly to the case $\delta J_1 < 0$, this optimizes the energy by relieving the frustration on the unweakened NNN bonds. However, according to Fig.~\ref{fig:analytics_sip}(c), the spiral contour only contains states with $\Theta_3 = 0$ when it crosses the line $k_y = 0$ or, equivalently, when $\Jrat \le 1/2$.
Thus, the response of the system once again differs between the SSL$_1$ and SSL$_2$ regimes. In the former, the selected states $\phiQ = 0,\pi$ are left undistorted. In the latter, the impurity favors the two configurations with $\QQ$ on a vertical edge of the Brillouin zone; since these do not fully frustrate the defective bond, they experience $\delta h_{m0}^{\perp} \ne 0$ and acquire a texture.

Conversely, a defect with $\delta J_2 > 0$ tends to maximize $\Theta_3$. From Figs.~\ref{fig:schems}(c) and \ref{fig:analytics_sip}(c), we see that, regardless of $\Jrat$, this condition is fulfilled by the points $\QQ$ on the spiral contour which belong to the first Brillouin zone and have $Q_x = 0$. Nevertheless, because the extreme case $\Theta_3 = \pi$ only occurs along the horizontal edges of the $\Jrat = 1/2$ spiral contour, a strong NNN bond defect induces textures in both SSL$_1$ and SSL$_2$ regimes.

By exploiting the $C_3$ symmetry of the system, one can readily generalize the arguments above to other orientations of bond defects. The corresponding results are summarized in Table \ref{tab:sbd} and importantly show that bond defects of the same kind, but with different orientations, favor distinct pairs of spin spirals. At a non-vanishing concentration of impurities, this will result in a competition between incompatible ObQD mechanisms, whose implications are discussed in Sec.~\ref{sec:mip} below.

\begin{table}
	\centering
	\newcolumntype{C}[1]{>{\centering \arraybackslash}m{#1}}
	\caption{Summary of the effects of different isolated bond defects. All $\mathbf{Q}$ are assumed to be in the first Brillouin zone. $\mathrm{BZ}_{\perp}$ and $\mathrm{BZ}_{\parallel}$ stand for the boundaries of the Brillouin zone which are perpendicular and parallel to the defect in question.}
	\begin{tabular}{C{1.6cm} C{2.1cm} C{2.1cm} C{1.05cm} C{1.05cm}}
		\hline
		\hline
		\multicolumn{1}{c}{Type} & \multicolumn{2}{c}{Selected states} & \multicolumn{2}{c}{Textures?} \\
		 & SSL$_1$ & SSL$_2$ & SSL$_1$ & SSL$_2$ \\
		\hline
		$\delta J_1 > 0$ & $\QQ \perp$ defect & $\QQ \perp$ defect & No & No \\
		$\delta J_1 < 0$ & $\QQ \parallel$ defect & $\QQ \in \mathrm{BZ}_{\perp}$ & Yes & No \\
		$\delta J_2 < 0$ & $\QQ \perp $ defect & $\QQ \in \mathrm{BZ}_{\parallel}$ & No & Yes\\
		$\delta J_2 > 0$ & $\QQ \parallel$ defect &  $\QQ \parallel$ defect & Yes & Yes \\
		\hline	
		\hline
	\end{tabular}
\label{tab:sbd}
\end{table}

%%%%%%%%%%%%%%%%%%%%%%%%%%%%%%%%%%%%%%%%%%%%%%%%%%%%%%%%%%%%%%%%%%%%%%%

\subsection{Single vacancy}
\label{subsec:svac}

Non-magnetic impurities, which are equivalent to spin vacancies, are another common type of defect in magnetic solids. Contrary to a bond defect, a single vacancy does not break the local lattice rotation symmetry, and thus cannot produce the same kind of state selection we encountered in Sec.~\ref{subsec:sbd}. In particular, it must preserve at least a sixfold ground-state degeneracy for the system at hand.

A vacancy eliminates all couplings connected to its site $I0$; equivalently, we may also set $\mathbf{S}_{I0} = \mathbf{0}$. Proceeding as in the previous sections, we find that the deviation fields at the vacancy site and at its nine neighbors yield the same energy correction $2h = 2\absval{\Egs}/N$ for every $\mathbf{Q}$ on the spiral contour. This, however, does not imply that the degeneracy of the spiral contour remains intact, as an energy splitting arises due to the impurity-induced spin texture, which we discuss now.

To this end, we calculate the full spin texture within linear-response theory. The procedure, which is explained in detail in Appendix \ref{app:linresp}, begins by identifying a combination of local fields that mimic the effect a given impurity (in this case, a vacancy). We then use linear-response theory to relate the texture to a susceptibility and compute this quantity to leading order in $1/S$ via linear spin-wave theory \cite{Wollny2011,Utesov2015,Dey2020}.

\begin{figure}
\includegraphics[width=\columnwidth]{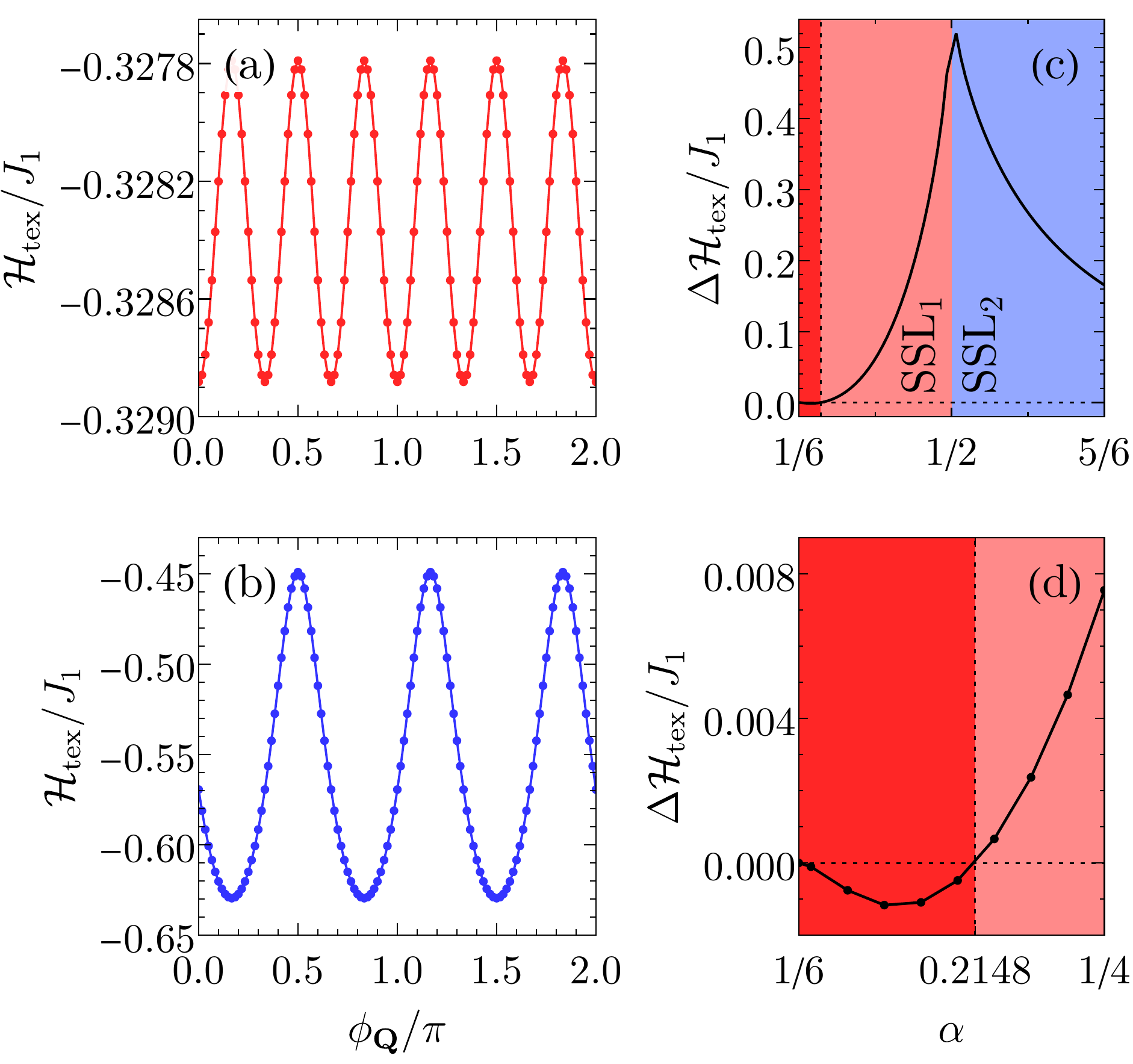}
\caption{
ObQD results for a single vacancy. (a,b) Variation along the spiral contour of the energy correction $\mathcal{H}_{\mathrm{tex}}/J_1$ due to the spin texture for $\Jrat = 0.2$ and $\Jrat = 0.8$, respectively. (c) Difference $\Delta \mathcal{H}_{\mathrm{tex}}$ between the energy corrections at different high-symmetry points as a function of $\Jrat$. (d) Same as (c), but zoomed into the vicinity of the first Lifshitz point $\Jrat=1/6$. The sign change at $\Jrat \approx 0.2148$ signals a shift in the wave vectors $\QQ$ selected by the vacancy.
}
\label{fig:svac_obd}
\end{figure}

As a byproduct, we obtain the necessary ingredients to derive the correction $\mathcal{H}_{\mathrm{tex}}$ to the ground-state energy due to the emergence of the texture. Figures~\ref{fig:svac_obd}(a) and (b) depict the variation of this quantity along the spiral contour for $\Jrat = 0.2$ and $0.8$, respectively.
In the first case, the data exhibits the expected sixfold symmetry and indicates that the vacancy selects the set of states $\phiQ = n\pi/3$ with $n\in\mathbb{Z}$. On the other hand, the second case displays a lower threefold symmetry because the angle $\phiQ\in\left[0,2\pi\right[$ only spans half of the spiral contour, corresponding to the pocket enclosing $\mathbf{K} = \left(0,4\pi/\left(3\sqrt{3}a\right)\right)$. Furthermore, the ObQD mechanism is seen to favor states with $\phiQ = \left(4n+1\right)\pi/6$, which live along line connecting the center of the Brillouin zone to its corners.

To verify whether the previous results are representative of the entire SSL$_1$ and SSL$_2$ regimes, we computed $\Delta \Htex = \Htex\left(\phi_{\mathbf{Q}1}\right) - \Htex\left(\phi_{\mathbf{Q}2}\right)$ with
\begin{equation}
	\left( \phi_{\mathbf{Q}1}, \phi_{\mathbf{Q}2} \right) =
	\begin{cases}
			\left(0, \pi/2\right) & \text{for $ \alpha$ in SSL$_1$},
			 \\
			 \left(\pi/2, -\pi/2\right) & \text{for $ \alpha$ in SSL$_2$},
		\end{cases}
\end{equation}
as a function of $\Jrat$. As shown Figs.~\ref{fig:svac_obd}(c,d), $\Delta \Htex$ is negative at $\Jrat = 0.2$, in full consistency with Fig.~\ref{fig:svac_obd}(a). However, it changes sign at $\Jrat^{*} \approx 0.2148$ and remains positive beyond that point. Thus, the selected states shift from $\phiQ = n\pi/3$ when $1/6 < \Jrat < \alpha^{*}$ to $\phiQ = \left(2n+1\right)n\pi/6$ when $\alpha^{*} < \Jrat < 1/2$. As suggested by Fig.~\ref{fig:schems}(c), the latter wave vectors evolve continuously into the points characterized by $\phiQ = \left(4n+1\right)\pi/6$ once $\Jrat>1/2$, which correspond precisely to the states selected at $\Jrat = 0.8$ and in the rest of the SSL$_2$ regime.
Finally, we note that the magnitude of $\Delta \Htex$ vanishes as $\Jrat \to 1/6^{+}$ and $\Jrat \to \infty$, revealing that the strength of the ObQD mechanism is proportional to the degree of anisotropy of the spiral contour as expected.

%%%%%%%%%%%%%%%%%%%%%%%%%%%%%%%%%%%%%%%%%%%%%%%%%%%%%%%%%%%%%%%%%%%%%%%

\subsection{Two-vacancy clusters}

In the previous subsection, we stressed that the selection mechanisms promoted by a bond defect and a vacancy must be fundamentally different due to symmetry. However, in situations with a finite concentration $p$ of vacancies there is a probability $\propto p^2$ for two-vacancy clusters to form. This type of defect lowers the symmetries of a single vacancy down to those of a bond defect, and can therefore have one of two effects: It can either further lift the sixfold ground-state degeneracy we encountered in Sec.~\ref{subsec:svac} or favor an entirely different pair of spiral states. The latter case gives rise to yet another instance where competing ObQD mechanisms coexist in the same system; we shall comment more on its implications in Sec.~\ref{subsec:vacmip}.

\begin{figure}
	\centering
	\includegraphics[width=\columnwidth]{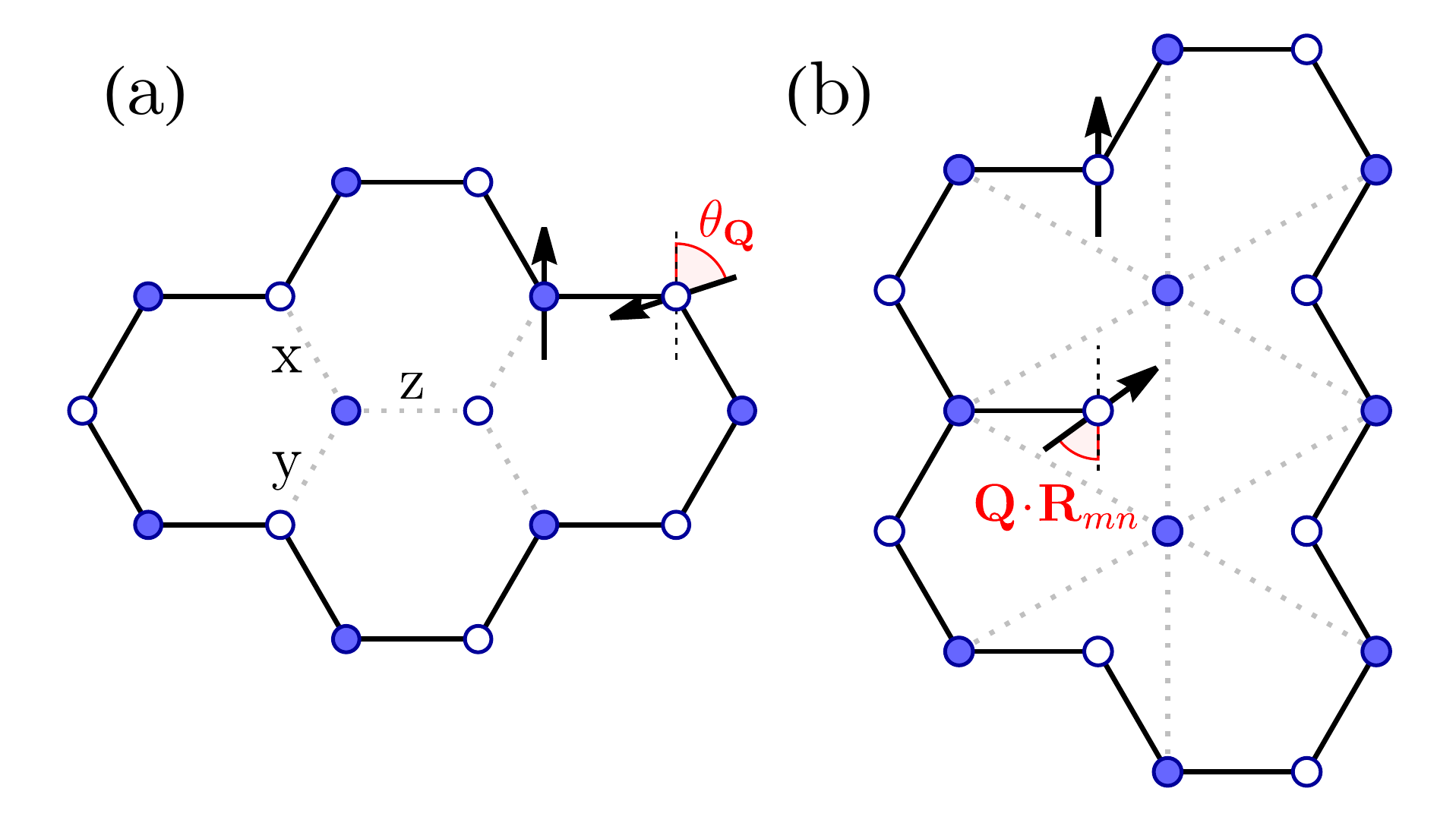}
	\caption{Different types of two-vacancy clusters on the honeycomb lattice with (a) NN and (b) NNN vacancies.
	}
	\label{fig:pair_vacs}
\end{figure}

Let us first consider the case where two vacancies are located on NN sites, $m0$ and $n1$. By considering the effect of the interaction between the deviation fields $\delta \mathbf{h}_{i\mu}$ and an undistorted spin spiral, one finds an energy correction
\begin{equation}
	\delta E_{\mathrm{NNvacs}} = 4h - J_1 \cos\left(\QQ \cdot \mathbf{R}_{nm} + \alphaQ \right),
\end{equation}
which is equal to the correction due to two isolated vacancies minus a term that eliminates the double-counting of the bond connecting $m0$ to $n1$. We thus see that a NN two-vacancy cluster favors the same states as a strong ($\delta J_1>0$) NN defect of the same orientation.
This result also has a simple visual explanation. If the two vacancies are connected by a z bond, as shown in Fig.~\ref{fig:pair_vacs}(a), then they eliminate two $x$ and two $y$ bonds, but only one z bond. Thus, configurations that minimize the energy on $z$ bonds ($\mathbf{Q} \perp$ to the defect) are preferred.

\begin{table}
\centering
\newcolumntype{C}[1]{>{\centering \arraybackslash}m{#1}}
\caption{Summary of the effects of a single vacancy or a pair of adjacent vacancies separated by a vector $\mathbf{d}$. $\phiQ$ is a polar angle parametrizing the entire spiral contour in the SSL$_1$ regime and the upward pocket centered at $\mathbf{K}$ in SSL$_2$. The ratio $\Jrat^{*} \approx 0.2148$ was obtained by analyzing the data in Fig.~\ref{fig:svac_obd}.
}
	\begin{tabular}{C{3.8cm} C{2.8cm} C{1.4cm} }
		\hline
		\hline
		Type \& regime & Selected states & Textures? \\
		\hline
		Vacancy $\left( 1/6 < \Jrat < \Jrat^{*} \right)$ & $\phi_{\QQ} = n\frac{\pi}{3}$ & Yes \\
		Vacancy $\left( \Jrat^{*} < \Jrat < 1/2 \right)$ & $\phi_{\QQ} = \left(2n+1\right)\frac{\pi}{6}$ & Yes \\
		Vacancy  (SSL$_2$) & $\phi_{\QQ} = \left(4n+1\right)\frac{\pi}{6}$ & Yes \\
		NN vacancies & $\QQ \perp$ $\mathbf{d}$ & Yes \\
		NNN vacancies & $\QQ \parallel $ $\mathbf{d}$ & Yes \\
		\hline	
		\hline
	\end{tabular}
\label{tab:vacs}
\end{table}

Analogously, the energy correction due to a pair of NNN vacancies that occupy sites $m0$ and $n0$ is
\begin{equation}
	\delta E_{\mathrm{NNNvacs}} = 4h + J_2 \cos\left(\QQ \cdot \mathbf{R}_{nm} \right).
\end{equation}
The $\mathbf{Q}$-dependent term here acts as a strong NNN bond defect, thereby favoring spirals for which $\mathbf{S}_{m0}$ and $\mathbf{S}_{n0}$ as misaligned as possible. From Fig.~\ref{fig:pair_vacs}(b), we see that this also has a simple rationalization: Vertically aligned vacancies, for instance, eliminate four of each of the different diagonal NNN couplings, but only three of the vertical ones. Hence, it is less energetically costly to frustrate interactions on the diagonal NNN bonds.

Finally, one can analyze the configuration of the deviation fields for the two types of vacancy clusters we considered above and verify that both cases have impurity-induced textures. This fact, along with the other results we obtained in connection to single vacancies and two-vacancy clusters, are summarized in Table~\ref{tab:vacs}.

%%%%%%%%%%%%%%%%%%%%%%%%%%%%%%%%%%%%%%%%%%%%%%%%%%%%%%%%%%%%%%%%%%%%%%%

\subsection{Spin textures from linear-response theory: Friedel-like oscillations and exceptionally slow decays}
\label{subsec:textures}

Having established that different types of isolated impurities induce distortions on top of selected spiral states, we now characterize the long-distance behavior of these spin textures. This is a relevant undertaking because, depending on the outcome, the interference between textures can drastically impact the stability of LRO in a system with a finite concentration of impurities \cite{Dey2020}.

\begin{figure}
\centering
\includegraphics[width=\columnwidth]{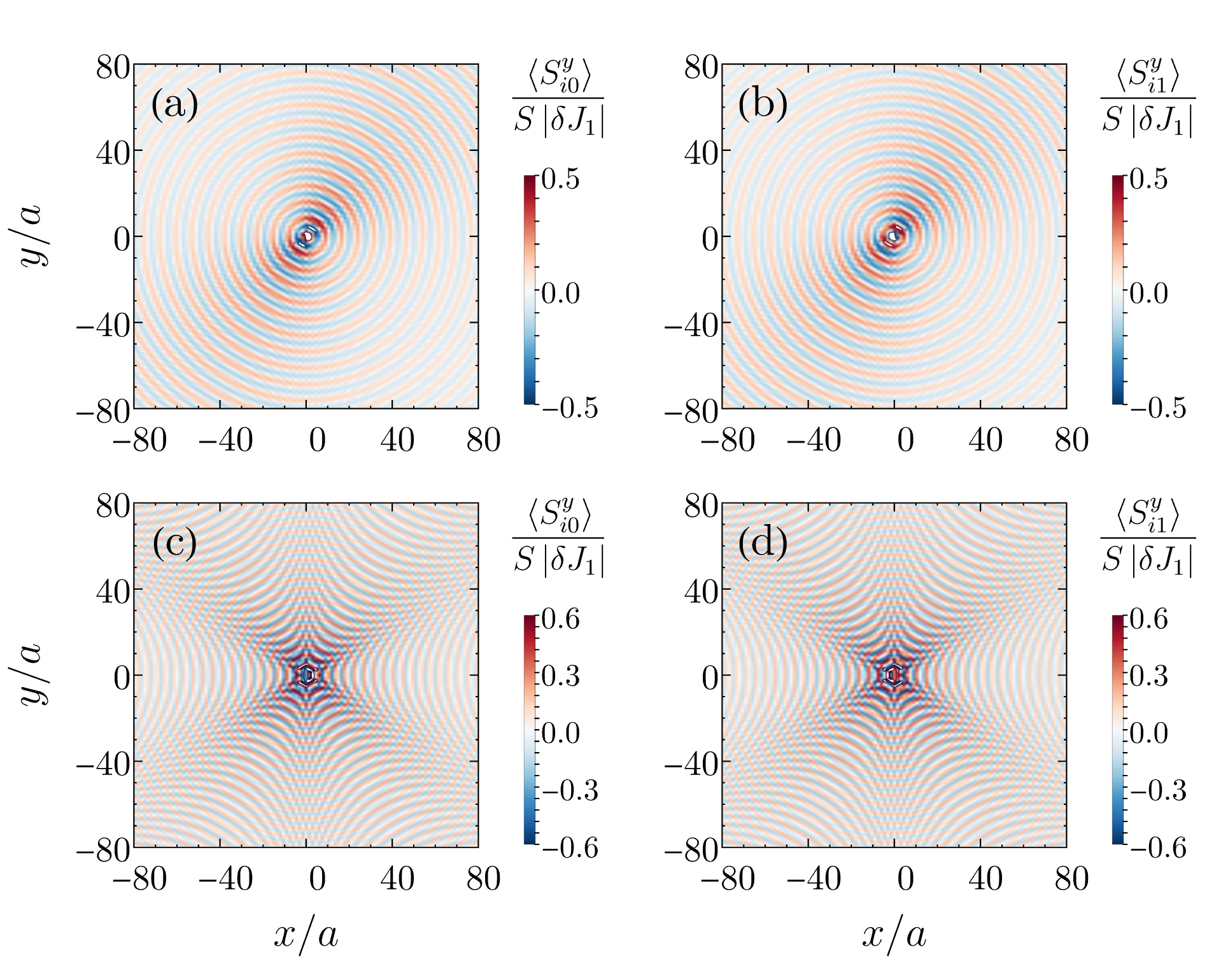}
\caption{
Friedel-like oscillations in the out-of-plane texture $\Braket{S_{i\mu}^y}/\left( S \absval{\delta J_1} \right)$ due to a weak NN bond defect in a system with (a,b) $\Jrat = 0.2$ and (c,d) $\Jrat = 0.3$, obtained from the calculation outlined in Appendix~\ref{app:linresp}. The left (right) column displays results for the $\mu=0$ $\left(\mu=1\right)$ sublattice, respectively. In all cases, the impurity is located at the origin and the data were obtained by performing the integral in Eq.~\eqref{eq:Ispc} numerically.
}
\label{fig:friedel}
\end{figure}

Using the same linear response procedure as outlined in Sec.~\ref{subsec:svac}, we computed the spin texture that an arbitrary impurity generates on top of a spiral state in the ground-state manifold. Upon doing so (see Appendix~\ref{app:linresp} for details), we were able to show that the presence of zero modes along the spiral contour generally disrupts the coplanarity of the spin configuration. As illustrated in Fig.~\ref{fig:friedel}, the texture acquires an out-of-plane component $\Braket{S_{i\mu}^y}$ which oscillates as it decays and, for this reason, is reminiscent of Friedel oscillations in an electron gas \cite{Friedel1958,FetterWalecka}.
If we denote the polar coordinates of unit cell $i$ as $\left(R,\theta\right)$, then
\begin{align}
	\Braket{S_{i\mu}^y} = \frac{S}{\sqrt{R}} \sum_{\phi_0 \in \Gamma_{\theta}}
	\!\! A_{\mu\theta}\left(\phi_0\right) \cos \left[ \mathbf{k}_{\alpha} \! \left(\phi_0\right) \! \cdot \mathbf{R}_i + \xi_{\mu\theta} \! \left(\phi_0\right) \right],
	\label{eq:asymptoop}
\end{align}
at large distances $R \gg 2\pi/Q$ from the impurity. Here, $\mathbf{k}_{\Jrat} : \left[0,\phi_{\mathrm{max}}\right[ \to \mathbb{R}^2$ is a function parametrizing the spiral contour, such that $\phi_{\mathrm{max}} = \pi$ ($2\pi$) in the SSL$_1$ (SSL$_2$) regime. The sum runs over the set $\Gamma_\theta$ of points $\phi_0$  for which the projection $\mathbf{k}_{\Jrat} \left(\phi_0\right) \cdot \mathbf{R}_i$ is stationary. Together, the amplitude $A_{\mu\theta}\left(\phi\right)$ and the phase shift $\xi_{\mu\theta}\left(\phi\right)$, which depend on $\QQ$ and whose precise definitions are given in Eq.~\eqref{eq:oopasympgen}, encode all the impurity-specific information. In the limit of small $\Jrat - 1/6$, where the spiral contour is approximately a circle of radius $k_\alpha$, Eq.~\eqref{eq:asymptoop} simplifies to
\begin{align}
	\Braket{S_{i\mu}^y} = \frac{S}{\sqrt{R}} A_{\mu} \cos \left( k_{\alpha} R + \xi_{\mu} \right),
\end{align}
which only carries a sublattice-dependent amplitude $A_{\mu}$ and phase shift $\xi_{\mu}$.

We have therefore shown that, whenever an impurity distorts a selected state in the spiral contour, it gives rise to Friedel-like oscillations in the out-of-plane component of the spin texture. Similarly to the electron gas \cite{Lounis2011}, the period of these oscillations in the asymptotic limit $R\to\infty$ is dictated by a discrete set of momenta on a surface of gapless modes. The degree of anisotropy of the corresponding ``Bose surface'' (i.e., the spiral contour) is directly reflected in the texture, as one can see by comparing the top and bottom rows of Fig.~\ref{fig:friedel}. For (a) and (b), the spiral contour is approximately circular and the resulting pattern is almost perfectly isotropic. In contrast, (c) and (d) display a nontrivial angular dependence because they correspond to a larger value of $\Jrat$, which, in the SSL$_1$ regime, accentuates the anisotropy of the spiral contour.
A close inspection of Fig.~\ref{fig:friedel} also reveals that the out-of-plane component of the texture has a staggered sublattice structure, i.e., spins in the same unit cell cant in opposite out-of-plane directions. This particular feature, however, is enforced by the sign of $J_1$; a  ferromagnetic $J_1$ leads to a uniform sublattice structure.
As we show in Appendix~\ref{subsec:Htex}, such a sublattice compensation allows the out-of-plane component of the texture to emerge at no energy cost (within linear response theory) and thereby relieve the frustration within the original ordering plane.

One aspect in which the present Friedel-like oscillations differ significantly from their analogue in the electron gas is their spatial decay rate. The $1/\sqrt{R}$ dependence in Eq.~\eqref{eq:asymptoop} is far slower than the $1/R^2$ power law expected for a two-dimensional electron gas \cite{Lau1978} or the $1/R^3$ behavior that applies specifically to graphene \cite{Cheianov2006}.
In a three-dimensional model, where the dimension $d$ of the spiral surface may be either 1 or 2 \cite{Yao2021}, the same methods we employed in Appendix~\ref{app:linresp} indicate that the texture decays as $R^{-d/2}$. This is again slower than the well-known $1/R^3$ power law in the three-dimensional electron gas \cite{Friedel1958,FetterWalecka}.

We note that the existence of an impurity-induced out-of-plane texture in a system with a spiral ground-state manifold was previously observed in numerical simulations on a diamond-lattice antiferromagnet with antisite disorder \cite{Savary2011}. However, the authors did not recognize the generality of this effect across the SSL regime, nor did they identify its long-range and oscillating nature.
Our results suggest that searching for such Friedel-like oscillations provides additional means to characterize spiral manifolds, complementary to neutron scattering, in the effort to identify new SSL candidates.

On the other hand, the in-plane component of the textures depends more sensitively on the microscopics of the impurity and on the ordering wave vector $\mathbf{Q}$ it selects through ObQD. In Appendix~\ref{app:linresp}, we provide a detailed compilation of linear-response results for $\Jrat = 0.2$. Our analysis shows that either type of bond defect (weak NN or strong NNN) produces a texture with $p$-wave-like symmetry (in the sense that it has a nodal line) and a remarkably slow power-law decay $r^{-\gamma}$, with $\gamma \approx 0.5$, for distances $r \gg 2\pi/Q$. Upon varying $\Jrat$, we found that the same qualitative behavior holds across the entire SSL$_1$ and over a wide range of the SSL$_2$ regime. Overall, the textures only become weaker as one approaches the second Lifshitz point $\Jrat = 1/2$.

The case of a single vacancy is perhaps even more striking. When $\Jrat = 0.2$, the texture displays $s$-wave-like symmetry (no nodal lines) and decays more slowly than \emph{any} power law. However, once $\Jrat > \Jrat^{*}$ and the outcome of ObQD changes (see Table~\ref{tab:vacs}), the texture acquires two orthogonal nodal lines ($d$-wave-like) and decays as a power law with exponent $\gamma \approx 1.5$ at large distances. Similarly to the case of bond defects, varying $\Jrat > \Jrat^{*}$ leads to a quantitative change in the strength of the texture, which decreases toward $\Jrat = 1/2$.
The recurrence of this observation may be related to $\Jrat$-dependence of the spin stiffness \cite{Savary2011}, which vanishes at the first Lifshitz point $\Jrat = 1/6$ and might reach a maximum at $\Jrat  = 1/2$.

To put the previous results into perspective, it is instructive to compare them to the structure of impurity-induced textures in another frustrated system: the NN triangular-lattice Heisenberg antiferromagnet.
Past studies have shown that, in this setting, a vacancy gives rise to an octupolar texture \cite{Wollny2011}, which has $f$-wave symmetry and decays as $1/r^3$, whereas a bond defect induces a dipolar texture \cite{Utesov2015}, which displays $p$-wave symmetry and decays as $1/r$. In the latter case, one can rigorously prove that any nonzero concentration of bond defects destroys LRO \cite{Dey2020}.
Although we refrain from pursuing a similar proof here due to the lower symmetry of our model and the higher complexity of the spin spirals, it is likely that the far-reaching textures of the SSL are equally detrimental to the stability of LRO. This applies in particular to the textures induced by bond defects, which have a similar angular symmetry as in the triangular lattice, but decay even more slowly. Such a destruction of LRO is consistent with the results in Sec.~\ref{sec:mip} below.

Given that the model we consider here reduces to two decoupled triangular-lattice antiferromagnets in the limit $\Jrat \to \infty$, we can expect consistency with the results of Refs.~\onlinecite{Utesov2015}, \onlinecite{Dey2020}, and \onlinecite{Wollny2011} in this limit. Indeed, our textures obtained for large $\Jrat$ (see Appendix~\ref{subsec:casetex}) show that the impurity behavior expected for the triangular lattice emerges at intermediate distances and crosses over to that of a SSL at a distance $R_{\mathrm{cross}} \left(\Jrat\right)$, which increases with $\Jrat$ and diverges as $\Jrat \to \infty$.

In broader terms, the above results underscore the fact that SSLs are, due to the abundance of low-energy modes, exceptionally soft systems, which are extremely susceptible to perturbations. Thus, taking the point of view of a ``swiss cheese model'' \cite{Savary2011}, where different impurities are regarded as independent, indiscriminately of the type of defect and the ratio $\Jrat$, may be dangerous even at small defect concentrations.

%%%%%%%%%%%%%%%%%%%%%%%%%%%%%%%%%%%%%%%%%%%%%%%%%%%%%%%%%%%%%%%%%%%%%%%
%%%%%%%%%%%%%%%%%%%%%%%%%%%%%%%%%%%%%%%%%%%%%%%%%%%%%%%%%%%%%%%%%%%%%%%
%%%%%%%%%%%%%%%%%%%%%%%%%%%%%%%%%%%%%%%%%%%%%%%%%%%%%%%%%%%%%%%%%%%%%%%

\section{Finite impurity concentration}
\label{sec:mip}

The wealth of different ObQD phenomena we encountered in Sec.~\ref{sec:sip}, along with the possibility that single impurities induce slowly decaying spin textures, implies that the response of a SSL to a finite concentration of impurities is rich and highly nontrivial. Here we explore this in three different settings.
In Sec.~\ref{subsec:sdmip}, we investigate whether the presence of multiple bond defects of the same type (NN or NNN, strong or weak) and orientation generically reinforces ordering tendencies, or if impurity-induced textures can destabilize LRO even at small to moderate disorder concentrations.
Next, in Sec.~\ref{subsec:admip}, we analyze a more realistic situation, where impurities of the same type are allowed to have different orientations. As we noted at the end of Sec.~\ref{subsec:sbd}, this gives rise to a competition between incompatible selection mechanisms.
Finally, Sec.~\ref{subsec:vacmip} discusses the case of a randomly distributed vacancies.

While our numerical results are specifically for $\Jrat = 0.2$, they will illustrate principles that are valid throughout the entire SSL regime and which bear relevance to other models hosting SSLs.

\begin{figure*}
\includegraphics[width=\textwidth]{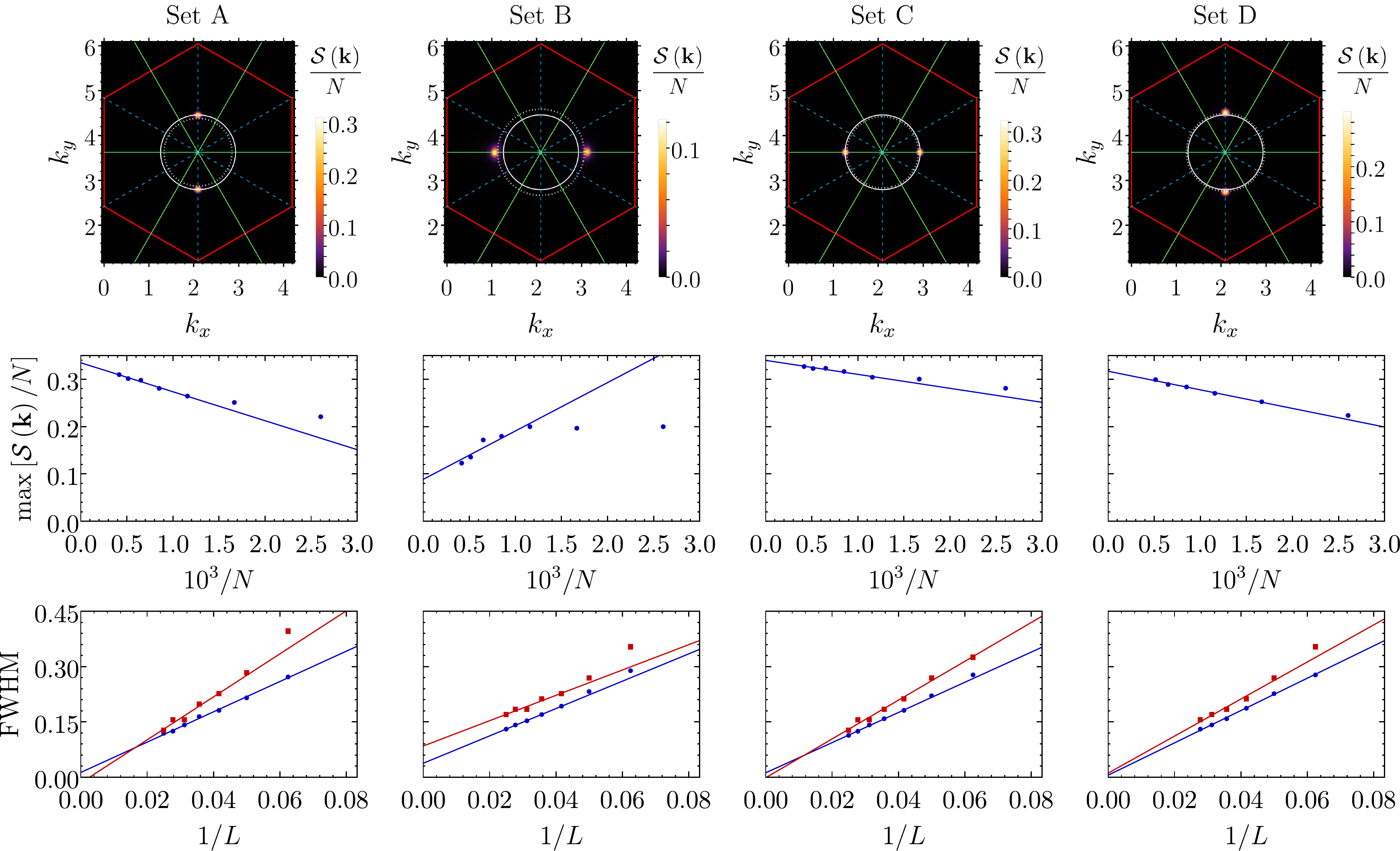}
\caption{
Numerical results for the four different parameter sets listed in Table ~\ref{tab:sbd}.
Top: Disorder-averaged static structure factors for systems of linear size $L=40$. The elements in the plots are the same as those in Fig.~\ref{fig:cleanzigzag}(b), except for the dotted white lines. These represent the spiral contour for the average ratio $\overline{\Jrat}$ in each parameter set.
Middle and bottom: Finite-size scaling of the normalized height and FWHM of the Bragg peaks. The linear fits shown discarded the data points corresponding to the two smallest system sizes.
Magnetic LRO is seen for sets A and C, but not for set B. For set D see text.
}
\label{fig:sdpanel}
\end{figure*}

\subsection{Bond defects of the same type and orientation}
\label{subsec:sdmip}

Using the energy minimization algorithm described in Appendix~\ref{app:algorithm}, we searched for the lowest-energy states of systems of linear size $L$ at a fixed ratio $p=10\%$ of the number of bond defects to the total number of sites. All bond defects were assigned the same deviations ($\delta J_1$ or $\delta J_2$) from reference values $J_1=1.0$ and $J_2=0.2$. From this, one can estimate an average coupling-constant ratio
\begin{equation}
	\overline{\Jrat} =
	\begin{cases}
		\Jrat \left( 1 - \frac{2p}{3} \frac{\delta J_1/J_1}{1 + \delta J_1/J_1} \right) & \text{for NN defects}\\
		\Jrat + \frac{p}{3} \frac{\delta J_2}{J_1} & \text{for NNN defects.}
	\end{cases}
	\label{eq:alphabar}
\end{equation}
For each disorder realization, we randomly distributed defects with the same orientation over the lattice and, after reaching converged solutions for $N_{\mathrm{init}}$ different initializations, selected the spin configuration with the minimal energy to compute the static structure factor $\mathcal{S} \left(\mathbf{k}\right)$. Finally, we averaged this quantity over $N_{\mathrm{dis}}$ different disorder realizations. The relevant parameter sets for this section are summarized in Table~\ref{tab:sdbd}, while the corresponding results for the disorder-averaged structure factor are shown in Fig.~\ref{fig:sdpanel}.

\begin{table}
	\centering
	\newcolumntype{C}[1]{>{\centering \arraybackslash}m{#1}}
	\caption{Parameters used in numerical simulations where defects were randomly distributed over a fraction $p = 10\%$ of bonds with the same direction. $\overline{\Jrat}$ refers to the average ratio defined in Eq.~\eqref{eq:alphabar}.}
	\begin{tabular}{
			%C{0.1\columnwidth} C{0.1\columnwidth} C{0.1\columnwidth} C{0.1\columnwidth} C{0.1\columnwidth} C{0.1\columnwidth} C{0.1\columnwidth} C{0.1\columnwidth}
			*{8}{C{0.11\columnwidth}}
		}
		\hline
		\hline
		\multicolumn{4}{c}{} & \multicolumn{2}{c}{$L\leq28$} & \multicolumn{2}{c}{$L>28$} \\
		Set & $\delta J_1$ & $\delta J_2$ & $\overline{\Jrat}$ & $\ndis$ & $\ninit$ & $\ndis$ & $\ninit$ \\
		\hline
		A & 1.0 & 0 & $0.19\overline{3}$ & 20 & 120 & 20 & 120 \\
		B & -0.5 & 0 & $0.21\overline{3}$ & 20 & 120 & 40 & 1000 \\
		C & 0 & -0.1 & $0.19\overline{6}$ & 40 & 150 & 40 & 150 \\
		D & 0 & 0.1 & $0.20\overline{3}$ & 40 & 150 & 40 & 150  \\
		\hline	
		\hline
	\end{tabular}
	\label{tab:sdbd}
\end{table}

According to our analysis in Sec.~\ref{subsec:sbd}, the cases covered by sets A (strong NN bonds) and C (weak NNN bonds) are special in that a single impurity does not distort the selected spin spirals in the thermodynamic limit. Based on this fact, we argued that an arbitrary concentration of defects of either kind stabilizes LRO with one of two states contained in the spiral contour of the clean system. This expectation is fully consistent with the numerical results presented in Fig.~\ref{fig:sdpanel}. In both cases, we see a pair of Bragg peaks that remain pinned to the $\Jrat=0.2$ spiral contour and lie along directions that are compatible with Table~\ref{tab:sbd}. Furthermore, the finite-size extrapolations of the height and width of the Bragg peaks show the correct behavior for systems with LRO.

In contrast, sets B (weak NN bonds) and D (strong NNN bonds) refer to cases where single impurities induce spin textures that decay slowly in space [see Sec.~\ref{subsec:textures}]. As such, it is conceivable that, even at small impurity concentrations, the interference between textures destroys LRO \cite{Dey2020}.
The data in Fig.~\ref{fig:sdpanel} shows that this happens for set B. Indeed, while Bragg peaks appear along the direction predicted in Table~\ref{tab:sbd}, they have drifted toward the reaveraged spiral contour (dotted white line) and seem broader than their counterparts in sets A and C. This is confirmed by the finite-size analysis, which reveals that the peak retains a non-zero width in the thermodynamic limit. Thus, the correlation length remains finite and the system only develops short-range magnetic order.
For set D we once again see Bragg peaks that are along the expected direction and detach from the $\Jrat=0.2$ spiral contour due to impurity-induced textures. However, there are no clear indications of the destruction of LRO up to the system sizes we could access in our numerics. We suspect that this is because the disorder, $\delta J_2=0.1$, is overall weak.

\subsection{Bond defects of random orientation: Emergent glassiness from competing ObQD mechanisms}
\label{subsec:admip}

\begin{figure*}
	\includegraphics[width=\textwidth]{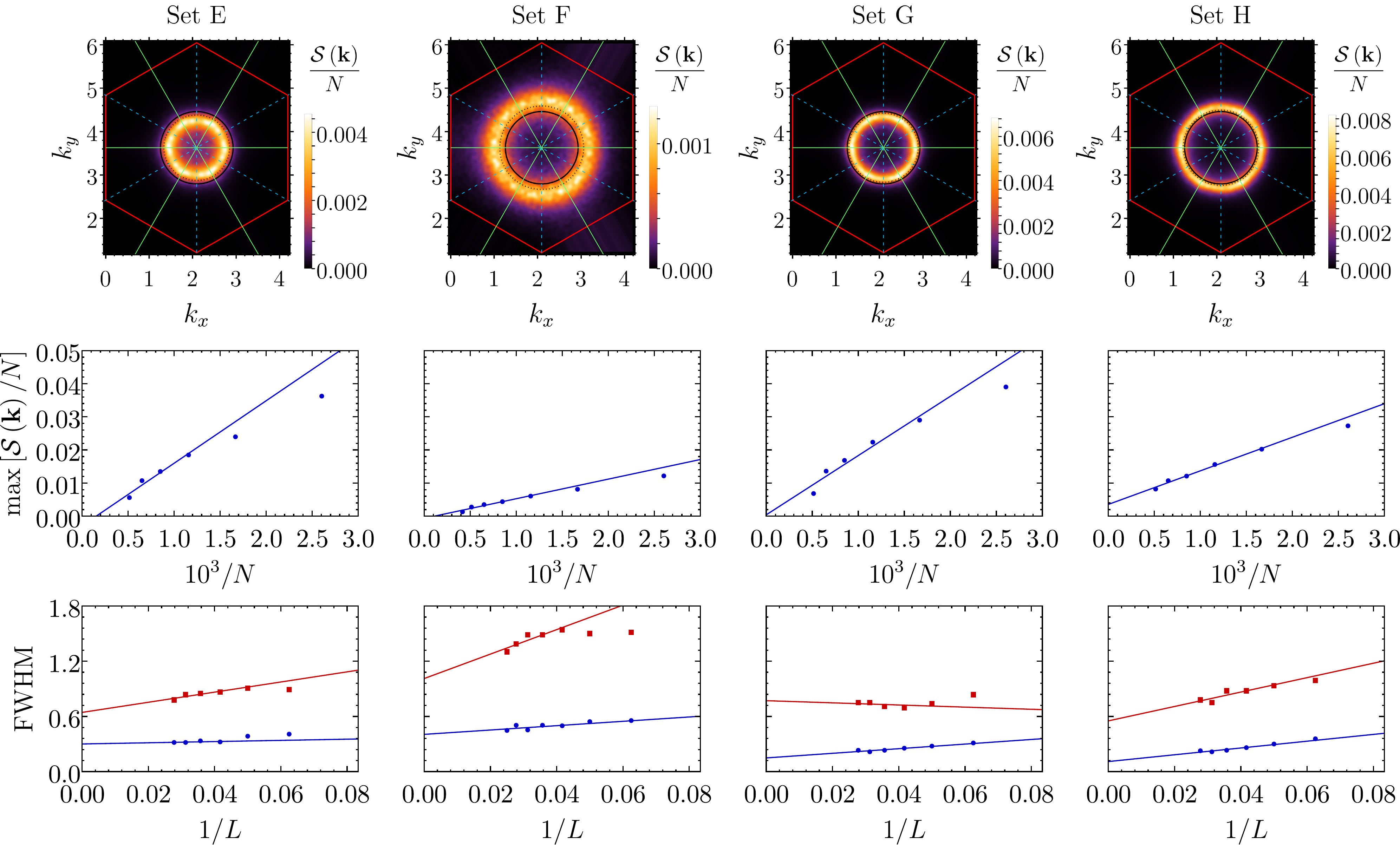}
	\caption{Same as Fig.~\ref{fig:sdpanel}, but for the parameter sets listed in Table ~\ref{tab:abd}. The disorder-averaged structure factors shown here are symmetrized by additionally averaging over $2\pi/3$ rotations. There is no magnetic LRO in any of the cases.}
	\label{fig:adpanel}
\end{figure*}

As a next step, we allowed the bond defects to have random orientations as well, while keeping the ratio $p=10\%$. For each disorder realization with NN (NNN) bond defects, the deviations $\delta J_{1,ij}$ ($\delta J_{2,ij\mu}$) were drawn from a Gaussian distribution with mean $\overline{\delta J_1}$ ($\overline{\delta J_2}$) and standard deviation $\sigma_{\delta J_1}$ ($\sigma_{\delta J_2}$). With this, we computed the disorder-averaged structure factor as before and symmetrized the result with respect to $2\pi/3$ rotations to restore the C$_3$ symmetry of the lattice.

\begin{table}
	\centering
	\newcolumntype{C}[1]{>{\centering \arraybackslash}m{#1}}
	\caption{Parameters used in numerical simulations where defects were randomly distributed over a fraction $p = 10\%$ of either NN or NNN bonds. We set $\left( \ndis, \ninit \right) =\left( 100, 500 \right)$ for $L \leq 28$ and $\left( \ndis, \ninit \right) =\left( 50, 1000 \right)100$ for $L>28$.
	}
	\begin{tabular}{ *{5}{C{0.18\columnwidth}} }
		\hline
		\hline
		\vspace{0.1cm}
		Set & $\overline{\delta J_1}$ & $\sigma_{\delta J_1}$ & $\overline{\delta J_2}$ & $\sigma_{\delta J_2}$ \\
		\hline
		E & 1.0 & 0.2 & 0 & 0 \\
		F & -0.5 & 0.2 & 0 & 0 \\
		G & 0 & 0 & -0.1 & 0.04 \\
		H & 0 & 0 & 0.1 & 0.04 \\
		\hline	
		\hline
	\end{tabular}
	\label{tab:abd}
\end{table}

Figure~\ref{fig:adpanel} shows a compilation of results for the parameter sets listed in Table~\ref{tab:abd}. From the finite-size extrapolations to the $L\to\infty$ limit, one sees that all four cases clearly lack long-range magnetic order. Their diffuse structure factors display no discernible Bragg peaks, and are much more reminiscent of glassy behavior than of short-range order found for set B (see Fig.~\ref{fig:sdpanel}). Given that the parameter sets E-H have the same concentration of defects as their counterparts in A-D, we conclude that the coexistence of the incompatible ObQD mechanisms promoted by bond defects with different orientations strongly suppresses magnetic order in favor of a spin glass phase.

\begin{figure}
	\centerline{
		\includegraphics[width=\columnwidth]{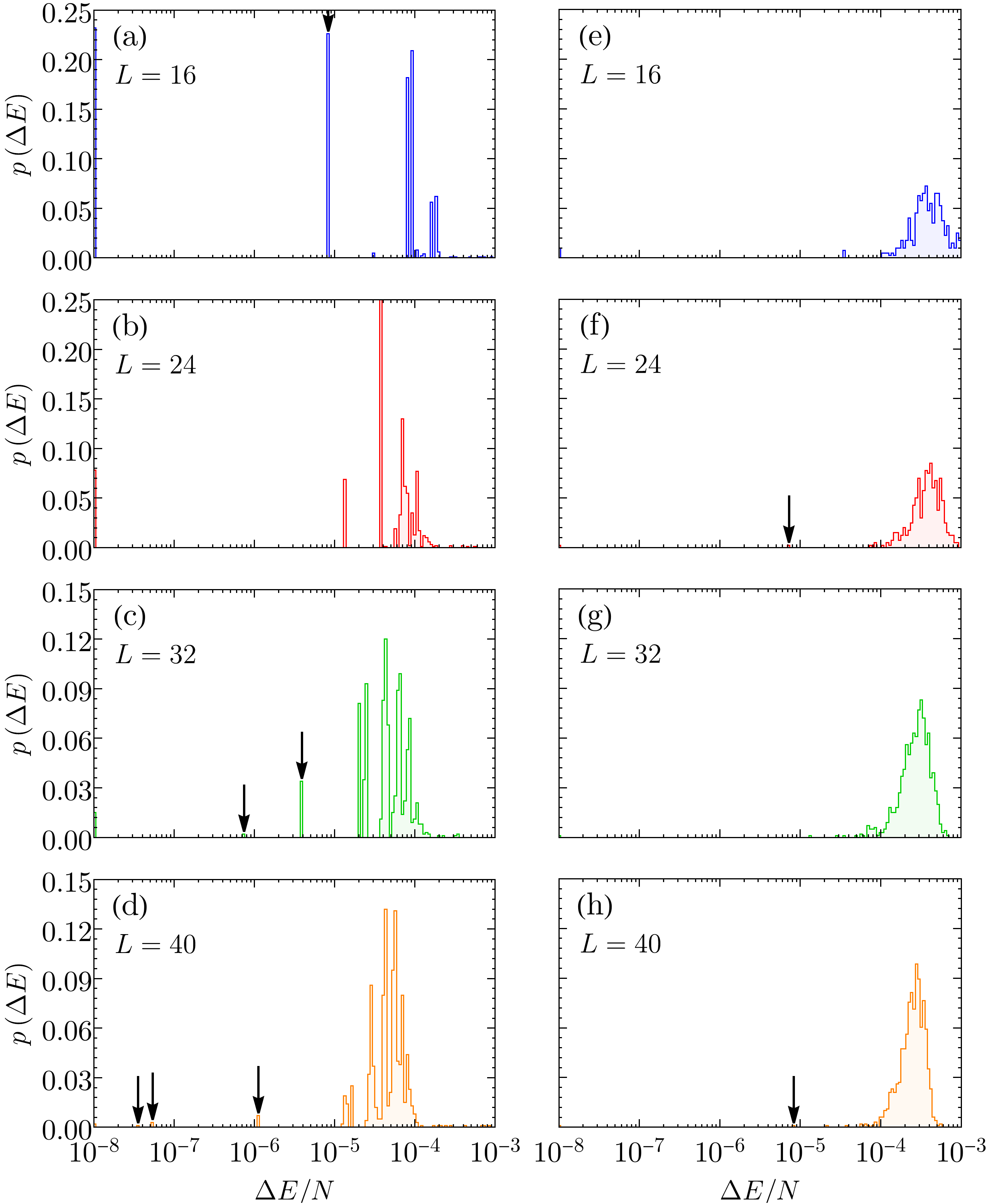}
	}
	\caption{Histograms of the energy per site, $E/N$, of converged solutions of the classical minimization algorithm. The horizontal axis represents deviations relative to the lowest energy, $\Delta E/N = \max\left[ 10^{-8}, \left(E - E_\mathrm{min}\right)/N \right]$, while the vertical axis corresponds to the frequency with which solutions in a given energy window were obtained after $10^3$ different initializations. The arrows highlight the positions of the peaks with $\Delta E/N < 10^{-5}$. These histograms map out the local minima of the energy landscape for finite clusters of (a-d) the clean system and (e-f) single disorder realizations of set F. In both cases, the clusters had open zigzag boundary conditions.}
	\label{fig:histograms}
\end{figure}

To further substantiate this statement, we exploited the fact that our iterative minimization algorithm samples different local minima in configuration space to gain information on the energy landscapes. More specifically, we tracked how the distribution of energies of converged solutions evolves with increasing $L$ for a clean system, and looked for qualitative differences in the results for parameter sets E-H. Figure~\ref{fig:histograms} illustrates this comparison for set F.

In the absence of disorder [Fig.~\ref{fig:histograms}(a-d)], we see that the excited states gradually move toward lower energy differences $\Delta E/N$ as $L$ grows and the boundary effects wear off. This is reflected not only in the appearance of isolated peaks at lower energies, but also in a noticeable shift of the dense group of states in the range $10^{-5} \lesssim \Delta E /N \lesssim 10^{-4}$. Although our simulations are limited to $L\le40$, it is clear that this trend is responsible for restoring the degeneracy of the spiral contour in the thermodynamic limit and must persist indefinitely. The low-lying states can thus be viewed as finite-size descendants of coplanar spiral states with values of $\phiQ$ that are slightly different from those selected by the boundaries.

However, a markedly different behavior emerges once disorder comes into play [Fig.~\ref{fig:histograms}(e-h)]. For a given system size $L$, the distribution of energies in the window $10^{-5} \lesssim \Delta E /N \lesssim 10^{-3}$ becomes both denser and smoother, while the frequency with which the lowest energy is reached (leftmost bin in the histograms) decreases significantly. This indicates that the energy landscape develops profuse barriers and local minima, which are, along with the diffuse structure factors, hallmarks of glassiness \cite{Fischer1991}. As $L$ increases, these features persist but we see no clear shift of energy levels toward $\Delta E = 0$.

We have also simulated other disorder distributions and found qualitatively similar results (not shown), re-enforcing the notion that random bond disorder efficiently destroy any selection of ordered magnetic states.

\subsection{Randomly distributed vacancies}
\label{subsec:vacmip}

Finally, we investigated the zero-temperature physics at a fixed concentration $p = N_{\mathrm{vac}}/N$ of vacancies. In light of Sec.~\ref{subsec:svac}, one could expect an infinitesimal $p$ to induce magnetic order by spontaneously selecting one of the six states $\phiQ = n\pi/3$, as listed in Table~\ref{tab:vacs}. However, any slight increase in $p$ simultaneously triggers up to \emph{four} different effects that conspire against the stability of this ordered phase.
The first is the interference between different vacancy-induced textures, which have long-range components both in and out of the original ordering plane.
The three remaining effects relate to the non-vanishing probability to form vacancy pairs. Depending on the $\Jrat$, one of these defects may favor configurations which are incompatible with the states selected by a single vacancy (see Table~\ref{tab:vacs}). Furthermore, it induces a texture with its own particular symmetries. Combined with the incompatibility between the ObQD mechanisms of vacancy pairs with different orientations, this leads to an enormous amount of frustration.

The fact that all of these mechanisms are active in the interval $1/6 < \Jrat \lesssim 0.2148$ where vacancy-induced textures are extremely long-ranged, indicates that magnetic order is particularly fragile to vacancies in the proximity of the first Lifshitz point. This is fully consistent with previous results obtained for diamond-lattice antiferromagnets \cite{Savary2011}, while offering further insight as to why that is true and reinforcing the expectation that the same conclusion holds for different SSL models.

\begin{figure}
	\includegraphics[width=\columnwidth]{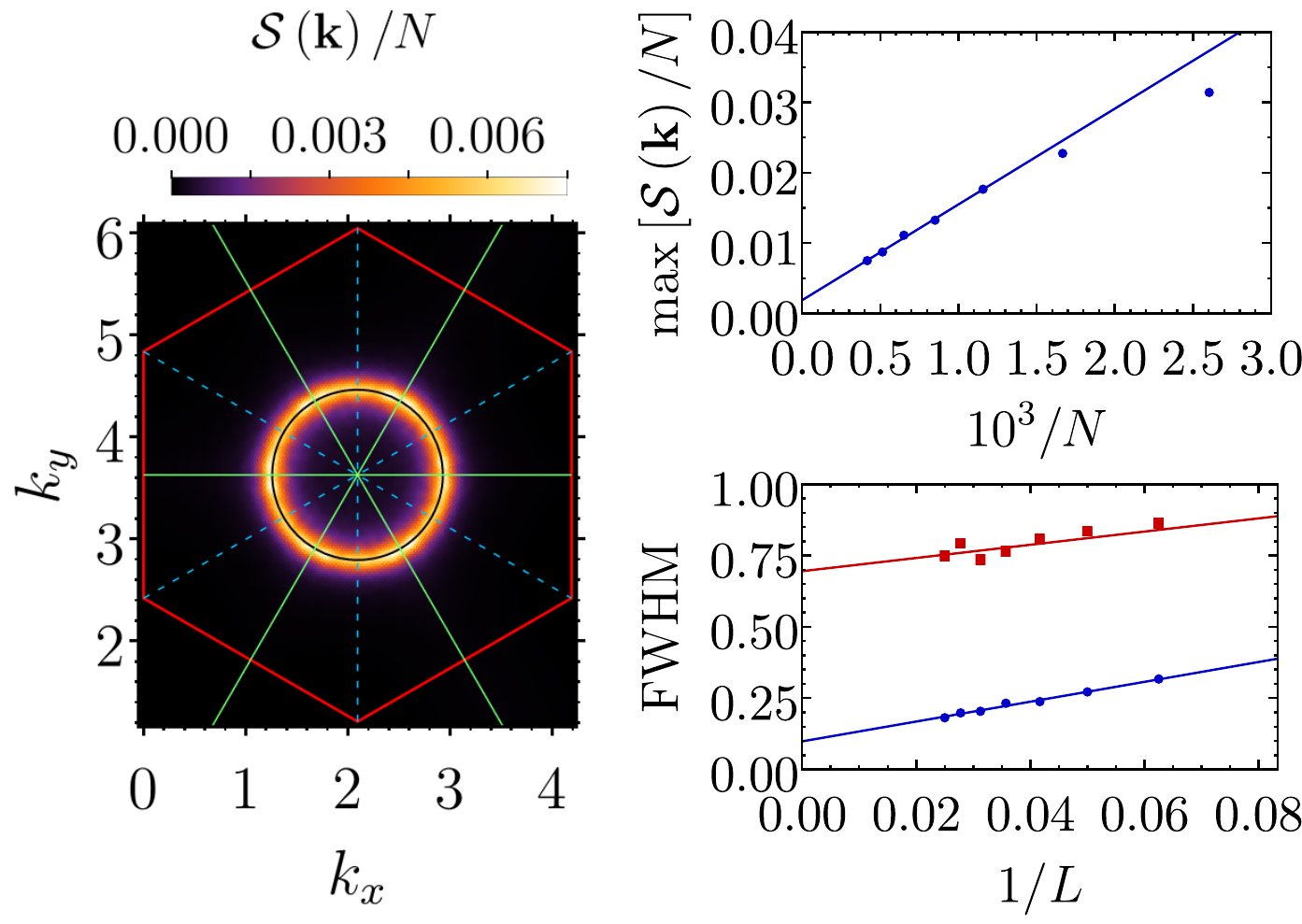}
	\caption{Same as Fig.~\ref{fig:sdpanel}, but for systems with $p=5\%$ of vacancies and a maximum system size of $L=40$. The disorder-averaged structure factors shown here are symmetrized by additionally averaging over $2\pi/3$ rotations.}
	\label{fig:vacpanel}
\end{figure}

We performed numerical simulations on systems of linear sizes $L \le 40$ with $\left(N_{\mathrm{dis}}, N_{\mathrm{init}}\right) = \left(100,500\right)$ for $L \le 28$ and $\left(N_{\mathrm{dis}}, N_{\mathrm{init}}\right) =  \left(50,1000\right)$ for $L > 28$, computed disorder-averaged structure factors and symmetrized them with respect to $2\pi/3$ rotations. As shown in Fig.~\ref{fig:vacpanel}, a system with $p = 5\%$ vacancies decidedly lacks LRO and presents the same glassy features as above; results for $p=10\%$ are similar. Given that simulations become increasingly expensive and unreliable for small $p$, we are not able to decide whether glassiness emerges at infinitesimal dilution or at a finite critical $p_c$; we leave it as an open problem for future research.

%%%%%%%%%%%%%%%%%%%%%%%%%%%%%%%%%%%%%%%%%%%%%%%%%%%%%%%%%%%%%%%%%%%%%%%
%%%%%%%%%%%%%%%%%%%%%%%%%%%%%%%%%%%%%%%%%%%%%%%%%%%%%%%%%%%%%%%%%%%%%%%
%%%%%%%%%%%%%%%%%%%%%%%%%%%%%%%%%%%%%%%%%%%%%%%%%%%%%%%%%%%%%%%%%%%%%%%

\section{Effects of thermal fluctuations and additional couplings}
\label{sec:disclaimers}

In this section we discuss physics aspects not covered by our numerical results, such as the effect of thermal fluctuations in $D=2$, the order of phase transitions, and generalizations to $D=3$, also connecting to results from the literature.

\subsection{$J_1$-$J_2$ honeycomb model}

At finite low temperatures $T>0$ and in the absence of disorder, the accidental degeneracy of the spiral contour is lifted, such that only a discrete set of six ground states remains. While the Hohenberg-Mermin-Wagner theorem \cite{Mermin1966,Hohenberg1967} forbids long-range spiral order at any $T>0$, the system can still spontaneously break the threefold rotational symmetry around a lattice site by realizing short-range-ordered patches with a single pair $\left\{ \QQ, -\QQ \right\}$ of opposite wave vectors. Importantly, the symmetry between $\pm \QQ$ cannot be broken spontaneosly because the corresponding spin spirals are related by a global spin rotation. This nematic phase has indeed been observed in previous numerical studies \cite{Mulder2010,Okumura2010}, and has a similar incarnation in the $J_1$-$J_2$ Heisenberg model on the square lattice \cite{Chandra1990,Weber2003}. Its thermal transition into the paramagnetic phase is equivalent to the three-state Potts transition in $D=2$, and is therefore continuous \cite{Baxter1973}.

The presence of impurities leads to a nontrivial interplay between ObQD and thermal order by disorder. Quenched disorder acts as a set of random spatial anisotropies, each of which favors a particular ordering ``axis'', i.e., a pair of wave vectors $\left\{ \QQ, -\QQ \right\}$ on the spiral contour (see Sec.~\ref{sec:sip}). Regardless of whether the system is in the SSL$_1$ or SSL$_2$ regime, bond disorder generally sets \emph{six} such axes. By symmetry, three of these must match the wave vectors selected by thermal fluctuations; the remaining three can be neglected as long as the disorder strength is small compared to the largest gap in the spiral contour. We thus infer that the effective low-energy description of the system corresponds to a three-state Potts model supplemented with random fields \cite{Mulder2010}. According to the Imry-Ma criterion \cite{Imry1975}, the lower critical dimension of the thermal transition between the ordered and paramagnetic phases is then $D_{c,\mathrm{RF}}^{-} = 2$. Therefore, at $T>0$, \emph{any} concentration $\Delta>0$ of bond defects destroys the long-range nematic order by inducing the proliferation of domains with different ordering axes. The resulting phase is, however, not a genuine spin glass, but rather a paramagnet, since the lower critical dimension of a Heisenberg spin-glass transition is larger than two \cite{Lee2007,Kawamura2011,Ogawa2020}, such that a glass phase only exists at $T=0$. The response of the system to disorder is thus similar to that of the $J_1$-$J_2$ Heisenberg model on the square lattice \cite{Miranda2021}; the main difference lies in the existence of a SSL regime, which one can expect to span an extended portion of the phase diagram [see Fig.~\ref{fig:pds}(a)].

The outcome is more subtle when only site disorder is present. As indicated by Table~\ref{tab:vacs}, the formation of vacancy pairs only leads to three -- as opposed to six -- inequivalent axes. Hence, the previous Imry-Ma argument only applies if these axes coincide with the wave vectors selected by thermal fluctuations. According to the results summarized in Sec.~\ref{subsec:ObTD}, this condition is violated over the majority of the SSL regime, such that nematic order can be expected to survive for small vacancy concentration.

Notably, the applicability of an Imry-Ma-type argument is not restricted to $T>0$, but in fact holds even at $T=0$ provided that the spiral degeneracy is lifted by additional terms in the Hamiltonian. At $T=0$ magnetic LRO is possible, and one such state has to break a $\mathbb{Z}_3 \otimes \mathrm{SO}\left(3\right)$ symmetry, corresponding to the direct product of a threefold lattice rotation and global spin rotation symmetries.
The presence of the discrete $\mathbb{Z}_{3}$ component implies that the low-energy theory of the system still includes a three-state Potts order parameter, which is again supplemented by disorder-induced random fields if another type of perturbation, such as a third-neighbor Heisenberg coupling, lowers the degeneracy to a discrete sixfold one. Hence, the system is equally unstable to weak bond disorder as for $T>0$ with no additional couplings. If only site disorder is present, the result depends on the sign of the supplemental coupling.

\subsection{Spiral spin liquids in three dimensions}

In three space dimensions, spiral LRO is stable at low temperatures $T$ and/or small disorder strength $\Delta$, since neither the Hohenberg-Mermin-Wagner theorem nor the Imry-Ma criterion apply. However, as suggested by Fig.~\ref{fig:pds}(b), thermal fluctuations and quenched disorder may favor different ordered phases, O$_2$ and O$_3$, which (without fine tuning) should be separated by a first-order transition. While the precise character of these phases depends on the lattice, ratios of couplings, and the type of defect present, they both spontaneously break spatial point-group symmetries in addition to the global SO(3) spin symmetry by selecting a particular ordering wave vector $\QQ$. One may thus expect their thermal transitions into the paramagnetic phase to be of the same order as a $q$-state Potts transition \cite{Henley1987}, with $q$ determined by lattice symmetries. For $\Delta = 0$ and $q>2$, this turns out to be of first order \cite{Blankschtein1984}, in agreement with Monte Carlo results for the $J_1$-$J_2$ Heisenberg model on the diamond lattice \cite{Bergman2007}. Nevertheless, the inclusion of disorder can render the phase transition continuous, which is known to happen, e.g., when the three-state Potts model is supplemented with moderate random-field disorder \cite{Blankschtein1984,Eichhorn1996}.

The increase of $\Delta$ at low $T$ eventually destroys LRO in favor of a spin glass, in what we expect to be a first-order phase transition. The freezing transition of the (Heisenberg) spin glass is, in contrast, most likely continuous. However, the nature of the latter remains a divisive issue to date with two opposing theories. One advocates that isotropic Heisenberg systems in fact display two glass transitions due to a decoupling between the scalar spin chirality and continuous spin degrees of freedom at large length scales. Upon cooling, the former would freeze before the latter, and an intermediate ``chiral glass phase" would arise above the spin glass phase \cite{Kawamura2011,Ogawa2020}. The second theory denies this spin-chirality decoupling and defends that both degrees of freedom freeze at the same temperature in what may be a Berezinskii-Kosterlitz-Thouless transition \cite{Campos2006,Lee2007,Fernandez2009}.

The fact that thermal fluctuations or additional couplings gap out the spiral contour also has implications for the Friedel-like oscillations we described in Sec.~\ref{subsec:textures} in both two and three dimensions. First, the finite energy of the excitations implies that the out-of-plane texture does not emerge for arbitrarily weak perturbations, but only beyond a finite strength of individual defects, see Appendix~\ref{app:linresp}. Second, we expect this texture to be damped and follow an exponential rather than a power-law decay. This, of course, does not preclude the experimental observation of such oscillations.

%%%%%%%%%%%%%%%%%%%%%%%%%%%%%%%%%%%%%%%%%%%%%%%%%%%%%%%%%%%%%%%%%%%%%%%
%%%%%%%%%%%%%%%%%%%%%%%%%%%%%%%%%%%%%%%%%%%%%%%%%%%%%%%%%%%%%%%%%%%%%%%
%%%%%%%%%%%%%%%%%%%%%%%%%%%%%%%%%%%%%%%%%%%%%%%%%%%%%%%%%%%%%%%%%%%%%%%

\section{Connections to experiments}
\label{sec:experiments}

Among the few known realizations of a SSL, FeCl$_3$ is the one which best fits the model we studied here. It is a van der Waals material where honeycomb planes of $S=5/2$ Fe$^{3+}$ ions are stacked with a three-layer periodicity. While the system orders in a commensurate magnetic phase below $T_{N} \sim 8$ K, a neutron scattering study reported the observation of a spiral contour for $T = 10-20$ K and attributed its existence to the competition between dominant intralayer interactions $J_1 < 0$ and $J_2 > 0$ \cite{Gao2022a}. The best fit of a minimal model to the data yielded a ratio $\Jrat = J_2/\absval{J_1} \approx 0.25$, which increases to $\Jrat \approx 0.36$ if further-neighbor couplings are included.

More recently, Ref.~\onlinecite{Cole2023} investigated how chemical substitution of Cl by Br affects the magnetic properties of FeCl$_3$ in a series of samples FeCl$_{3-x}$Br$_x$ with $0.08 \le x \le 3$. Besides changing the lattice spacing and the average ratio $\Jrat$, introducing small $x$ is expected to enhance the level of bond randomness.
By performing magnetization measurements as a function of temperature and magnetic field, Cole \textit{et al.} \cite{Cole2023} showed that even doping as small as $x = 0.08$ is sufficient to change the ground state of FeCl$_{3-x}$Br$_x$ to an antiferromagnet with alternating ferromagnetic layers, a collinear state incompatible with a spiral contour. Based on first-principles calculations, the authors associated such an abrupt transition with the proximity of FeCl$_3$ to the Lifshitz point $\Jrat = 1/6$ and a rapid decrease of the average $\Jrat$ as a function of doping. However, we believe that the previous interpretation may be incomplete.
In Ref.~\onlinecite{Gao2022a}, Gao \textit{et al.} suggested that a weak single-ion anistropy, which tends to lock an incommensurate $\QQ$ to a nearby commensurate position \cite{Lee2008}, was responsible for the mismatch between the ordering wave vectors of the $x = 0$ ordered phase and the ground state of the model that produced the best fit to their data.
By a similar token, the action of the same lock-in mechanism on a set of $\QQ$ favored by ObQD may accelerate the transition to the $x \ge 0.08$ ground state. In other words, the extreme sensitivity of FeCl$_{3-x}$Br$_x$ to doping may be a combined effect of ObQD and weak single-ion anisotropy in the proximity of the Lifshitz point $\Jrat = 1/6$.

A related honeycomb-lattice compound, ZnMnO$_{3}$, has also been proposed as a SSL candidate. According to Ref.~\onlinecite{Haraguchi2019}, this material orders below $T_{N} \approx 17.4$ K in an $S=3/2$ zigzag state, which can be described by a $J_1$-$J_2$ Heisenberg model supplemented by a weak antiferromagnetic $J_3$. The ratio of its Curie-Weiss temperature by its saturation field produces an estimate $\Jrat \sim 0.44$, which is in the correct range for the emergence of SSL behavior at intermediate temperatures. However, there has been no report to date on the observation of a spiral contour in the material. This calls for further experimental measurements of both clean and disordered samples.

Early investigations on the influence of disorder in SSLs were largely motivated by the unusual properties of CoAl$_2$O$_4$, an A-site spinel which freezes into a collinear short-ranged ordered state at low temperatures and has an estimated ratio $\Jrat \approx 0.110$ \cite{Zaharko2011,MacDougall2011,MacDougall2016}.
Previous theoretical work \cite{Savary2011} showed that such a propensity toward glassiness could be understood as a consequence of the proximity to the first Lifshitz point $\Jrat = 0.125$, where the spin stiffness in the ordered phase of a clean sample vanishes.
As mentioned in Secs.~\ref{subsec:textures} and \ref{subsec:vacmip}, our findings here are consistent with this statement and show that spin vacancies, as may arise from Co--Al site exchange in CoAl$_2$O$_4$, can lead to pronounced responses. In particular, larger disorder is likely to produce a conventional spin glass in the compound.
Returning to FeCl$_{3}$, we expect that substituting Fe by a nonmagnetic ion could similarly destabilize LRO and cause the emergence of an extended SSL or SSG regime.

Finally, as mentioned in Sec.~\ref{subsec:textures}, we expect Friedel-like oscillations in the out-of-plane component of impurity-induced textures to occur in any SSL system, both in two and three dimensions. It would be extremely interesting to search for this behavior in SSL candidates. This applies particularly to the compounds {\MSS} \cite{Gao2017}, FeCl$_3$ \cite{Gao2022a}, and {\LYO} \cite{Graham2023}, for which there is clear evidence of SSL behavior.

%%%%%%%%%%%%%%%%%%%%%%%%%%%%%%%%%%%%%%%%%%%%%%%%%%%%%%%%%%%%%%%%%%%%%%%
%%%%%%%%%%%%%%%%%%%%%%%%%%%%%%%%%%%%%%%%%%%%%%%%%%%%%%%%%%%%%%%%%%%%%%%
%%%%%%%%%%%%%%%%%%%%%%%%%%%%%%%%%%%%%%%%%%%%%%%%%%%%%%%%%%%%%%%%%%%%%%%

\section{Conclusion and outlook}
\label{sec:concl}

We have provided an in-depth investigation of disorder effects at $T=0$ in a prototypical two-dimensional model realizing a SSL. This includes the derivation of ObQD results, the characterization of impurity-induced textures, and the identification of conceptually distinct mechanisms that destabilize magnetic order to different degrees. Based on the association of one of these mechanisms with random fields, we propose that disorder can provide a route to expand the intermediate-temperature SSL regime, but also give rise to a ``spiral spin glass".

By cataloguing different ObQD results, we have identified a number of situations in which quenched disorder selects spiral states other than those favored by quantum and/or thermal fluctuations. Given that such disparities should exist in any SSL, one can expect the interplay between incompatible selection mechanisms to generate a nontrivial competition between ordered phases, as sketched in Fig.~\ref{fig:pds}(b). Similar behavior is known to occur in XY pyrochlore magnets \cite{Maryasin2014,Ross2016,Sarkar2017,Andrade2018}, and we have argued that it may well be of experimental relevance to FeCl$_{3-x}$Br$_x$ \cite{Cole2023}. A detailed theoretical investigation of disorder-induced ordered phases in three space dimensions is left for future work.

In the context of neutron scattering signatures of SSLs, i.e., the emergence of a spiral line or surface, we have found that these also arise in frozen SSG states, see Fig.~\ref{fig:pds}. Our results therefore prompt to search for glassiness, e.g., by ac susceptibility measurements, in SSL candidate compounds.

Arguably, our most surprising result is that, whenever an impurity distorts a selected spin configuration from the ground-state manifold of a SSL, the ensuing spin texture has a component out of the original ordering plane which displays Friedel-like oscillations. We believe that searching for this effect in material candidates can be highly worthwhile, as it provides a distinctive signature of the classical ground-state degeneracy and may complement neutron scattering in the task of mapping out spiral manifolds. Magnetic imaging techniques based, e.g., on transmission electron microscopy or hard-X-ray tomography may be suited for this purpose, especially in materials with ferromagnetic NN couplings.
A more fundamental, but equally interesting, question is whether these non-coplanar textures can have nontrivial topological properties. We leave this as an open topic for future research.

Finally, we recall that our investigations have been restricted to the semiclassical limit of large spins. Quantum effects can destroy both LRO and SSG phases in favor of quantum spin liquids, valence-bond glasses or other paramagnetic states dominated by quantum fluctuations.

%%%%%%%%%%%%%%%%%%%%%%%%%%%%%%%%%%%%%%%%%%%%%%%%%%%%%%%%%%%%%%%%%%%%%%%

\acknowledgments

We thank Eric C. Andrade, Bernhard Frank, and David J. Moser for helpful discussions.
P.M.C. is especially grateful to José A. Hoyos for the short series of lectures ``Der Schmutz der Schmutzphysik", taught in Dresden in the Fall of 2022.
Financial support from the Deutsche Forschungsgemeinschaft through SFB 1143 (project-id 247310070) and the W\"urzburg-Dresden Cluster of Excellence on Complexity and Topology in Quantum Matter -- \textit{ct.qmat} (EXC 2147, project-id 390858490) is gratefully acknowledged.

%%%%%%%%%%%%%%%%%%%%%%%%%%%%%%%%%%%%%%%%%%%%%%%%%%%%%%%%%%%%%%%%%%%%%%%

\appendix

\section{Energy minimization algorithm}
\label{app:algorithm}

The energy minimization algorithm we used in this paper is based on the notion of local exchange fields, which we introduced at the beginning of Sec.~\ref{sec:sip}. The algorithm starts by generating an initial spin configuration, which we constructed by adding small random in-plane and out-of-plane deviations to a coplanar spiral state with either fixed or random $\phiQ$. Next, it computes the set of $N$ local exchange fields defined by Eq.~\eqref{eq:hloc}.
The algorithm thus initiates an iterative scheme where each step consists of three actions:
\begin{enumerate}
	\itemsep=0pt
	\item Randomly select $n_\mathrm{upd} \sim 3L$ spins;
	\item Align the selected spins with their local fields;
	\item Update the subset of mean fields that were affected by the changes.
\end{enumerate}
Furthermore, once every $\tau_\mathrm{targ} \sim 10^5$ steps, we perform a ``targeted'' update, whereby action 1 in the list above is replaced by selecting the $n_{\mathrm{upd}}$ spins that are most misaligned with their mean fields.

This procedure is repeated until the maximum change in a single spin, $\max \absval{\mathbf{S}_{i\mu}^\mathrm{new} - \mathbf{S}_{i\mu}^\mathrm{old}}$, is less than a certain convergence parameter $\varepsilon_\mathrm{conv}$ for $n_\mathrm{conv}$ consecutive iterations. For most simulations, we used $\varepsilon_\mathrm{conv} = 10^{-12}$ and $n_\mathrm{conv} = 15$. Finally, we checked the global convergence of our solutions by ensuring that the angle between each spin and its local mean field did not exceed $\varepsilon_{\mathrm{conv}}$. For $L=28$, the algorithm required roughly $10^{6}$ steps to reach convergence, a number which increases to $\sim 10^{8}$ for $L=40$.

%%%%%%%%%%%%%%%%%%%%%%%%%%%%%%%%%%%%%%%%%%%%%%%%%%%%%%%%%%%%%%%%%%%%%%%
%%%%%%%%%%%%%%%%%%%%%%%%%%%%%%%%%%%%%%%%%%%%%%%%%%%%%%%%%%%%%%%%%%%%%%%
%%%%%%%%%%%%%%%%%%%%%%%%%%%%%%%%%%%%%%%%%%%%%%%%%%%%%%%%%%%%%%%%%%%%%%%

\section{Details on the boundary-selected states for clean finite systems}
\label{app:clean}

\begin{figure}
\includegraphics[width=\columnwidth]{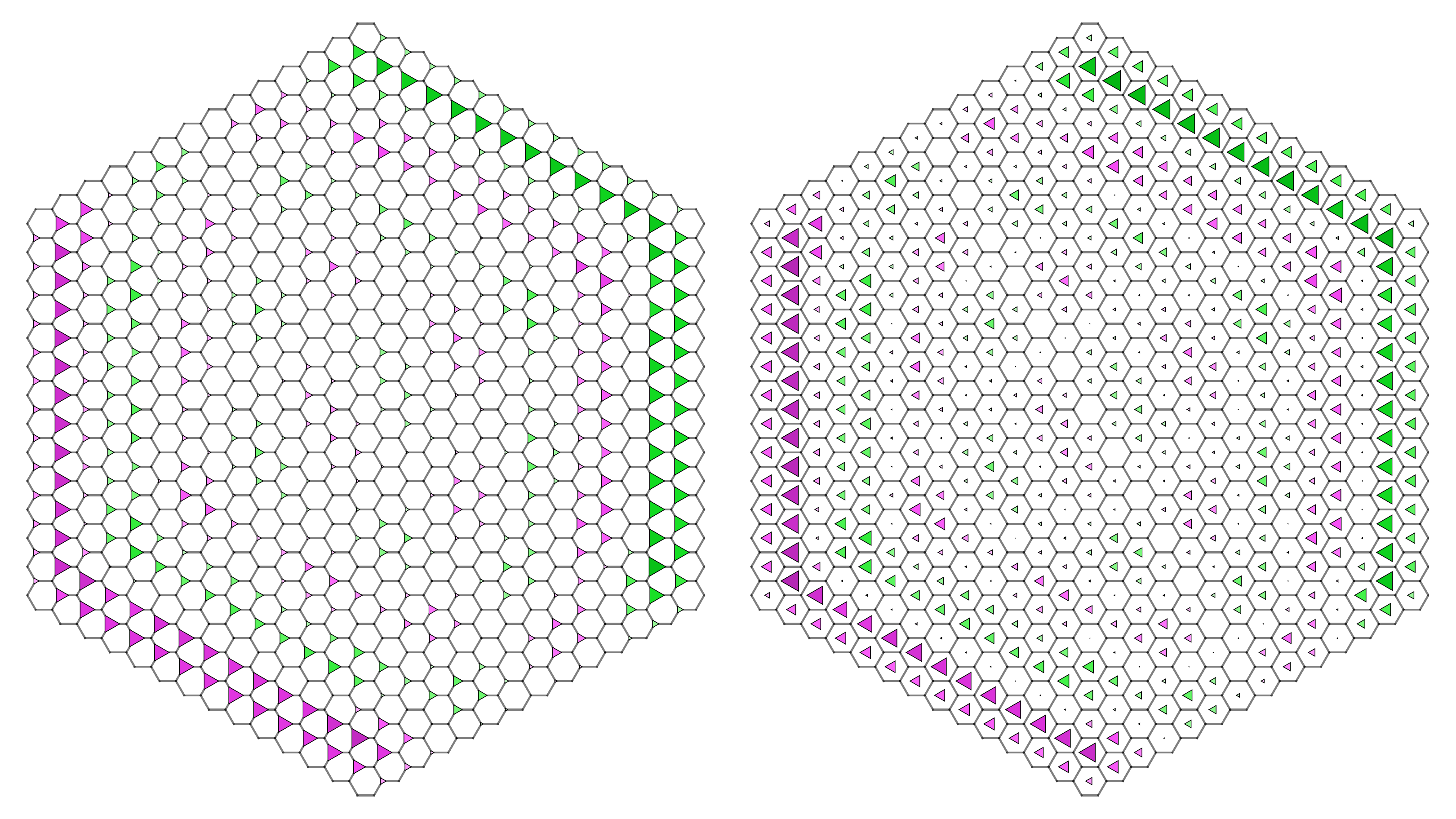}
\caption{
Illustration of scalar chiralities $\chi_{ijk}^1$ in a ground state of a clean system with $\Jrat = 0.2$, open zigzag boundary conditions, and linear system size $L = 28$. The left and right figures correspond to two different choices for triplets of sites such that $i1$, $j1$ and $k1$ are next-nearest neighbors.
		In both cases, the values of $\chi_{ijk}^1$ are represented by triangles whose vertices point toward these three sites. The size of each triangle is proportional to the absolute values of $\chi_{ijk}^1$, while the binary color code represents its sign.}
	\label{fig:chirality}
\end{figure}

In Sec.~\ref{subsec:clean}, we analyzed numerical data that led us to conclude that open zigzag boundary conditions lift the degeneracy of the spiral contour and stabilize LRO. This appendix is dedicated to providing further details on the structure of the resulting ground states and elucidating the nature of the boundary-induced selection mechanism.
To do so, we computed the scalar spin chirality
\begin{equation}
	\chi^{\mu}_ {ijk} = \mathbf{S}_{i\mu} \cdot \left(\mathbf{S}_{j\mu} \times \mathbf{S}_{k\mu} \right)
	\label{eq:chirality}
\end{equation}
for ground states on finite clusters with different sizes. As before, $\mu$ is a sublattice index, while $i$, $j$ and $k$ label the unit cells of three next-nearest-neighbor sites. Importantly, the absolute value of Eq.~\eqref{eq:chirality} is restricted to the interval $\left[0,1\right]$ and reflects the degree of non-coplanarity of a triplet of spins. Thus, it can be used to diagnose deviations from coplanar order.

Figure~\ref{fig:chirality} illustrates the spatial dependence of $\chi_{ijk}^{1}$ for one of the six ground states on an $L=28$ cluster. There, the chirality of each triplet of spins is depicted as a triangle whose vertices point in the direction of the three relevant sites. The size and color of a triangle encode the absolute value and sign of $\chi_{ijk}^{1}$, respectively.
When we compare the data in Fig.~\ref{fig:chirality} to the structure factor in Fig.~\ref{fig:cleanzigzag}(b), we verify that both results are (up to symmetry-allowed transformations) fully consistent. Indeed, the latter indicates that, in the boundary-selected configurations, spins connected by lines which are parallel to a pair of opposite edges of the cluster have the same orientation. While this statement is only strictly true in the thermodynamic limit, it explains why the chirality nearly vanishes around the top left and bottom right edges of the cluster in Fig.~\ref{fig:chirality}.

The triangles adjacent to the four remaining edges are likewise quite small, but are shielded from the bulk by walls of large triangles. This remarkable feature reflects that the finite-size ground state in fact realizes a compromise between bulk- and boundary-preferred configurations. More specifically, it spontaneously adopts the ordering wave vector favored by any of the three pairs of opposite edges and accommodates the other four edges with nearly parallel NNN spins by distorting the perfect coplanar spin spiral. Such a distortion occurs at a minimal energy cost when it rotates spins out of the ordering plane, which is why we observe the aforementioned strips of large chirality.

The comparison between the energies of non-coplanar ground states and truncated coplanar spin spiral in Fig.~\ref{fig:cleanzigzag}(a) shows that the boundaries supply a significant energetic contribution for all system sizes we accessed in our numerics. As we noted in Sec.~\ref{subsec:cleanfs}, this hinders a systematic numerical analysis of most single-impurity effects. Meaningful conclusions about impurity-induced textures, for instance, can only be drawn from clusters that are large enough for the bulk to be sufficiently insensitive to the boundary-induced textures. However, in this case, the energy landscape is already so complex that convergence to the exact ground state is not guaranteed.

%%%%%%%%%%%%%%%%%%%%%%%%%%%%%%%%%%%%%%%%%%%%%%%%%%%%%%%%%%%%%%%%%%%%%%%
%%%%%%%%%%%%%%%%%%%%%%%%%%%%%%%%%%%%%%%%%%%%%%%%%%%%%%%%%%%%%%%%%%%%%%%
%%%%%%%%%%%%%%%%%%%%%%%%%%%%%%%%%%%%%%%%%%%%%%%%%%%%%%%%%%%%%%%%%%%%%%%

\section{Linear spin-wave theory}
\label{app:lswt}

To establish notation and lay the groundwork for the linear-response calculations discussed in Sec.~\ref{subsec:textures} and Appendix~\ref{app:linresp}, we devote this appendix to a brief exposition of the linear spin-wave theory for the clean $J_1$-$J_2$ model. Different aspects of the theory were previously explored in Refs.~\onlinecite{Rastelli1979}, \onlinecite{Fouet2001} and \onlinecite{Mulder2010}.

To this end, we promote the Hamiltonian in Eq.~\eqref{eq:H} to a quantum spin model and consider an arbitrary spiral state described by a wave vector $\QQ$ in the spiral contour. In the original reference frame, each spin has components $\left\{S_{i\mu}^x, S_{i\mu}^y, S_{i\mu}^z\right\}$ given by Eqs.~\eqref{eq:Spar} and \eqref{eq:alphaQ}. We then perform rotations in spin space that map $\left\{\hat{\mathbf{x}}, \hat{\mathbf{y}}, \hat{\mathbf{z}} \right\}$ onto a set of site-dependent basis vectors $\left\{\hat{\mathbf{e}}_{1,i\mu}, \hat{\mathbf{e}}_{2,i\mu}, \hat{\mathbf{e}}_{3,i\mu} \right\}$, defined such that $\hat{\mathbf{e}}_{3,i\mu} \parallel \mathbf{S}_{i\mu}$ and $\hat{\mathbf{e}}_{2,i\mu} = \hat{\mathbf{y}}$. The components in this rotated frame, $\left\{S_{i\mu}^1, S_{i\mu}^2, S_{i\mu}^3\right\}$, are related to those in the original frame via
\begin{equation}
	\begin{pmatrix}
		S_{i\mu}^x \\
		S_{i\mu}^y \\
		S_{i\mu}^z
	\end{pmatrix}
	= \left(-1\right)^{\mu}
	\begin{pmatrix}
		\cos\varphi_{i\mu} & 0 & \sin\varphi_{i\mu} \\
		0 & \left(-1\right)^{\mu} & 0 \\
		-\sin\varphi_{i\mu} & 0 & \cos\varphi_{i\mu} \\
		
	\end{pmatrix}
	\begin{pmatrix}
		S_{i\mu}^1 \\
		S_{i\mu}^2 \\
		S_{i\mu}^3
	\end{pmatrix},
\end{equation}
where $\varphi_{i\mu} \left(\QQ\right) = \QQ\cdot\mathbf{R}_i + \delta_{\mu1}\alphaQ$ as before.
The resulting Hamiltonian reads
\begin{widetext}
\begin{equation}
	\mathcal{H} = \sum_{ij} \sum_{\mu\nu} \left[
	\frac{L_{ij}^{\mu\nu}}{2} \left( S_{i\mu}^{3}S_{j\nu}^{3} + S_{i\mu}^{1}S_{j\nu}^{1} \right)
	+ \frac{J_{ij}^{\mu\nu}}{2} S_{i\mu}^{2}S_{j\nu}^{2}
	+ T_{ij}^{\mu\nu} S_{i\mu}^{3}S_{j\nu}^{1}
	\right],
	\label{eq:Hrot}
\end{equation}
with coupling matrices given by
\begin{align}
	J_{ij}^{\mu\nu} &= J_1 \delta_{\absval{\mathbf{r}_{j\nu} - \mathbf{r}_{i\mu}},a} + J_2 \delta_{\absval{\mathbf{r}_{j\nu} - \mathbf{r}_{i\mu}},\sqrt{3}a},
	\label{eq:Jij}
	\\
	L_{ij}^{\mu\nu} &= -J_1 \cos \left[ \left(-1\right)^\mu \QQ\cdot \mathbf{R}_{ji} + \alphaQ \right] \delta_{\absval{\mathbf{r}_{j\nu} - \mathbf{r}_{i\mu}},a}
	+ J_2 \cos \left(\QQ \cdot \mathbf{R}_{ji} \right) \delta_{\absval{\mathbf{r}_{j\nu} - \mathbf{r}_{i\mu}},\sqrt{3}a},
	\label{eq:Lij}
	\\
	T_{ij}^{\mu\nu} &= \left(-1\right)^{\mu} J_1 \sin \left[ \left(-1\right)^\mu \QQ\cdot \mathbf{R}_{ji} + \alphaQ \right] \delta_{\absval{\mathbf{r}_{j\nu} - \mathbf{r}_{i\mu}},a}
	- J_2 \sin \left(\QQ \cdot \mathbf{R}_{ji} \right) \delta_{\absval{\mathbf{r}_{j\nu} - \mathbf{r}_{i\mu}},\sqrt{3}a},
	\label{eq:Tij}
\end{align}
\end{widetext}
Thus, the original coupling matrix in the laboratory frame, $J_{ij}^{\mu\nu}$, sets the interaction between out-of-plane components that the spin texture might acquire. Meanwhile, Eqs.~\eqref{eq:Lij} and \eqref{eq:Tij} can be interpreted as the components of the local field that $\mathbf{S}_{i\mu}$ generates parallel and perpendicular to $\mathbf{S}_{j\nu}$, respectively, in an unperturbed spin spiral. Also, $a$ denotes the lattice constant, such that the Kronecker deltas enforce that matrix elements are only nonzero when $i\mu$ and $j\nu$ are NN or NNN sites.

Being in the rotated frame enables us to employ the Holstein-Primakoff transformation
\begin{equation}
	S_{j\mu}^3 = S - a_{j\mu}^{\dagger} a_{j\mu},\,
	S_{j\mu}^1 + iS_{j\mu}^2 = \sqrt{2S - a_{j\mu}^\dagger a_{j\mu}} a_{j\mu},
	\label{eq:HP}
\end{equation}
and expand the Hamiltonian in powers of $S^{-1/2}$. After applying a Fourier transform, $a_{j\mu} = \sqrt{2/N} \sum_{\mathbf{k}} e^{i\mathbf{k}\cdot\mathbf{R}_j} a_{\mathbf{k}\mu}$, and truncating the series at $\mathcal{O}\left(S\right)$, we obtain the linear spin-wave Hamiltonian
\begin{equation}
	\mathcal{H}_{\mathrm{LSW}} = S\left(S+1\right)\Egs + \frac{S}{2} \sum_{\mathbf{k}\mu} \Psi_{\mathbf{k}}^\dagger
	\begin{pmatrix}
		\mathbb{A}_{\mathbf{k}} & \mathbb{B}_{\mathbf{k}} \\
		\mathbb{B}_{\mathbf{k}}^\dagger & \mathbb{A}_{\mathbf{k}}
	\end{pmatrix}
	\Psi_{\mathbf{k}}
	\label{eq:HLSW1}
\end{equation}
where $\Psi_{\mathbf{k}}^\dagger = \left(a_{\mathbf{k}0}^\dagger, a_{\mathbf{k}1}^\dagger, a_{-\mathbf{k}0}, a_{-\mathbf{k}1} \right)$. If we set the lattice constant $a=1$, introduce the shorthand notation $k_0 = 0$ and $k_n = \mathbf{k}\cdot\mathbf{a}_n$, and recall the definition $h = \absval{E_{\mathrm{gs}}}/N$, we can write their matrix elements of the $2\times2$ Hermitian matrices $\mathbb{A}_{\mathbf{k}}$ and $\mathbb{B}_{\mathbf{k}}$ as
\begin{align}
	A_{\mathbf{k},00} &= A_{\mathbf{k},11} = 2h + J_2 \sum_{a=1}^{3} \cos k_a \left(1 + \cos Q_a\right),
	\notag \\
	A_{\mathbf{k},01} &= A_{\mathbf{k},10}^{*} = \frac{J_1}{2} \sum_{a=0}^{2} \left[1- \cos\left(\alphaQ - Q_a\right)\right] e^{-ik_a},
	\notag \\
	B_{\mathbf{k},00} &= B_{\mathbf{k},11} = -J_2 \sum_{a=1}^{3} \cos k_a \left(1 - \cos Q_a\right),
	\notag \\
	B_{\mathbf{k},01} &= B_{\mathbf{k},10}^{*} = -\frac{J_1}{2} \sum_{a=0}^{2} \left[1 + \cos\left(\alphaQ - Q_a\right)\right] e^{-ik_a}.
	\label{eq:LSWmatelements}
\end{align}

The quadratic Hamiltonian in Eq.~\eqref{eq:HLSW1} can be diagonalized via a Bogoliubov transformation and recast in the form
\begin{equation}
	\mathcal{H}_{\mathrm{LSW}} = S\left(S+1\right)\Egs + S \sum_{\mathbf{k}\mu}
	\epsilon_{\mathbf{k}\mu} \left(b_{\mathbf{k}\mu}^\dagger b_{\mathbf{k}\mu} + \frac{1}{2} \right),
	\label{eq:HLSW2}
\end{equation}
with $\epsilon_{\mathbf{k}\mu} \ge 0$ being the excitation energy of a single magnon created by the bosonic operator $b_{\mathbf{k}\mu}^\dagger$. The relation between Eqs.~\eqref{eq:HLSW1} and \eqref{eq:HLSW2} is established by diagonalizing the non-Hermitian matrix
\begin{align}
	\sigma_3 \mathbb{M}_{\mathbf{k}} &=
	\begin{pmatrix}
		\mathbb{A}_{\mathbf{k}} & \mathbb{B}_{\mathbf{k}} \\
		-\mathbb{B}_{\mathbf{k}}^\dagger & -\mathbb{A}_{\mathbf{k}}
	\end{pmatrix},
	\label{eq:sigma3Mk}
\end{align}
where $\sigma_3 = \mathrm{diag} \left(1,1,-1,-1\right)$ \cite{Blaizot1986}. The eigenvalues of $\sigma_3 \mathbb{M}_{\mathbf{k}}$ are $\left\{\epsilon_{\mathbf{k}0}, \epsilon_{\mathbf{k}1}, -\epsilon_{-\mathbf{k}0}, -\epsilon_{-\mathbf{k}1} \right\}$, with
\begin{equation}
	\epsilon_{\mathbf{k}\mu} = \sqrt{P_{\mathbf{k}} + 2\left(-1\right)^\mu \sqrt{R_{\mathbf{k}}} }
	\label{eq:LSWdisp}
\end{equation}
and
\begin{align}
	P_{\mathbf{k}} &= A_{\mathbf{k},00}^2 - B_{\mathbf{k},00}^2 + \absval{A_{\mathbf{k},01}}^2 - \absval{B_{\mathbf{k},01}}^2, \\
	R_{\mathbf{k}} &= \absval{A_{\mathbf{k},11}A_{\mathbf{k},01} - B_{\mathbf{k},00}B_{\mathbf{k},01} }^2 - \mathrm{Im}\left(B_{\mathbf{k},01} A_{\mathbf{k},01}^*\right). \notag
\end{align}
The four eigenvalues listed above correspond to right eigenvectors we shall denote as $\left\{V_{\mathbf{k}0}, V_{\mathbf{k}1}, W_{\mathbf{k}0}, W_{\mathbf{k}1}\right\}$, respectively. With the normalizations
\begin{align}
	V_{\mathbf{k}\mu}^\dagger \sigma_{3} V_{\mathbf{k}\mu} &= 1,
	&
	W_{\mathbf{k}\mu}^\dagger \sigma_{3} W_{\mathbf{k}\mu} &= -1,
\end{align}
the Bogoliubov transformation reads
\begin{equation}
	\begin{pmatrix}
		a_{\mathbf{k}\mu} \\
		a_{-\mathbf{k}\mu}^{\dagger}
	\end{pmatrix}
	=
	\sum_{\nu=0}^{1}
	\begin{pmatrix}
		V_{\mathbf{k}\nu,\mu} & W_{\mathbf{k}\nu,\mu} \\
		V_{\mathbf{k}\nu,2+\mu} & W_{\mathbf{k}\nu,2+\mu}
	\end{pmatrix}
	\begin{pmatrix}
		b_{\mathbf{k}\nu} \\
		b_{-\mathbf{k}\nu}^{\dagger}.
	\end{pmatrix}.
	\label{eq:Bogoliubov}
\end{equation}
For the case at hand, we were able to derive the eigenvectors of $\sigma_{3} \mathbb{M}_{\mathbf{k}}$ analytically. Although we shall not reproduce the expressions here, we note that this task is facilitated by the property $W_{\mathbf{k}\mu} = \sigma_1 V_{-\mathbf{k}\mu}^*$ \cite{Blaizot1986}, where $\sigma_1$ is a $4\times4$ generalization of the Pauli matrix $\sigma_x$.

\begin{figure}
	\centerline{
		\includegraphics[width=\columnwidth]{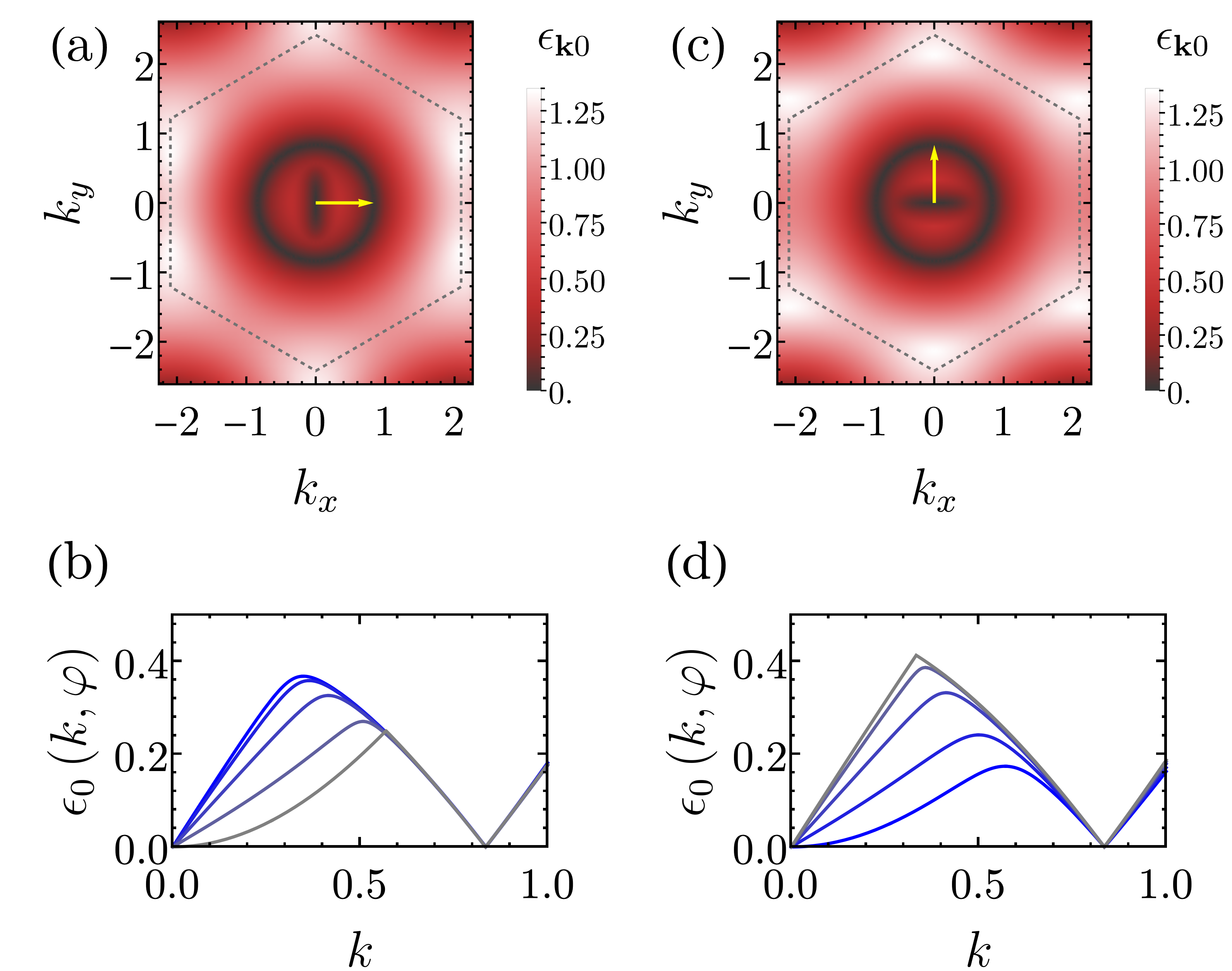}
	}
	\caption{Plots of the lower branch $\epsilon_{\mathbf{k}0}$ of the linear spin-wave dispersion for $\Jrat=0.2$ with (a,b) $\phiQ=0$ and (c,d) $\phiQ=\pi/2$. In upper row, the dashed gray hexagons indicate the boundary of the first Brillouin zone, while the yellow vector corresponds to the ordering wave vector (a) $\mathbf{Q} = \left(2/3\right) \cos^{-1} \left[1/\left(4\Jrat\right)^2 - 5/4\right] \hat{\mathbf{x}} \approx 0.8353 \hat{\mathbf{x}}$ and (c) $\mathbf{Q} = \left(2/\sqrt{3}\right) \cos^{-1} \left[1/\left(4\Jrat\right) - 1/2\right] \hat{\mathbf{y}} \approx 0.8345 \hat{\mathbf{y}}$. Plots (b) and (d) depict the radial dependence of the dispersion along rays with polar angles $\varphi$ ranging from $0$ (blue) to $\pi/2$ (gray).}
	\label{fig:lswdisp}
\end{figure}

The linear spin-wave spectrum given by Eq.~\eqref{eq:LSWdisp} has a few remarkable properties. To illustrate them, we present plots of the lower branch $\epsilon_{\mathbf{k}0}$ of the $\Jrat=0.2$ spectrum for two different ordering wave vectors, corresponding to $\phiQ=0$ and $\phiQ=\pi/2$, in Fig.~\ref{fig:lswdisp}. Both cases exhibit a Goldstone mode at $\mathbf{k}=\boldsymbol{0}$. However, the dispersion around this point is highly anisotropic and even changes from linearly (for $\mathbf{k} \parallel \QQ$) to quadratically dispersing (for $\mathbf{k} \perp \QQ$) [see Figs.~\ref{fig:lswdisp}(b,d)]. This anisotropy of the Goldstone mode may be the underlying cause of the pinch-point singularities we find Appendix~\ref{app:linresp}, when inspecting susceptibilities that describe the response of the system to different types of impurities.

While both spectra exhibit zero modes at the expected locations, a comparison between the left and right columns of Fig.~\ref{fig:lswdisp} reveals that they are not equivalent under 90° rotations. This, however, is by no means a contradiction, since the Hamiltonian is not invariant under such a rotation. It merely reflects the fact that the degeneracy of the spiral contour is not protected by symmetry, and is therefore prone to be lifted by fluctuations. By computing the leading correction to the ground-state energy within linear spin-wave theory, the authors of Ref.~\onlinecite{Mulder2010} showed that quantum fluctuations select the same set of spiral states as a single vacancy (see Table~\ref{tab:vacs}): $\phiQ = n\pi/3$ in the SSL$_1$ regime and $\phiQ = \left(4n+3\right)\pi/6$ in SSL$_2$.

%%%%%%%%%%%%%%%%%%%%%%%%%%%%%%%%%%%%%%%%%%%%%%%%%%%%%%%%%%%%%%%%%%%%%%%
%%%%%%%%%%%%%%%%%%%%%%%%%%%%%%%%%%%%%%%%%%%%%%%%%%%%%%%%%%%%%%%%%%%%%%%
%%%%%%%%%%%%%%%%%%%%%%%%%%%%%%%%%%%%%%%%%%%%%%%%%%%%%%%%%%%%%%%%%%%%%%%

\section{Long-distance behavior of spin textures via linear-response theory}
\label{app:linresp}

This appendix presents a detailed derivation of the linear-response results discussed in Sec.~\ref{subsec:textures}. We start by outlining the procedure in full generality and later, while focusing on the SSL$_1$ regime, address the cases of a NN bond defect, a NNN bond defect and single vacancy separately.

The first step of the calculation has already been provided in Appendix~\ref{app:lswt}, where we solved the unperturbed (clean) Hamiltonian at the level of linear spin-wave theory. Thus, our next task is to construct a pertubation $\delta\mathcal{H}$ which can be added to the Hamiltonian to mimic the effect of a weak impurity that preserves the form of Eq.~\eqref{eq:H}, but is otherwise arbitrary.
To do so, we return to the rotated frame introduced in Appendix~\ref{app:lswt} and consider modifications to the coupling matrices appearing in Eq.~\eqref{eq:Jij}-\eqref{eq:Tij}, which is to say $T_{ij}^{\mu\nu} \to T_{ij}^{\mu\nu} + \delta T_{ij}^{\mu\nu}$ and so on. Assuming that this induces small deviations from the original spin spiral, we use the normalization constraint of classical spins to expand the different terms in powers of the tranverse components $S_{i\mu}^1$ and $S_{i\mu}^2$. By truncating the expansion at next-to-leading order, we obtain
\begin{align}
	\delta\mathcal{H} &= \sum_{ij}\sum_{\mu\nu} \left(
	\frac{S^2}{2} \delta L_{ij}^{\mu\nu} + S\, \delta T_{ij}^{\mu\nu} S_{j\nu}^1
	\right)
	\notag \\
	& = S^2 \delta \mathcal{H}_0 \left(\mathbf{Q}\right) - S \sum_{j\nu} B_{j\nu} S_{j\nu}^{1}.
	\label{eq:dHgen}
\end{align}
The first term, $S^2 \delta \mathcal{H}_0 \left(\mathbf{Q}\right)$, represents the energy correction due to the change in the longitudinal component of the local exchange fields. Thus, it encapsulates the same principles that guided our considerations in Sec.~\ref{subsec:sbd}.
The second term, in contrast, corresponds to a transverse in-plane field $B_{j\nu}$ acting only in the immediate vicinity of the impurity. As such, it is responsible for the possible emergence of a spin texture.

The expectation values of the spin operators with respect to the ground state of the perturbed Hamiltonian $\left(\mathcal{H}+\delta\mathcal{H}\right)$ can be expressed as
\begin{equation}
	\begin{pmatrix}
		\Braket{S_{i\mu}^1} \\
		\Braket{S_{i\mu}^2} \\
		\Braket{S_{i\mu}^3}
	\end{pmatrix}
	= S
	\begin{pmatrix}
		\cos \vartheta_{i\mu} \sin \delta\varphi_{i\mu} \\
		\sin \vartheta_{i\mu} \\
		\cos \vartheta_{i\mu} \cos \delta\varphi_{i\mu}
	\end{pmatrix}.
	\label{eq:linrespexp}
\end{equation}
Here, $\delta\varphi_{i\mu}$ and $\vartheta_{i\mu}$ denote the correction to the in-plane angle of the spin spiral, $\varphi_{i\mu} \to \varphi_{i\mu} + \delta\varphi_{i\mu}$, and a canting angle out of the original ordering plane, respectively. Naively, one could be tempted to conclude that all $\vartheta_{i\mu}$ must be zero, since the impurity does not couple to the out-of-plane components $S_{i\mu}^2$ directly (through $\delta\mathcal{H}$) nor indirectly ($\mathcal{H}$ does not have a $S_{i\mu}^1 S_{j\nu}^2$ cross term). However, as we will show below, this assumption turns out to be incorrect due to the presence of zero modes along the spiral contour.
Within linear-response theory \cite{FetterWalecka,Sachdev},
\begin{align}
	\delta \Braket{S_{i\mu}^a \left(t\right)}
	&= \Braket{S_{i\mu}^a \left(t\right)} - \Braket{S_{i\mu}^a }_0
	\notag \\
	&= -i \int_{-\infty}^t dt' \Braket{ \left[ S_{i\mu}^a \left(t\right), \delta\mathcal{H} \left(t'\right) \right] }_0
	\notag \\
	&= \int_{-\infty}^{\infty} dt'' \chi_{i\mu}^{a 1} \left(t''\right)
	= \tilde{\chi}_{i\mu}^{a 1} \left(\omega=0\right),
	\label{eq:linresp}
\end{align}
where $a=1,2,3$. Furthermore,
\begin{equation}
	i\chi_{i\mu}^{a 1} \left(t\right) = \sum_{\ell \lambda}\sum_{j\nu} \delta T_{\ell j}^{\lambda\nu}
	\Braket{ \left[ S_{i\mu}^a \left(t\right), S_{j\nu}^1 \left(0\right) \right] }_0 \Theta\left(t\right)
	\label{eq:chirt}
\end{equation}
is a real-space retarded correlation function, $\tilde{\chi}_{i\mu}^{a1} \left(\omega=0\right)$ is its Fourier transform at zero frequency, and $\Braket{\cdots}_0$ denotes the expectation value with respect to the ground state of the unperturbed Hamiltonian. The time dependence in Eq.~\eqref{eq:linresp} follows from the fact that both operators appearing in the correlation function are in the interaction picture.

As usual, the susceptibility $\tilde{\chi}_{i\mu}^{a1} \left(\omega\right)$ is most conveniently evaluated by first computing the Matsubara correlation function
\begin{equation}
	\tilde{\chi}_{\mathbf{k}\mu}^{a1} \left(i\omega_n\right) = \sum_{\nu} \beta_{\mathbf{k}\nu}
	\int_{0}^{1/T} \!\!\! d\tau e^{i\omega_n \tau} \Braket{S_{\mathbf{k}\mu}^{a} \left(\tau\right) S_{-\mathbf{k}\nu}^{1} \left(0\right) }_0
	\label{eq:chikwn}
\end{equation}
at a nonzero temperature $T$. Here, $\omega_{n} = 2\pi n /T$ are bosonic Matsubara frequencies, $\Braket{\cdots}_0$ is the thermal average in the Gibbs ensemble of the unperturbed Hamiltonian, and
\begin{equation}
	\beta_{\mathbf{k}\nu} = -\sum_{j \ell \lambda} \delta T_{\ell j}^{\lambda \nu} e^{-i \mathbf{k} \cdot \mathbf{R}_j}
	\label{eq:formfactordef}
\end{equation}
is a sublattice-dependent form factor. Using the Holstein-Primakoff transfomation in Eq.~\eqref{eq:HP} and the Bogoliubov transformation in Eq.~\eqref{eq:Bogoliubov}, it is easy to see that Eq.~\eqref{eq:chikwn} vanishes for $a=3$, given that it only involves averages of odd numbers of bosons. The two remaining cases ($a=1,2$) can be calculated straighforwardly to leading order in $1/S$. After performing analytic continuation to real frequencies, we obtain
\begin{equation}
	\tilde{\chi}_{\mathbf{k}\mu}^{a1} \left(\omega=0\right) =
	\frac{S}{2i^{a-1}} \sum_{\nu\lambda} \beta_{\mathbf{k}\nu} \left[ \frac{F_{\mu\nu}^{\pm} \left( V_{\mathbf{k}\lambda} \right)}{\epsilon_{\mathbf{k}\lambda} - i0^+} + \frac{F_{\mu\nu}^{\pm} \left(W_{\mathbf{k}\lambda}\right)}{\epsilon_{\mathbf{k}\lambda} + i0^+} \right],
	\label{eq:chik}
\end{equation}
where $\epsilon_{\mathbf{k}\lambda} = \epsilon_{-\mathbf{k}\lambda}$ is the spin-wave spectrum given by Eq.~\eqref{eq:LSWdisp} and
\begin{equation}
	F_{\mu\nu}^{\pm} \left( V \right) = \left(V_{\mu} \pm V_{2+\mu}\right)
	\left(V_{\nu} + V_{2+\nu}\right)^{*}
	\label{eq:Fmunu}
\end{equation}
appear as functions of the eigenvectors (i.e., the Bogoliubov coefficients) of the spin-wave matrix $\sigma_{3} \mathbb{M}_{\mathbf{k}}$ defined in Eq.~\eqref{eq:sigma3Mk}. The upper (lower) sign applies for $a=1$ ($a=2$).

Equation~\eqref{eq:chik} can, however, be cast in a more transparent form if one exploits the fact that the linear spin-wave matrix obeys $\mathbb{M}_{\mathbf{k}} = \mathbb{M}_{\mathbf{-k}}^*$. This property, which is a consequence of the time-reversal symmetry of $\mathcal{H}$, implies that $V_{-\mathbf{k}\lambda}^* = V_{\mathbf{k}\lambda}$ and $W_{\mathbf{k}\lambda} = \sigma_1 V_{\mathbf{k}\lambda}$. By combining this result with Eqs.~\eqref{eq:linresp} and \eqref{eq:chik}, we conclude that
\begin{align}
	\Braket{S_{i\mu}^1} &= \frac{S}{N_\mathrm{c}} \sum_{\mathbf{k}} e^{i \mathbf{k} \cdot \mathbf{R}_i}
	\sum_{\nu\lambda} \beta_{\mathbf{k}\nu} F_{\mu\nu}^{+} \left(V_{\mathbf{k}\lambda}\right) \mathcal{P} \left( \frac{1}{\epsilon_{\mathbf{k}\lambda}} \right),
	\label{eq:linrespav1}
	\\
	\Braket{S_{i\mu}^2} &= \frac{\pi S}{N_\mathrm{c}} \sum_{\mathbf{k}} e^{i \mathbf{k} \cdot \mathbf{R}_i}
	\sum_{\nu} \beta_{\mathbf{k}\nu} F_{\mu\nu}^{-} \left(V_{\mathbf{k}0}\right) \delta \left( \epsilon_{\mathbf{k}0} \right),
	\label{eq:linrespav2}
\end{align}
where $\mathcal{P}$ stands for the Cauchy principal value and $N_{c} = N/2$ is the number of unit cells in the system. It is worth noting that, although the eigenvectors $V_{\mathbf{k}0}$ diverge at the points where $\epsilon_{\mathbf{k}0} = 0$, the functions $\sum_{\nu} \beta_{\mathbf{k}\nu} F_{\mu\nu}^{\pm}(V_{\mathbf{k}0})$ will turn out to be analytic over the entire Brillouin zone for all three cases we consider below.

\subsection{Energy correction due to the texture}
\label{subsec:Htex}

With the results outlined above, one can compute the energy correction due the distortions an isolated impurity causes to the spin configuration. To achieve this, we employ the normalization constraint for classical spins,
\begin{equation}
	S_{i\mu}^{3} \approx S - \frac{1}{2S} \left[ \left(S_{i\mu}^{1}\right)^{2} + \left(S_{i\mu}^{2}\right)^{2} \right],
\end{equation}
and express Eq.~\eqref{eq:Hrot} solely in terms of transverse spin components. As a result, the Hamiltonian naturally decomposes into three separate contributions, $\mathcal{H} = E_{\mathrm{gs}} + \delta \mathcal{H}_0 \left(\mathbf{Q}\right) + \mathcal{H}_{\mathrm{tex}}\left(\mathbf{Q}\right)$. The first two, which have appeared previously in Eqs.~\eqref{eq:Egs} and \eqref{eq:dHgen}, represent the ground-state energy of the clean system and a trivial correction due to changes in the local fields. The third term, on the other hand, encompasses the influence of the impurity-induced spin texture and reads
\begin{align}
	\mathcal{H}_{\mathrm{tex}} &= \frac{1}{2} \sum_{ij}\sum_{\mu\nu}
	\left\{
	L_{ij}^{\mu\nu} \left[ S_{i\mu}^{1} S_{j\nu}^{1} - \left(S_{i\mu}^{1}\right)^2 - \left(S_{i\mu}^{2}\right)^2 \right]
	\right.
	\notag \\
	& \left.
	+ J_{ij}^{\mu\nu} S_{i\mu}^{2} S_{j\nu}^{2}
	\right\}
	- S \sum_{i\mu} B_{i\mu} S_{i\mu}^{1}.
	\label{eq:Htex}
\end{align}
It depends on the ordering wave vector $\QQ$ through the coupling matrices and contributes to the ObQD mechanism at a higher order in response theory than $\delta \mathcal{H}_0$. However, as demonstrated in Sec.~\ref{subsec:svac}, there are situations in which $\delta \mathcal{H}_0$ is independent of $\phiQ$ and the state selection is determined \emph{exclusively} by $\mathcal{H}_{\mathrm{tex}}$. This justifies an explicit evaluation of Eq.~\eqref{eq:Htex}, which is most conveniently carried out in reciprocal space.

Since we are interested in the classical limit $S\to\infty$, we can drop the ground-state expectation values in Eqs.~\eqref{eq:linrespav1} and \eqref{eq:linrespav2}, and substitute these expressions directly into Eq.~\eqref{eq:Htex}. After converting the discrete sum over momenta into an integral over the Brillouin zone, we obtain
\begin{equation}
	\mathcal{H}_{\mathrm{tex}} =
	\frac{S^2 v_{uc}}{8\pi^2} \underset{\mathrm{BZ}}{\int} d^{2}k \;
	\left\{ f_{1} \!\left(\mathbf{k}\right) + f_{2} \!\left(\mathbf{k}\right) \left[\delta\left(\epsilon_{\mathbf{k}0}\right)\right]^{2} \right\},
\end{equation}
where $v_\mathrm{uc}$ denotes the area of a unit cell in real space. The functions
\begin{align}
	f_{1} \!\left(\mathbf{k}\right) &=
	\beta_{\mathbf{k}}^{\dagger} \;
	\mathbb{G}_{\mathbf{k}}
	\left[ \left( \mathbb{L}_{\mathbf{k}} + 2h\mathds{1} \right) \mathbb{G}_{\mathbf{k}} -2 \mathds{1} \right]
	\beta_{\mathbf{k}}
	\label{eq:f1k}
	\\
	f_{2} \!\left(\mathbf{k}\right) &=
	\pi^{2} \, \beta_{\mathbf{k}}^{\dagger}
	\left(\mathbb{F}_{\mathbf{k}0}^{-}\right)^{\dagger} \left( \mathbb{J}_{\mathbf{k}} + 2h\mathds{1} \right) \mathbb{F}_{\mathbf{k}0}^{-} \;
	\beta_{\mathbf{k}}
	\label{eq:f2k}
\end{align}
carry information about the magnon modes through
\begin{align}
	\mathbb{F}^{\pm}_{\mathbf{k}\lambda} &=
	\begin{pmatrix}
		F_{00}^{\pm} \left(V_{\mathbf{k}\lambda} \right) & F_{01}^{\pm} \left(V_{\mathbf{k}\lambda} \right)
		\\
		F_{10}^{\pm} \left(V_{\mathbf{k}\lambda} \right) & F_{11}^{\pm} \left(V_{\mathbf{k}\lambda} \right)
	\end{pmatrix},
	&
	\mathbb{G}_{\mathbf{k}} &= \sum_{\lambda} \frac{\mathbb{F}^{+}_{\mathbf{k}\lambda}}{\epsilon_{\mathbf{k}\lambda}},
\end{align}
which are matrices defined in terms of the functions given in Eq.~\eqref{eq:Fmunu} and the spin-wave dispersion in Eq.~\eqref{eq:LSWdisp}. Equations~\eqref{eq:f1k} and \eqref{eq:f2k} also depend on $\mathbb{L}_{\mathbf{k}}$ and $\mathbb{J}_{\mathbf{k}}$, the Fourier transforms of the $2\times2$ coupling matrices $L_{ij}^{\mu\nu} = L_{\mu\nu}\left(\mathbf{R}_{ji}\right)$ and $J_{ij}^{\mu\nu} = J_{\mu\nu}\left(\mathbf{R}_{ji}\right)$. Finally, all impurity-specific information is encoded in the form factors $\beta_{\mathbf{k}} = \left(\beta_{\mathbf{k}0} , \beta_{\mathbf{k}1}\right)^{\top}$.

With Eq.~\eqref{eq:Htex}, we have expressed $\mathcal{H}_\mathrm{tex}$ in terms of two integrals over the Brillouin zone, one related to the in-plane and the other to the out-of-plane component of the spin texture. However, the latter requires a careful examination due to the presence of the square of a Dirac delta. We thus start by noting that only momenta hosting zero modes can contribute to $I_{2} = \int_{\mathrm{BZ}} d^{2}k \, f_{2}\left(\mathbf{k}\right) \left[\delta\left(\epsilon_{\mathbf{k}0}\right)\right]^2$. This allows us to split the integration domain into disjoint sets consisting of a disk of infinitesimal radius centered at the origin and closed portions of the spiral contour. Let us focus on one of the latter contours for now, which we denote as $\gamma$. It can be parametrized as $\mathbf{k}_{\Jrat}\left(\phi\right) = \mathbf{k}_0 + \mathbf{q}_{\Jrat}\left(\phi\right)$ with a polar angle $\phi \in \left[0,2\pi\right[$ measured with respect to an origin $\mathbf{k}_0 = \mathbf{0}$ ($\mathbf{k}_0 = \mathbf{K}$) if $\Jrat$ is in the SSL$_1$ (SSL$_2$) regime.
Given that the lower spin-wave branch has a linear dispersion around the spiral contour, i.e.,
\begin{equation}
	\epsilon_{\mathbf{k}_0+\mathbf{q},0} = v\left(\phi\right) \absval{q - q_{\Jrat}\left(\phi\right)} + \mathcal{O}\left(\absval{q - q_{\Jrat}\left(\phi\right)}^2\right)
	\label{eq:dispexpsc}
\end{equation}
for $\mathbf{q} \approx \mathbf{q}_{\Jrat}\left(\phi\right)$ (see Fig.~\ref{fig:lswdisp}), the corresponding contribution to $I_2$ is
\begin{align}
	I_{2}' &= \int_{\gamma} \! d^{2}q \; f_{2} \left(\mathbf{k}_0 + \mathbf{q}\right) \left[ \frac{\delta \left( q - q_{\Jrat}\left(\phi\right)\right)}{v\left(\phi\right)} \right]^2
	\notag \\
	&= \int_{0}^{2\pi} \!\!\! \frac{d\phi}{v^2\left(\phi\right)} \int_{-\varepsilon}^{\varepsilon} dx \; g_{\phi} \left(x\right) \left[\delta\left(x\right)\right]^2,
	\label{eq:I2prime}
\end{align}
with $g_{\phi} \left(x - q_{\Jrat}\left(\phi\right) \right) = x f_{2}\left( \mathbf{k}_0 + x \left(\cos\phi, \sin\phi\right) \right)$.

We are thus left with the task of evaluating an integral of the form $I = \int_{-\infty}^{\infty} dx \; g \left(x\right) \left[\delta\left(x\right)\right]^2$. We claim that if
\begin{enumerate}[label=(\roman*)]
	\itemsep=0pt
	\item $g\left(x\right)$ is analytic at $x=0$,
	\item $g\left(0\right) = 0$,
	\item $g'\left(0\right) = 0$,
\end{enumerate}
then $I=0$. To prove this claim, we shall consider the representation of the Dirac delta as the limit of the sequence of functions $\delta_{\varepsilon} \left(x\right) = \Theta\left(\varepsilon - \absval{x}\right)/\left(2\varepsilon\right)$ when $\varepsilon \to 0^{+}$. Due to Assumption (i), there exists an $\varepsilon^{*} > 0$ for which $g\left(x\right)$ is continuous in the interval $\left[-\varepsilon^{*}, \varepsilon^{*}\right]$. The mean value theorem then implies that, if $\varepsilon \le \varepsilon^{*}$, $I_{\varepsilon} = \int_{-\varepsilon}^{\varepsilon} \! dx \; g \left(x\right) \left[\delta_{\varepsilon} \left(x\right) \right]^2 = g\left(x_0\right)/\left(2\varepsilon\right)$ for some $x_0 \in \left[-\varepsilon,\varepsilon\right]$. Together with Assumption (ii), continuity also guarantees that $\absval{g\left(x_0\right)} \le \absval{g\left(\varepsilon\right)}$ for sufficiently small $\varepsilon$. Finally, when supplemented by Assumption (iii), these results yield
\begin{equation}
	\left| I \right|
	\le \underset{\varepsilon \to 0^{+}}{\lim} \frac{\left| g\left(\varepsilon\right) \right|}{2\varepsilon}
	= \frac{\absval{g{''} \left(0\right)}}{4} \underset{\varepsilon \to 0^{+}}{\lim} \varepsilon = 0,
\end{equation}
which concludes the proof.

Hence, if every function $g_{\phi} \left(x\right)$ satisfies properties (i)-(iii), then $I_{2}' = 0$. We will argue below in three steps that this is indeed the case. Before doing so, however, we note that Eq.~\eqref{eq:f2k} can be -- and in fact is -- finite along the entire spiral contour even though $V_{\mathbf{k}0}$ itself diverges as $\absval{k-k_{\Jrat}\left(\phi\right)}^{-1/2}$. This is possible because the components of such eigenvectors appear in combinations that eliminate the divergences.

As a first step, let us prove that $g_{\phi} \left(x\right) \ge 0$ for every $x \ge -q_{\Jrat} \left(\phi\right)$. The key insight behind the proof is that $\mathbb{J}_{\mathbf{k}}$ is the Fourier transform of the coupling matrix of the clean $J_1$-$J_2$ Hamiltonian in its original laboratory frame. Thus, by the Luttinger-Tisza method, its minimum eigenvalue as a function of $\mathbf{k}$ is $E_{\mathrm{gs}}/N_c = -2h$. This implies that $\left(\mathbb{J}_{\mathbf{k}} + 2h\mathds{1}\right)$ is a positive semidefinite matrix with a unique square root, such that Eq.~\eqref{eq:f2k} can be rewritten as
\begin{equation}
	f_{2} \left(\mathbf{k}\right) = \left| \pi \left(\mathbb{J}_{\mathbf{k}} + 2h\mathds{1}\right)^{1/2} \mathbb{F}_{\mathbf{k}0}^{-} \; \beta_{\mathbf{k}} \right|^2 \ge 0.
	\label{eq:f2k positive}
\end{equation}
The stated result then follows straightforwardly from the definition of $g_{\phi} \left(x\right)$, which was given below Eq.~\eqref{eq:I2prime}.

Second, we would like to argue that $g_{\phi} \left(0\right) = 0$. In light of Eq.~\eqref{eq:f2k positive}, this can only hold for an arbitrary defect if $\left(\mathbb{J}_{\mathbf{k}} + 2h\mathds{1}\right)^{1/2} \mathbb{F}_{\mathbf{k}0}^{-} = 0$ along the entire spiral contour. By using the unitary transformation that diagonalizes $\mathbb{J}_\mathbf{k}$, one can show that this is equivalent to the simpler set of conditions $e^{i\theta_{\mathbf{k}}} F_{0\nu}^{-}\left(V_{\mathbf{k}0}\right) + F_{1\nu}^{-}\left(V_{\mathbf{k}0}\right) = 0$. Although $V_{\mathbf{k}0}$ itself diverges as $\absval{k-k_{\Jrat}\left(\phi\right)}^{-1/2}$, we find by numerical inspection that its components not only enter the functions $F_{\mu\nu}^{-} \left( V_{\mathbf{k}0} \right)$ in combinations that cancel out the divergence, but that also satisfy the above condition, yielding compelling evidence that $g_{\phi} \left(0\right) = 0$.

Finally, given that $g_{\phi} \left(0\right) = 0$ and $g_{\phi} \left(x\right) \ge 0$ in the vicinity of $x=0$, Assumption (iii) could only be violated if $g$ were non-analytic at $x=0$. However, this scenario is incompatible with our numerical observations, which suggest that $g\left(x\right)$ vanishes quadratically in $x$ since $\left[ e^{i\theta_{\mathbf{k}}} F_{0\nu}^{-}\left(V_{\mathbf{k}0}\right) + F_{1\nu}^{-}\left(V_{\mathbf{k}0}\right) \right] \propto \absval{k-k_{\Jrat}\left(\phi\right)}$.

We therefore conclude that Eq.~\eqref{eq:I2prime} is zero. Similar considerations indicate that the contribution coming from $\mathbf{k} = \mathbf{0}$ also vanishes, such that
\begin{equation}
	\mathcal{H}_{\mathrm{tex}} =
	\frac{S^2 v_{uc}}{8\pi^2} \underset{\mathrm{BZ}}{\int} d^{2}k \; f_{1} \!\left(\mathbf{k}\right)
	\label{eq:Htex final}
\end{equation}
only depends on the in-plane component of the texture at leading order in response theory. The texture develops an out-of-plane component merely to relieve the frustration of its in-plane part.

\subsection{Out-of-plane texture: General considerations and Friedel-like oscillations}

In Sec.~\ref{sec:sip}, we considered several examples of how a single impurity lifts the degeneracy of the spiral contour and can induce a texture on top of a selected coplanar spiral state. With Eq.~\eqref{eq:linrespav2}, we have proven that the existence of gapless spin-wave modes allows such a texture to acquire an out-of-plane component, even though the impurity acts as an effective magnetic field applied within the original ordering plane.

To show that this unintuitive possibility does generally come to fruition in a SSL, let us consider Eq.~\eqref{eq:linrespav2} in closer detail. Similarly to previous subsection, after taking the thermodynamic limit and converting the sum over momenta into an integral over the Brillouin zone, we can split the latter into two separate contributions, which originate from the $\mathbf{k} = \boldsymbol{0}$ Goldstone mode and the spiral contour. Explicitly, if we denote the polar coordinates of unit cell $i$ by $\left(R_i, \theta_i \right)$,
\begin{equation}
\Braket{S_{i\mu}^2} = \frac{S v_\mathrm{uc}}{4\pi} \left[ I_{\mathbf{k}=\boldsymbol{0}} + I_{\mathrm{spc}}\left(R_i, \theta_i \right) \right].
\label{eq:S2avsplit}
\end{equation}
The first term reads
\begin{equation}
	I_{\mathbf{k}=\boldsymbol{0}} =
	\int_{\Omega} d^{2}k \; \delta\left(\epsilon_{\mathbf{k}0} \right) \sum_{\nu} \beta_{\boldsymbol{0}\nu} F_{\mu\nu}^{-} \left(V_{\boldsymbol{0}\nu}\right),
	\label{eq:Ik0}
\end{equation}
where the integration is over a disk $\Omega$ of infinitesimal radius centered at the origin. For the system at hand, one can show that $F_{\mu\nu}^{-} \left(V_{\boldsymbol{0}\nu}\right)$ is independent of $\nu$.
Paired with the fact that $\beta_{\mathbf{0}0} = -\beta_{\mathbf{0}1}$ due to the antisymmetry of $\delta T_{ij}^{\mu\nu}$, this implies that the spatially uniform term $I_{\mathbf{k}=\boldsymbol{0}}$ vanishes and that the out-of-plane component of the texture is dictated \emph{exclusively} by the spiral contour.

To evaluate the second term in Eq.~\eqref{eq:S2avsplit}, it is convenient to split the spiral contour into two separate branches related by inversion symmetry. We can then choose one of these branches arbitrarily and parametrize it in terms of the polar angle $\phi \in \left[0, \phi_{\mathrm{max}} \right[$ in a coordinate system whose origin lies at a point $\mathbf{k}_0$. This results in a function $\mathbf{k}_{\Jrat} \left(\phi\right) = \mathbf{k}_0 + \mathbf{q}_{\Jrat} \left(\phi\right)$ with
\begin{equation}
	\left( \mathbf{k}_0, \phi_{\mathrm{max}}\right) =
	\begin{cases}
		\left( \boldsymbol{0}, \pi \right) & \text{for $1/6<\Jrat<1/2$},
		\\
		\left( \mathbf{K}, 2\pi \right) & \text{for $\Jrat>1/2$},
	\end{cases}
\end{equation}
which describes half of the closed contour in the SSL$_1$ regime and the upper pocket surrounding $\mathbf{K} = \left(0, 4\pi\left(3\sqrt{3}a\right) \right)$ in SSL$_2$ [see Fig.~\ref{fig:schems}(c)]. By additionally using Eq.~\eqref{eq:dispexpsc}, we arrive at
\begin{align}
	I_{\mathrm{spc}} \left(R,\theta\right) &= 2\,\mathrm{Re} \int_{0}^{\phi_{\mathrm{max}}} \!\!\!\! d\phi \; e^{ i f_{\theta}\left(\phi\right) R } z_{\mu}\left(\phi\right),
	\label{eq:Ispc}
\end{align}
where $f_{\theta}\left(\phi\right) = \mathbf{k}_{\Jrat}\left(\phi\right) \cdot \hat{\mathbf{R}}$ is the projection of a point $\mathbf{k}_{\Jrat}\left(\phi\right)$ on the spiral contour along $\hat{\mathbf{R}} = \mathbf{R}/R$ and
\begin{equation}
	z_{\mu}\left(\phi\right) = \frac{q_{\Jrat} \left(\phi\right)}{v\left(\phi\right)} \sum_{\nu} \beta_{\mathbf{k}_{\Jrat}\left(\phi\right) \nu} F_{\mu\nu}^{-} \left(V_{\mathbf{k}_{\Jrat}\left(\phi\right) 0}\right).
\end{equation}

The asymptotic form of $I_{\mathrm{spc}}$ in the limit $R \to \infty$, which determines the long-distance behavior of the out-of-plane texture, can be derived by means of the stationary phase approximation \cite{Bender1999,Lounis2011}. This approach is based on the notion that, in the stated limit, the complex exponential in Eq.~\eqref{eq:Ispc} is a rapidly oscillating function of $\phi$. Consequently, terms in a small neighborhood around a point $\phi_0$ interfere destructively and average to zero unless $f_\theta' \left(\phi_0\right) = 0$. To derive the desired asymptotic behavior, it then suffices to expand $z_{\mu}\left(\phi\right)$ and $f_\theta \left(\phi\right)$ to zeroth and second order, respectively, around the stationary points of $f_\theta \left(\phi\right)$. By following this prescription, we obtain
\begin{align}
	\Braket{S_{i\mu}^2} \approx \frac{S v_{uc}}{\sqrt{8\pi R_i}} \sum_{\phi_0 \in \Gamma}
	\frac{ \absval{z_{\mu} \! \left(\phi_0\right)} }{  \absval{f_{\theta_i}'' \! \left(\phi_0\right)}^\frac{1}{2} } \cos \left[ f_{\theta_i} \! \left(\phi_0\right)R_i + \xi \! \left(\phi_0\right) \right].
	\label{eq:oopasympgen}
\end{align}
Here, $\Gamma$ is the set of stationary points of $f_{\theta} \left(\phi\right)$ in the interval $\left[0,\phi_{\mathrm{max}}\right[$, while $\xi\left(\phi\right) = \arg z_{\mu} \! \left(\phi\right) + \mathrm{sgn} \left[ f_{\theta}'' \left(\phi\right) \right] \pi/4$ is a phase shift.

In the limit of small $\left( \Jrat-1/6 \right)$, where the spiral contour is nearly circular, $k_{\Jrat}\left(\phi\right) \approx k_{\Jrat}$, one can easily show that $\Gamma = \left\{\theta \!\! \mod \! \pi\right\}$ and that Eq.~\eqref{eq:oopasympgen} becomes
\begin{equation}
	\Braket{S_{i\mu}^2} \approx \frac{ S v_{uc} \absval{z_{\mu}\left(\theta_i\right)} }{\sqrt{8\pi k_{\Jrat} R_i}}
	\cos\left[ k_{\Jrat} R_i + \arg z_{\mu}\left(\theta_i\right) - \frac{\pi}{4} \right].
	\label{eq:oopasympex}
\end{equation}
In fact, in this special case, it is even possible to evaluate Eq.~\eqref{eq:Ispc} exactly. To do so, one can exploit the property $z_{\mu}\left(\phi+\pi\right) = z_{\mu}\left(\phi\right)^*$ to conclude that Fourier series of the real (imaginary) part of $z_{\mu}\left(\phi\right)$ only has nonzero coefficients for sines and cosines of even (odd) multiples of $\phi$. Then, one can perform the integrals and express $I_{\mathrm{spc}}$ as a series of Bessel functions. By taking the asymptotic limit of the latter, one recovers Eq.~\eqref{eq:oopasympex}.

Given that the out-of-plane texture arises from zero modes, we need to ask what happens if these modes acquire a small finite gap, either by additional couplings or by interaction effects beyond $T=0$ linear spin-wave theory. While Eq.~\eqref{eq:linrespav2} shows that the linear out-of-plane response is then strictly zero, continuity demands that an out-of-plane texture still emerges from defects of sufficient strength to overcome the gap. Hence, the relevant response will be strongly non-linear, but for stronger defects our results remain qualitatively valid.

\subsection{In-plane textures: A case study}
\label{subsec:casetex}

To obtain the in-plane component $\Braket{S_{i\mu}^1}$ of the texture, one must compute the Fourier transform in Eq.~\eqref{eq:linrespav1}. However, this step turns out to be highly nontrivial due to presence of pinch points and (integrable) singularities in $\tilde{\chi}_{\mathbf{k}\mu}^{11}$ [see Eqs.~\eqref{eq:chikNNbd} and \eqref{eq:chikvac} below]. Being unable to compute the Fourier integrals that arise in the thermodynamic limit to a satisfactory degree of accuracy, we instead performed discrete Fourier transforms on grids of $M^2$ momenta corresponding to finite systems of $N=2M^2$ sites under periodic boundary conditions. To warrant this choice, we slightly adjusted the desired $\Jrat$ to the nearest value $\tilde{\Jrat}$ which guaranteed that the spin spirals used as reference states for the spin-wave expansion were commensurate with our smallest cluster ($M_\mathrm{min} = 2^6$). The results shown in Figs.~\ref{fig:linrespNN1}, \ref{fig:linrespNN2} and \ref{fig:linrespVAC} below refer to use $\Jrat = 0.2$ and $\tilde{\Jrat} \approx 0.201$. We then ensured that the same condition held for larger systems by only considering values of $M$ that were multiples of $M_\mathrm{min}$.

In practice, before performing the discrete Fourier transform itself, we mapped the spin model onto a graph-equivalent brickwall lattice with primitive vectors $\mathbf{a}_1 = \sqrt{2}a \left( 1,0 \right)$ and $\mathbf{a}_2 = \sqrt{2}a \left( 0,1 \right)$. This resulted in a rectangular equispaced momentum grid and enabled us to employ a standard fast Fourier transform routine \cite{NumericalRecipes}, with which we could easily access system sizes up to $M = 2^{14}$.
Whenever a momentum $\mathbf{q}$ coincided with the position of a zero mode $\epsilon_{\mathbf{q} \lambda} = 0$, we replaced the result obtained from the expression in Eq.~\eqref{eq:chik} with an average over the susceptibilities at the four nearest momenta. This regularization procedure is innocuous in the cases where $\mathbf{q}$ lies on top of the spiral contour, since $\lim_{\mathbf{k} \to \mathbf{q}} \tilde{\chi}_{\mathbf{k}\mu}^{11}$ is then well defined. However, it leads to artifacts in the Fourier transform when applied to the singular $\mathbf{q} = \boldsymbol{0}$ point. Fortunately, these features only appear at distances on the order of the inverse of the spacing between points in the momentum grid, such that we can still extract meaningful results from the calculation.
Finally, we carried out the reverse mapping to restore the geometry of the honeycomb lattice.

\subsubsection{Nearest-neighbor bond defect}
\label{subsubsec:NN1iptex}

For a NN bond defect that changes the coupling between sites $m0$ and $n1$ by an amount $\delta J_1$, the corrections to the coupling matrix $\delta T_{ij}^{\mu\nu}$ are given by $\delta T_{ij}^{00} = \delta T_{ij}^{11} = 0$ and
\begin{equation}
	 \delta T_{ij}^{01} =  -\delta T_{ji}^{10} = \delta_{im} \delta_{jn}  \delta J_1 \sin\left(\QQ\cdot\mathbf{R}_{nm} + \alphaQ\right).
\end{equation}
Thus, the form factors in Eq.~\eqref{eq:formfactordef} read
\begin{align}
	\beta_{\mathbf{k}0} &= -\absval{\delta J_1} \sin\left(\QQ\cdot\mathbf{R}_{nm} + \alphaQ\right),
	\notag \\
	\beta_{\mathbf{k}1} &= \absval{\delta J_1} \sin\left(\QQ\cdot\mathbf{R}_{nm} + \alphaQ\right) e^{i\mathbf{k}\cdot\mathbf{R}_{nm}}.
	\label{eq:NNbetas}
\end{align}
Note that the $\QQ$-dependent prefactor is (by no mistake) proportional to the tranverse component of the deviation field in Eq.~\eqref{eq:dhJ1bd}. Therefore, it vanishes at the pair of ordering wave vectors selected by $\delta J_1>0$, in agreement with our previous conclusion that no texture arises in this case (see Table~\ref{tab:sbd}).

\begin{figure}
	\centering{
		\includegraphics[width=\columnwidth]{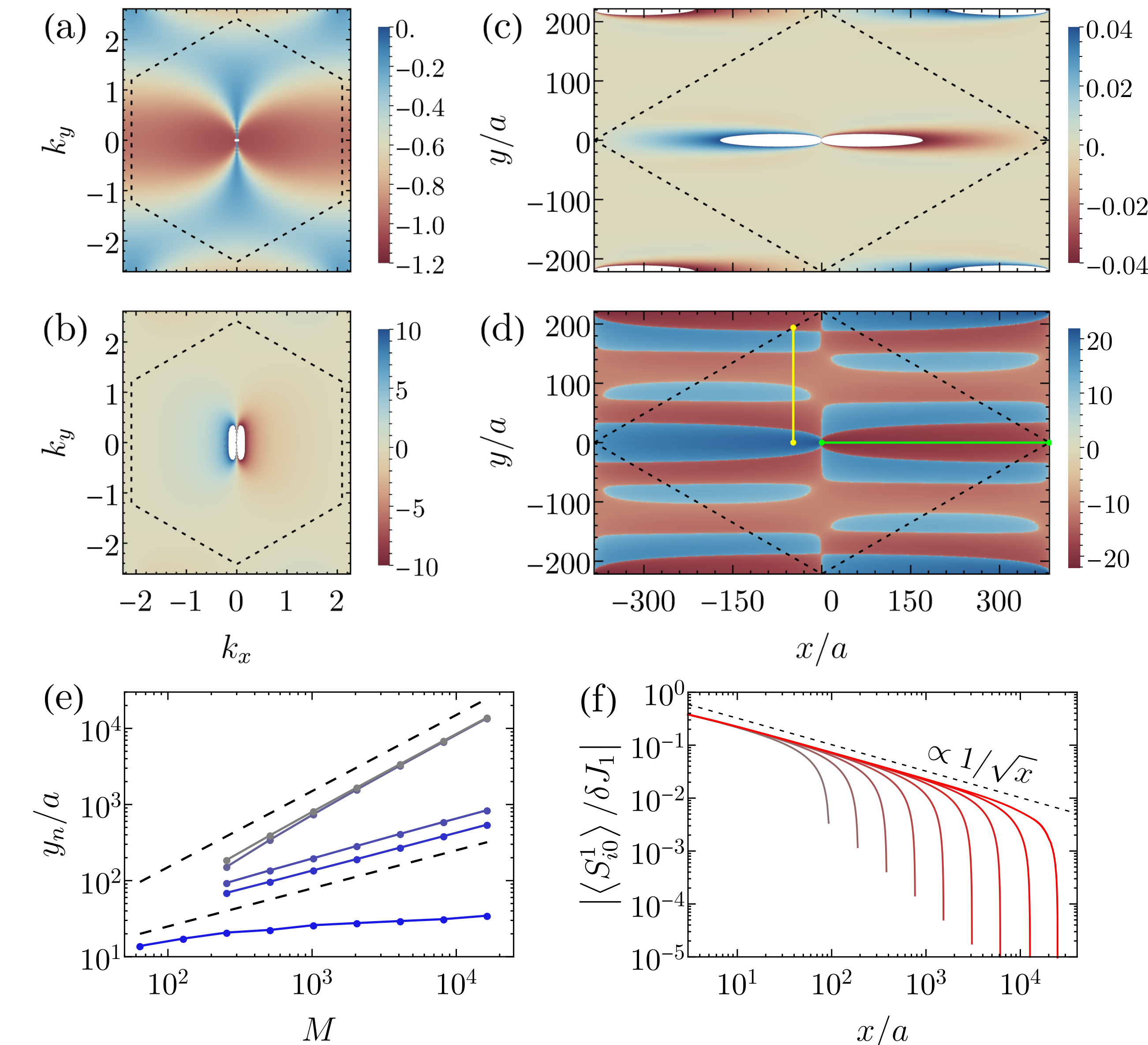}
	}
	\caption{
		Linear-response results for a horizontal ($\mathbf{R}_{nm}=\boldsymbol{0}$) NN bond defect with $\delta J_1 < 0$, obtained by using the $\phiQ =0$ spin spiral as a reference state for the spin-wave expansion.
		(a,b) Real and imaginary parts, respectively, of $\tilde{\chi}^{11}_{\mathbf{k}0} \left(\omega=0\right)/\left( \absval{\delta J_1}S \right)$. The dashed hexagons indicate the boundaries of the Brillouin zone.
		(c,d) Real-space configuration of $\Braket{S_{i0}^1}/ \absval{\delta J_1}$ on a linear and logarithmic scale $\mathrm{sgn}\left(z\right) \ln \left(\absval{z/z_\mathrm{min}}\right)$, respectively. These in-plane textures correspond to a periodic system with $M = 256$ and are depicted such that an impurity lies at the center of the unit cell outlined in black.
		(e) Finite-size scaling of the position $y_n$ of the $n$-th zero along the yellow line $x = 48a$ in (d). The dashed black lines indicate the slope of power laws proportional to $\sqrt{M}$ and $M$.
		(f) Decay of the texture along the green high-symmetry line in (d). The color code was selected such that the brighteness of the shade of red increases with system size.
		In (b) and (c), white colors occur when the susceptibility exceeds the limits of the plot legends.
	}
	\label{fig:linrespNN1}
\end{figure}

By combining Eqs.~\eqref{eq:chik} and \eqref{eq:NNbetas}, we computed the static susceptibility for a horizontal defect ($\mathrm{R}_{nm} = \boldsymbol{0}$) with $\delta J_1 < 0$. To keep consistency with our ObQD analysis, we chose the coplanar spiral with $\phiQ = 0$ as the classical reference state for the spin-wave expansion. The result for the $\mu = 0$ sublattice is depicted in Figs.~\ref{fig:linrespNN1}(a,b). There, we see that the (a) real and (b) imaginary parts of $\tilde{\chi}_{\mathbf{k}0}^{11}$ are both even under a reflection with a mirror plane parallel to $\QQ$, yet differ in parity when the mirror plane is taken perpendicularly to $\QQ$. After taking the observed symmetry properties into account and performing fits to numerical results, we find that the susceptibility behaves as
\begin{equation}
	\frac{ \tilde{\chi}_{0}^{11} \left( k,\phi \right) }{S \absval{\delta J_1}} =
	 a + \frac{b}{\cos^2 \phi + ck^2} \left(k^2 + i \frac{d\cos\phi}{k} \right)
	\label{eq:chikNNbd}
\end{equation}
in the long-wavelength limit $k\to0$. Here, $\phi$ denotes the polar angle respect to $\QQ$, whereas $a$, $b$, $c$, and $d$ are real constants.  Hence, the susceptibility displays both a pinch point and a simple pole at $\mathbf{k} = \boldsymbol{0}$.

Figures~\ref{fig:linrespNN1}(c,d) show the spatial dependence of $\Braket{S_{i0}^1}/\absval{\delta J_1}$ for a system with $M = 256$, as obtained by taking a discrete Fourier transform of the full momentum-space susceptibility depicted in Figs.~\ref{fig:linrespNN1}(a,b). Though the texture has the same mirror symmetries as a dipole, it is strongly confined along the high-symmetry direction parallel to $\QQ$. In the orthogonal direction, it displays a curious oscillatory behavior which gives rise to nodal lines (i.e., lines of zeros).
To determine whether the latter persist in the thermodynamic limit, we tracked how the position $y_n$ of the $n$-th zero and the height $h_n$ of the $n$-th peak along a cut with fixed $x$ [see yellow line highlighted in Fig.~\ref{fig:linrespNN1}(d)] evolve with $M$.
As shown in Fig.~\ref{fig:linrespNN1}(e), $y_1$ displays a slow (possibly logarithmic) growth, while $y_n \propto \sqrt{M}$ for $n=2,3$ and $y_n \propto M$ for $n=4,5$. Moreover, we verified that $h_1$ converges to a nonzero constant, but all other peaks vanish according to $h_n \propto M^{-3/2}$.
When combined, these results indicate that all $n\ge2$ nodal lines are finite-size artifacts which disappear as $M\to\infty$. On the other hand, the $n=1$ nodal line collapses onto $x = 0$ as the two lobes crossing $y=0$ in Fig.~\ref{fig:linrespNN1}(d) expand into half-planes and the texture acquires a $p$-wave-like symmetry.

Given that the finite-size effects are weakest parallel to $\QQ$, we resorted to a cut along this high-symmetry direction to determine the decay rate of the texture at distances $r$ far from the impurity. From the data shown in Fig.~\ref{fig:linrespNN1}(f), we extracted a power-law behavior of $r^{-\gamma}$, with $\gamma \approx 0.5$, as stated in Sec.~\ref{subsec:textures}.

\subsubsection{Next-nearest-neighbor bond defect}
\label{subsubsec:NN2iptex}

Alternatively, if we consider a bond defect between NNN sites $m0$ and $n0$, we have $\delta T_{ij}^{01} = \delta T_{ij}^{11} = 0$ and
\begin{equation}
	\delta T_{ij}^{00} = \delta J_2 \left( \delta_{in}\delta_{jm} - \delta_{im}\delta_{jn} \right) \sin\left(\QQ\cdot\mathbf{R}_{nm}\right),
\end{equation}
which leads to the form factors
\begin{align}
	\beta_{\mathbf{k}0} &= \delta J_2 e^{-i \mathbf{k} \cdot \mathbf{R}_m} \left( e^{-i \mathbf{k} \cdot \mathbf{R}_{nm}} - 1 \right) \sin\left(\QQ \cdot \mathbf{R}_{nm}\right),
	\notag \\
	\beta_{\mathbf{k}1} &= 0.
\end{align}
As before, these expressions are fully consistent with our ObQD results. The fact that $\delta J_2 < 0$ selects states with $\QQ \perp \mathbf{R}_{nm}$ implies that $\beta_{\mathbf{k}0} = 0$ and, consequently, that the defect only induces a texture if $\delta J_2 > 0$ (see Table~\ref{tab:sbd}).

\begin{figure}
	\centering{
		\includegraphics[width=\columnwidth]{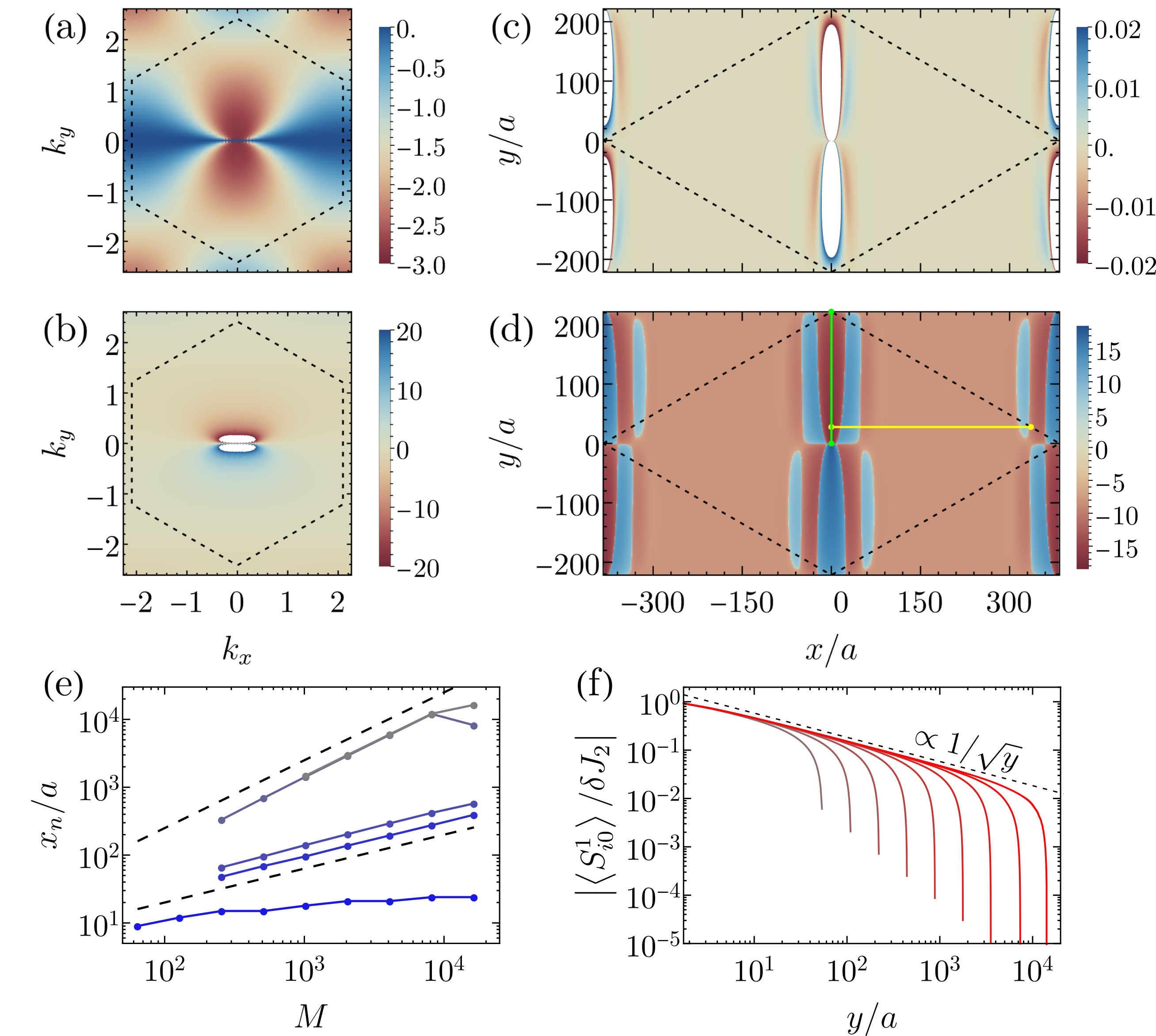}
	}
	\caption{Similar to Fig.~\ref{fig:linrespNN1}, but for a vertical ($\mathbf{R}_{nm} = \mathbf{a}_3$) NNN bond defect on the $\mu=0$ sublattice. To keep consistency with the ObQD result, the spin-wave calculations were performed by using the $\phiQ = \pi/2$ coplanar spiral as a reference state.}
	\label{fig:linrespNN2}
\end{figure}

By using the disorder-selected $\phiQ = \pi/2$ spin spiral as a reference state for the spin-wave expansion and assuming that the defect is vertical ($\mathbf{R}_{nm} = \mathbf{a}_3$), we obtained the momentum-dependent $\mu = 0$ susceptibility shown in Figs.~\ref{fig:linrespNN2}(a,b). Apart from a rotation by 90°, which is due to the change in $\phiQ$, this result has the same symmetries as the susceptibility for the NN bond defect. Upon closer inspection, one finds that it also behaves as Eq.~\eqref{eq:chikNNbd} at small $k$, albeit with a different set of constants.
Given such similarities, it is not entirely surprising that, after going through the same type of analysis we described above, we conclude that a strong NNN bond defect also induces a texture with $p$-wave-like symmetry which decays as $r^{-\gamma}$, with $\gamma \approx 0.5$, at long distances [see Figs.~\ref{fig:linrespNN2}(c-f)].

\subsubsection{Vacancy}
\label{subsubsec:VACiptex}

\begin{figure}
\centering
\includegraphics[width=\columnwidth]{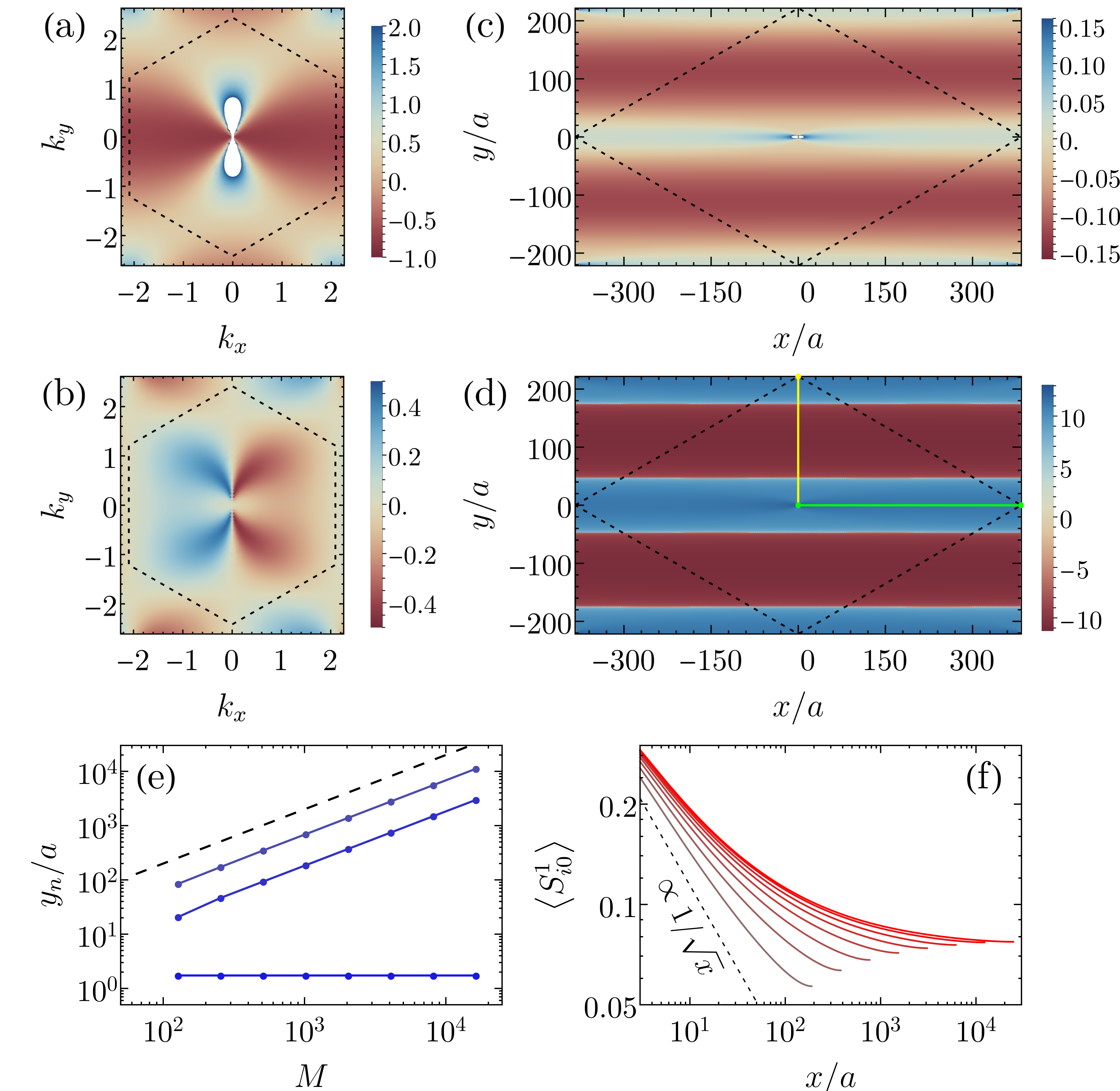}

\caption{Same as Fig.~\ref{fig:linrespNN1}, but for a single vacancy on the $\mu=0$ sublattice.}
\label{fig:linrespVAC}
\end{figure}

Finally, we turn to the case of a single vacancy located at site $m0$, which we can take as the origin without loss of generality. In this case, the only changes to the coupling matrix $T_{ij}^{\mu\nu}$ come from
\begin{align}
	\delta T_{mj}^{00} &= J_2 \delta_{\absval{\mathbf{r}_{j0}},\sqrt{3}a} \sin\left(\QQ \cdot \mathbf{R}_j \right),
	\notag \\
	\delta T_{mj}^{01} &= -J_1 \delta_{\absval{\mathbf{r}_{j1}},a} \sin\left(\QQ \cdot \mathbf{R}_j + \alphaQ \right),
\end{align}
and the corresponding transpose elements. Using the same shorthand notation introduced before Eq.~\eqref{eq:LSWmatelements}, we can express the resulting form factors as
\begin{align}
	\beta_{\mathbf{k}0} &= 2i J_2 \sum_{a=1}^{3} \sin Q_a \sin k_a,
	\notag \\
	\beta_{\mathbf{k}1} &= J_1 \sum_{a=0}^{2} e^{ik_a} \sin\left(\alphaQ - Q_a\right).
\end{align}

After performing the spin-wave expansion with respect to the $\phiQ = 0$ spiral state (as suggested by our ObQD results discussed Sec.~\ref{subsec:svac}), one obtains the susceptibility $\tilde{\chi}_{\mathbf{k}0}^{11}$ depicted in Figs.~\ref{fig:linrespVAC}(a,b). Once again, the real and imaginary parts are even under a reflection with respect to a plane parallel to $\QQ$, but have opposite parities when the mirror plane is chosen perpendicularly to $\QQ$. However, apart from these symmetries, the vacancy susceptibility bears little resemblance to the bond-defect susceptibilities. In particular, its long-wavelength behavior is markedly different from Eq.~\eqref{eq:chikNNbd} and can be well approximated as
\begin{equation}
	\tilde{\chi}_{0}^{11} \left( k,\phi \right) =
	a + \frac{b}{\cos^2 \phi + ck^2} \left[ 1 + i k \left(d \cos\phi + \bar{d} \cos^3 \phi\right) \right],
	\label{eq:chikvac}
\end{equation}
with real constants $a$, $b$, $c$, $d$, and $\bar{d}$. This expression captures the fact that $\mathrm{Re} \tilde{\chi}_{\mathbf{k}0}^{11}$ diverges as $1/k^2$ when $\phi \to \pi/2$, but is well behaved when one approaches the origin from other directions.

The corresponding spin texture for a system of size $M = 256$ is plotted in Figs.~\ref{fig:linrespVAC}(c,d). One of its most prominent features is that $\Braket{S_{i0}^1}$ is sizeable across almost the entire unit cell. The only exceptions occur around horizontal nodal lines, which are parallel to $\QQ$. Proceeding as in the previous cases, we tracked the finite-size evolution of the position $y_n$ of the $n$-th zero and the height $h_n$ of the $n$-th peak along on the vertical yellow line in Figs.~\ref{fig:linrespVAC}(d).
The result shows that the positions all zeros but the first, which is located in the immediate vicinity of the vacancy, scale as $M$ and thus disappear in the limit of a single impurity [see Figs.~\ref{fig:linrespVAC}(e)]. The heights $h_n$ vary only weakly, without any discernible trend.

Hence, we sought to determine the long-distance behavior of the $M\to\infty$ texture by analyzing the scaling of the data along the high-symmetry green line in Fig.~\ref{fig:linrespVAC}(d), given that it belongs to the region that is the least affected by finite-size effects. Figure~\ref{fig:linrespVAC}(f) shows that the outcome is quite remarkable: Within linear-response theory, the texture induced by a single vacancy decays more slowly than \emph{any} power law. We note that the pinch-point structure in Eq.~\eqref{eq:chikvac} provides an intrinsic length scale, such that the unconventional texture is compatible with dimensional analysis.

\subsubsection{Changes in textures with increasing $\Jrat$}

\begin{figure}
\centering
\includegraphics[width=\columnwidth]{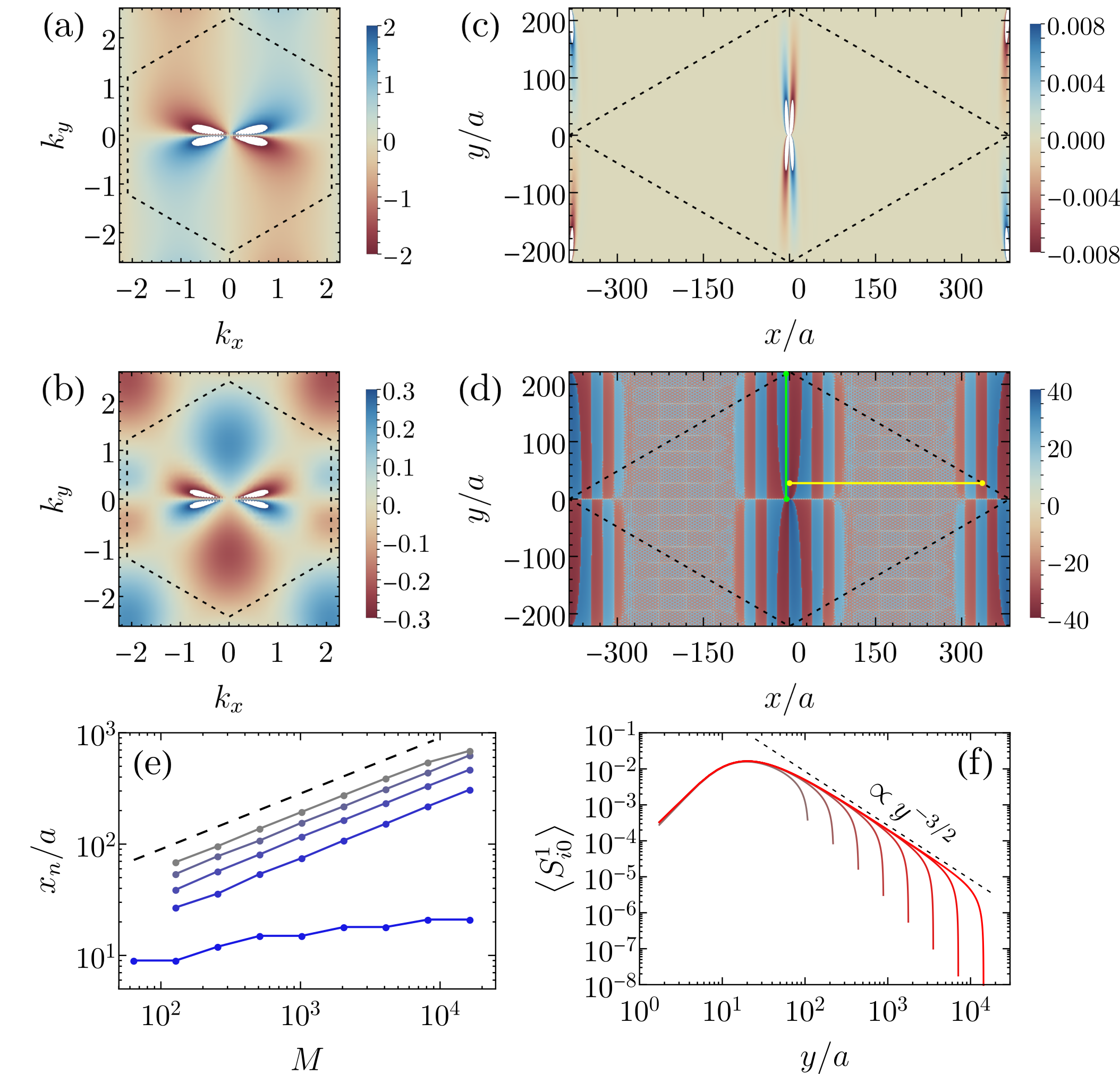}
\caption{Same as Fig.~\ref{fig:linrespVAC}, but for $\Jrat = 0.3$.}
\label{fig:linrespVAC2}
\end{figure}

\begin{figure}
	\centering{
		\includegraphics[width=0.7\columnwidth]{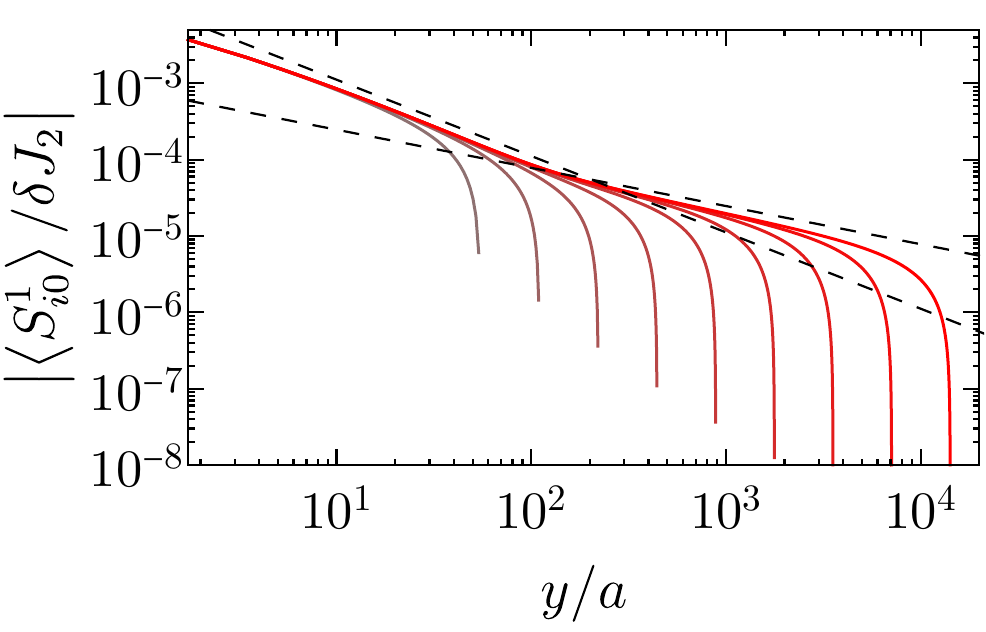}
	}
	\caption{Spatial decay of the texture generated by a weak NNN bond defect for $\Jrat = 30$. The two dashed lines correspond to power laws $1/y$ and $1/\sqrt{y}$, indicating that the behavior expected for a triangular-lattice antiferromagnet ($\Jrat \to \infty$) emerges below a characteristic distance $R_{\mathrm{cross}} \left(\Jrat\right)$ from the impurity.}
	\label{fig:textcross}
\end{figure}

As mentioned in Sec.~\ref{subsec:textures}, the qualitative form of the texture induced by a vacancy is affected by the change in the states selected by ObQD once $\Jrat \gtrsim 0.2148$. To demonstrate this, we present in Fig.~\ref{fig:linrespVAC2} numerical results which were obtained for $\Jrat = 0.3$ and a classical reference state $\phiQ = \pi/2$.
When compared to Fig.~\ref{fig:linrespVAC}, we observe clear differences in the symmetries of the momentum-state susceptibility (a,b), which are directly reflected in the real space (c,d). Based of our previous analyses, Figs.~\ref{fig:linrespVAC2}(c,d) suggests that the $\Jrat = 0.3$ texture displays a $d$-wave-like symmetry, with two orthogonal nodal lines.
Proceeding as before, we sought to confirm this by tracking the positions $y_{n}$ of the zeros along the yellow cut in Fig.~\ref{fig:linrespVAC2}(d). The results, which are depicted in Fig.~\ref{fig:linrespVAC2}(e), indicate that $y_{1}$ grows logarithmically with $M$, while the rest scale as $\sqrt{M}$. Combined with the fact that the heights of secondary peaks along the same cut vanish as $1/\sqrt{M}$ (not shown), we confirm our initial hypothesis.
The data in Fig.~\ref{fig:linrespVAC2}(f), which corresponds to the variation of texture along the green vertical cut highlighted in Fig.~\ref{fig:linrespVAC2}(d), furthermore shows that the texture decays as $r^{-3/2}$ at large distances from the vacancy.

We also pointed out in Sec.~\ref{subsec:textures} that the results we derived above differ from those expected for the limit $\Jrat \to \infty$, where the system becomes two decoupled triangular-lattice antiferromagnets. However, we explained that this is not a contradiction, because the behavior appropriate for the triangular lattice emerges as a short-distance effect and crosses over to our results at a characteristic length $R_{\mathrm{cross}} \left(\Jrat\right)$.
This is exemplified in Fig.~\ref{fig:textcross}, which illustrates the spatial decay of the texture generated by a weak NNN bond defect for $\Jrat = 30$. There, one sees a clear crossover between a power law $1/y$ to $1/\sqrt{y}$ upon increasing $y$, as we claimed.

%%%%%%%%%%%%%%%%%%%%%%%%%%%%%%%%%%%%%%%%%%%%%%%%%%%%%%%%%%%%%%%%%%%

\bibliography{ssldis}

\begin{thebibliography}{67}%
\makeatletter
\providecommand \@ifxundefined [1]{%
 \@ifx{#1\undefined}
}%
\providecommand \@ifnum [1]{%
 \ifnum #1\expandafter \@firstoftwo
 \else \expandafter \@secondoftwo
 \fi
}%
\providecommand \@ifx [1]{%
 \ifx #1\expandafter \@firstoftwo
 \else \expandafter \@secondoftwo
 \fi
}%
\providecommand \natexlab [1]{#1}%
\providecommand \enquote  [1]{``#1''}%
\providecommand \bibnamefont  [1]{#1}%
\providecommand \bibfnamefont [1]{#1}%
\providecommand \citenamefont [1]{#1}%
\providecommand \href@noop [0]{\@secondoftwo}%
\providecommand \href [0]{\begingroup \@sanitize@url \@href}%
\providecommand \@href[1]{\@@startlink{#1}\@@href}%
\providecommand \@@href[1]{\endgroup#1\@@endlink}%
\providecommand \@sanitize@url [0]{\catcode `\\12\catcode `\$12\catcode
  `\&12\catcode `\#12\catcode `\^12\catcode `\_12\catcode `\%12\relax}%
\providecommand \@@startlink[1]{}%
\providecommand \@@endlink[0]{}%
\providecommand \url  [0]{\begingroup\@sanitize@url \@url }%
\providecommand \@url [1]{\endgroup\@href {#1}{\urlprefix }}%
\providecommand \urlprefix  [0]{URL }%
\providecommand \Eprint [0]{\href }%
\providecommand \doibase [0]{https://doi.org/}%
\providecommand \selectlanguage [0]{\@gobble}%
\providecommand \bibinfo  [0]{\@secondoftwo}%
\providecommand \bibfield  [0]{\@secondoftwo}%
\providecommand \translation [1]{[#1]}%
\providecommand \BibitemOpen [0]{}%
\providecommand \bibitemStop [0]{}%
\providecommand \bibitemNoStop [0]{.\EOS\space}%
\providecommand \EOS [0]{\spacefactor3000\relax}%
\providecommand \BibitemShut  [1]{\csname bibitem#1\endcsname}%
\let\auto@bib@innerbib\@empty
%</preamble>
\bibitem [{\citenamefont {Bergman}\ \emph {et~al.}(2007)\citenamefont
  {Bergman}, \citenamefont {Alicea}, \citenamefont {Gull}, \citenamefont
  {Trebst},\ and\ \citenamefont {Balents}}]{Bergman2007}%
  \BibitemOpen
  \bibfield  {author} {\bibinfo {author} {\bibfnamefont {D.}~\bibnamefont
  {Bergman}}, \bibinfo {author} {\bibfnamefont {J.}~\bibnamefont {Alicea}},
  \bibinfo {author} {\bibfnamefont {E.}~\bibnamefont {Gull}}, \bibinfo {author}
  {\bibfnamefont {S.}~\bibnamefont {Trebst}},\ and\ \bibinfo {author}
  {\bibfnamefont {L.}~\bibnamefont {Balents}},\ }\bibfield  {title} {\bibinfo
  {title} {Order-by-disorder and spiral spin-liquid in frustrated
  diamond-lattice antiferromagnets},\ }\href {https://doi.org/10.1038/nphys622}
  {\bibfield  {journal} {\bibinfo  {journal} {Nat. Phys.}\ }\textbf {\bibinfo
  {volume} {3}},\ \bibinfo {pages} {487} (\bibinfo {year} {2007})}\BibitemShut
  {NoStop}%
\bibitem [{\citenamefont {Gao}\ \emph {et~al.}(2017)\citenamefont {Gao},
  \citenamefont {Zaharko}, \citenamefont {Tsurkan}, \citenamefont {Su},
  \citenamefont {White}, \citenamefont {Tucker}, \citenamefont {Roessli},
  \citenamefont {Bourdarot}, \citenamefont {Sibille}, \citenamefont
  {Chernyshov}, \citenamefont {Fennell}, \citenamefont {Loidl},\ and\
  \citenamefont {R{\"u}egg}}]{Gao2017}%
  \BibitemOpen
  \bibfield  {author} {\bibinfo {author} {\bibfnamefont {S.}~\bibnamefont
  {Gao}}, \bibinfo {author} {\bibfnamefont {O.}~\bibnamefont {Zaharko}},
  \bibinfo {author} {\bibfnamefont {V.}~\bibnamefont {Tsurkan}}, \bibinfo
  {author} {\bibfnamefont {Y.}~\bibnamefont {Su}}, \bibinfo {author}
  {\bibfnamefont {J.~S.}\ \bibnamefont {White}}, \bibinfo {author}
  {\bibfnamefont {G.}~\bibnamefont {Tucker}}, \bibinfo {author} {\bibfnamefont
  {B.}~\bibnamefont {Roessli}}, \bibinfo {author} {\bibfnamefont
  {F.}~\bibnamefont {Bourdarot}}, \bibinfo {author} {\bibfnamefont
  {R.}~\bibnamefont {Sibille}}, \bibinfo {author} {\bibfnamefont
  {D.}~\bibnamefont {Chernyshov}}, \bibinfo {author} {\bibfnamefont
  {T.}~\bibnamefont {Fennell}}, \bibinfo {author} {\bibfnamefont
  {A.}~\bibnamefont {Loidl}},\ and\ \bibinfo {author} {\bibfnamefont
  {C.}~\bibnamefont {R{\"u}egg}},\ }\bibfield  {title} {\bibinfo {title}
  {{Spiral spin-liquid and the emergence of a vortex-like state in
  MnSc$_2$S$_4$}},\ }\href {https://doi.org/10.1038/nphys3914} {\bibfield
  {journal} {\bibinfo  {journal} {Nat. Phys.}\ }\textbf {\bibinfo {volume}
  {13}},\ \bibinfo {pages} {157} (\bibinfo {year} {2017})}\BibitemShut
  {NoStop}%
\bibitem [{\citenamefont {Gao}\ \emph {et~al.}(2022{\natexlab{a}})\citenamefont
  {Gao}, \citenamefont {McGuire}, \citenamefont {Liu}, \citenamefont
  {Abernathy}, \citenamefont {Cruz}, \citenamefont {Frontzek}, \citenamefont
  {Stone},\ and\ \citenamefont {Christianson}}]{Gao2022a}%
  \BibitemOpen
  \bibfield  {author} {\bibinfo {author} {\bibfnamefont {S.}~\bibnamefont
  {Gao}}, \bibinfo {author} {\bibfnamefont {M.~A.}\ \bibnamefont {McGuire}},
  \bibinfo {author} {\bibfnamefont {Y.}~\bibnamefont {Liu}}, \bibinfo {author}
  {\bibfnamefont {D.~L.}\ \bibnamefont {Abernathy}}, \bibinfo {author}
  {\bibfnamefont {C.~d.}\ \bibnamefont {Cruz}}, \bibinfo {author}
  {\bibfnamefont {M.}~\bibnamefont {Frontzek}}, \bibinfo {author}
  {\bibfnamefont {M.~B.}\ \bibnamefont {Stone}},\ and\ \bibinfo {author}
  {\bibfnamefont {A.~D.}\ \bibnamefont {Christianson}},\ }\bibfield  {title}
  {\bibinfo {title} {{Spiral Spin Liquid on a Honeycomb Lattice}},\ }\href
  {https://doi.org/10.1103/PhysRevLett.128.227201} {\bibfield  {journal}
  {\bibinfo  {journal} {Phys. Rev. Lett.}\ }\textbf {\bibinfo {volume} {128}},\
  \bibinfo {pages} {227201} (\bibinfo {year} {2022}{\natexlab{a}})}\BibitemShut
  {NoStop}%
\bibitem [{\citenamefont {Graham}\ \emph {et~al.}(2023)\citenamefont {Graham},
  \citenamefont {Qureshi}, \citenamefont {Ritter}, \citenamefont {Manuel},
  \citenamefont {Wildes},\ and\ \citenamefont {Clark}}]{Graham2023}%
  \BibitemOpen
  \bibfield  {author} {\bibinfo {author} {\bibfnamefont {J.~N.}\ \bibnamefont
  {Graham}}, \bibinfo {author} {\bibfnamefont {N.}~\bibnamefont {Qureshi}},
  \bibinfo {author} {\bibfnamefont {C.}~\bibnamefont {Ritter}}, \bibinfo
  {author} {\bibfnamefont {P.}~\bibnamefont {Manuel}}, \bibinfo {author}
  {\bibfnamefont {A.~R.}\ \bibnamefont {Wildes}},\ and\ \bibinfo {author}
  {\bibfnamefont {L.}~\bibnamefont {Clark}},\ }\bibfield  {title} {\bibinfo
  {title} {{Experimental Evidence for the Spiral Spin Liquid in LiYbO$_2$}},\
  }\href {https://doi.org/10.1103/PhysRevLett.130.166703} {\bibfield  {journal}
  {\bibinfo  {journal} {Phys. Rev. Lett.}\ }\textbf {\bibinfo {volume} {130}},\
  \bibinfo {pages} {166703} (\bibinfo {year} {2023})}\BibitemShut {NoStop}%
\bibitem [{\citenamefont {Zaharko}\ \emph {et~al.}(2011)\citenamefont
  {Zaharko}, \citenamefont {Christensen}, \citenamefont {Cervellino},
  \citenamefont {Tsurkan}, \citenamefont {Maljuk}, \citenamefont {Stuhr},
  \citenamefont {Niedermayer}, \citenamefont {Yokaichiya}, \citenamefont
  {Argyriou}, \citenamefont {Boehm},\ and\ \citenamefont
  {Loidl}}]{Zaharko2011}%
  \BibitemOpen
  \bibfield  {author} {\bibinfo {author} {\bibfnamefont {O.}~\bibnamefont
  {Zaharko}}, \bibinfo {author} {\bibfnamefont {N.~B.}\ \bibnamefont
  {Christensen}}, \bibinfo {author} {\bibfnamefont {A.}~\bibnamefont
  {Cervellino}}, \bibinfo {author} {\bibfnamefont {V.}~\bibnamefont {Tsurkan}},
  \bibinfo {author} {\bibfnamefont {A.}~\bibnamefont {Maljuk}}, \bibinfo
  {author} {\bibfnamefont {U.}~\bibnamefont {Stuhr}}, \bibinfo {author}
  {\bibfnamefont {C.}~\bibnamefont {Niedermayer}}, \bibinfo {author}
  {\bibfnamefont {F.}~\bibnamefont {Yokaichiya}}, \bibinfo {author}
  {\bibfnamefont {D.~N.}\ \bibnamefont {Argyriou}}, \bibinfo {author}
  {\bibfnamefont {M.}~\bibnamefont {Boehm}},\ and\ \bibinfo {author}
  {\bibfnamefont {A.}~\bibnamefont {Loidl}},\ }\bibfield  {title} {\bibinfo
  {title} {{Spin liquid in a single crystal of the frustrated diamond lattice
  antiferromagnet CoAl$_2$O$_4$}},\ }\href
  {https://doi.org/10.1103/PhysRevB.84.094403} {\bibfield  {journal} {\bibinfo
  {journal} {Phys. Rev. B}\ }\textbf {\bibinfo {volume} {84}},\ \bibinfo
  {pages} {094403} (\bibinfo {year} {2011})}\BibitemShut {NoStop}%
\bibitem [{\citenamefont {MacDougall}\ \emph {et~al.}(2011)\citenamefont
  {MacDougall}, \citenamefont {Gout}, \citenamefont {Zarestky}, \citenamefont
  {Ehlers}, \citenamefont {Podlesnyak}, \citenamefont {McGuire}, \citenamefont
  {Mandrus},\ and\ \citenamefont {Nagler}}]{MacDougall2011}%
  \BibitemOpen
  \bibfield  {author} {\bibinfo {author} {\bibfnamefont {G.~J.}\ \bibnamefont
  {MacDougall}}, \bibinfo {author} {\bibfnamefont {D.}~\bibnamefont {Gout}},
  \bibinfo {author} {\bibfnamefont {J.~L.}\ \bibnamefont {Zarestky}}, \bibinfo
  {author} {\bibfnamefont {G.}~\bibnamefont {Ehlers}}, \bibinfo {author}
  {\bibfnamefont {A.}~\bibnamefont {Podlesnyak}}, \bibinfo {author}
  {\bibfnamefont {M.~A.}\ \bibnamefont {McGuire}}, \bibinfo {author}
  {\bibfnamefont {D.}~\bibnamefont {Mandrus}},\ and\ \bibinfo {author}
  {\bibfnamefont {S.~E.}\ \bibnamefont {Nagler}},\ }\bibfield  {title}
  {\bibinfo {title} {{Kinetically inhibited order in a diamond-lattice
  antiferromagnet}},\ }\bibfield  {journal} {\bibinfo  {journal} {Proc. Natl.
  Acad. Sci. USA}\ }\textbf {\bibinfo {volume} {108}},\ \href
  {https://doi.org/https://doi.org/10.1073/pnas.1107861108}
  {https://doi.org/10.1073/pnas.1107861108} (\bibinfo {year}
  {2011})\BibitemShut {NoStop}%
\bibitem [{\citenamefont {MacDougall}\ \emph {et~al.}(2016)\citenamefont
  {MacDougall}, \citenamefont {Aczel}, \citenamefont {Su}, \citenamefont
  {Schweika}, \citenamefont {Faulhaber}, \citenamefont {Schneidewind},
  \citenamefont {Christianson}, \citenamefont {Zarestky}, \citenamefont {Zhou},
  \citenamefont {Mandrus},\ and\ \citenamefont {Nagler}}]{MacDougall2016}%
  \BibitemOpen
  \bibfield  {author} {\bibinfo {author} {\bibfnamefont {G.~J.}\ \bibnamefont
  {MacDougall}}, \bibinfo {author} {\bibfnamefont {A.~A.}\ \bibnamefont
  {Aczel}}, \bibinfo {author} {\bibfnamefont {Y.}~\bibnamefont {Su}}, \bibinfo
  {author} {\bibfnamefont {W.}~\bibnamefont {Schweika}}, \bibinfo {author}
  {\bibfnamefont {E.}~\bibnamefont {Faulhaber}}, \bibinfo {author}
  {\bibfnamefont {A.}~\bibnamefont {Schneidewind}}, \bibinfo {author}
  {\bibfnamefont {A.~D.}\ \bibnamefont {Christianson}}, \bibinfo {author}
  {\bibfnamefont {J.~L.}\ \bibnamefont {Zarestky}}, \bibinfo {author}
  {\bibfnamefont {H.~D.}\ \bibnamefont {Zhou}}, \bibinfo {author}
  {\bibfnamefont {D.}~\bibnamefont {Mandrus}},\ and\ \bibinfo {author}
  {\bibfnamefont {S.~E.}\ \bibnamefont {Nagler}},\ }\bibfield  {title}
  {\bibinfo {title} {{Revisiting the ground state of CoAl$_2$O$_4$: Comparison
  to the conventional antiferromagnet MnAl$_2$O$_4$}},\ }\href
  {https://doi.org/10.1103/PhysRevB.94.184422} {\bibfield  {journal} {\bibinfo
  {journal} {Phys. Rev. B}\ }\textbf {\bibinfo {volume} {94}},\ \bibinfo
  {pages} {184422} (\bibinfo {year} {2016})}\BibitemShut {NoStop}%
\bibitem [{\citenamefont {Bai}\ \emph {et~al.}(2019)\citenamefont {Bai},
  \citenamefont {Paddison}, \citenamefont {Kapit}, \citenamefont {Koohpayeh},
  \citenamefont {Wen}, \citenamefont {Dutton}, \citenamefont {Savici},
  \citenamefont {Kolesnikov}, \citenamefont {Granroth}, \citenamefont
  {Broholm}, \citenamefont {Chalker},\ and\ \citenamefont
  {M.~Mourigal}}]{Bai2019}%
  \BibitemOpen
  \bibfield  {author} {\bibinfo {author} {\bibfnamefont {X.}~\bibnamefont
  {Bai}}, \bibinfo {author} {\bibfnamefont {J.~A.~M.}\ \bibnamefont
  {Paddison}}, \bibinfo {author} {\bibfnamefont {E.}~\bibnamefont {Kapit}},
  \bibinfo {author} {\bibfnamefont {S.~M.}\ \bibnamefont {Koohpayeh}}, \bibinfo
  {author} {\bibfnamefont {J.-J.}\ \bibnamefont {Wen}}, \bibinfo {author}
  {\bibfnamefont {S.~E.}\ \bibnamefont {Dutton}}, \bibinfo {author}
  {\bibfnamefont {A.~T.}\ \bibnamefont {Savici}}, \bibinfo {author}
  {\bibfnamefont {A.~I.}\ \bibnamefont {Kolesnikov}}, \bibinfo {author}
  {\bibfnamefont {G.~E.}\ \bibnamefont {Granroth}}, \bibinfo {author}
  {\bibfnamefont {C.~L.}\ \bibnamefont {Broholm}}, \bibinfo {author}
  {\bibfnamefont {J.~T.}\ \bibnamefont {Chalker}},\ and\ \bibinfo {author}
  {\bibfnamefont {M.}~\bibnamefont {M.~Mourigal}},\ }\bibfield  {title}
  {\bibinfo {title} {{Magnetic Excitations of the Classical Spin Liquid
  MgCr$_2$O$_4$}},\ }\href {https://doi.org/10.1103/PhysRevLett.122.097201}
  {\bibfield  {journal} {\bibinfo  {journal} {Phys. Rev. Lett.}\ }\textbf
  {\bibinfo {volume} {122}},\ \bibinfo {pages} {097201} (\bibinfo {year}
  {2019})}\BibitemShut {NoStop}%
\bibitem [{\citenamefont {Rastelli}\ \emph {et~al.}(1979)\citenamefont
  {Rastelli}, \citenamefont {Tassi},\ and\ \citenamefont
  {Reatto}}]{Rastelli1979}%
  \BibitemOpen
  \bibfield  {author} {\bibinfo {author} {\bibfnamefont {E.}~\bibnamefont
  {Rastelli}}, \bibinfo {author} {\bibfnamefont {A.}~\bibnamefont {Tassi}},\
  and\ \bibinfo {author} {\bibfnamefont {L.}~\bibnamefont {Reatto}},\
  }\bibfield  {title} {\bibinfo {title} {{Non-simple magnetic order for simple
  Hamiltonians}},\ }\href {https://doi.org/10.1016/0378-4363(79)90002-0}
  {\bibfield  {journal} {\bibinfo  {journal} {Physica B+C}\ }\textbf {\bibinfo
  {volume} {97}},\ \bibinfo {pages} {1} (\bibinfo {year} {1979})}\BibitemShut
  {NoStop}%
\bibitem [{\citenamefont {Niggemann}\ \emph {et~al.}(2020)\citenamefont
  {Niggemann}, \citenamefont {Hering},\ and\ \citenamefont
  {Reuther}}]{Niggemann2020}%
  \BibitemOpen
  \bibfield  {author} {\bibinfo {author} {\bibfnamefont {N.}~\bibnamefont
  {Niggemann}}, \bibinfo {author} {\bibfnamefont {M.}~\bibnamefont {Hering}},\
  and\ \bibinfo {author} {\bibfnamefont {J.}~\bibnamefont {Reuther}},\
  }\bibfield  {title} {\bibinfo {title} {Classical spiral spin liquids as a
  possible route to quantum spin liquids},\ }\href
  {https://doi.org/10.1088/1361-648X/ab4480} {\bibfield  {journal} {\bibinfo
  {journal} {J. Phys.: Condens. Matter}\ }\textbf {\bibinfo {volume} {32}},\
  \bibinfo {pages} {024001} (\bibinfo {year} {2020})}\BibitemShut {NoStop}%
\bibitem [{\citenamefont {Yao}\ \emph {et~al.}(2021)\citenamefont {Yao},
  \citenamefont {Liu}, \citenamefont {Huang}, \citenamefont {Wang},\ and\
  \citenamefont {Chen}}]{Yao2021}%
  \BibitemOpen
  \bibfield  {author} {\bibinfo {author} {\bibfnamefont {X.-P.}\ \bibnamefont
  {Yao}}, \bibinfo {author} {\bibfnamefont {J.~Q.}\ \bibnamefont {Liu}},
  \bibinfo {author} {\bibfnamefont {C.-J.}\ \bibnamefont {Huang}}, \bibinfo
  {author} {\bibfnamefont {X.}~\bibnamefont {Wang}},\ and\ \bibinfo {author}
  {\bibfnamefont {G.}~\bibnamefont {Chen}},\ }\bibfield  {title} {\bibinfo
  {title} {Generic spiral spin liquids},\ }\href
  {https://doi.org/10.1007/s11467-021-1074-9} {\bibfield  {journal} {\bibinfo
  {journal} {Front. Phys.}\ }\textbf {\bibinfo {volume} {16}},\ \bibinfo
  {pages} {53303} (\bibinfo {year} {2021})}\BibitemShut {NoStop}%
\bibitem [{\citenamefont {Gao}\ \emph {et~al.}(2022{\natexlab{b}})\citenamefont
  {Gao}, \citenamefont {Pokharel}, \citenamefont {May}, \citenamefont
  {Paddison}, \citenamefont {Pasco}, \citenamefont {Liu}, \citenamefont
  {Taddei}, \citenamefont {Calder}, \citenamefont {Mandrus}, \citenamefont
  {Stone},\ and\ \citenamefont {Christianson}}]{Gao2022b}%
  \BibitemOpen
  \bibfield  {author} {\bibinfo {author} {\bibfnamefont {S.}~\bibnamefont
  {Gao}}, \bibinfo {author} {\bibfnamefont {G.}~\bibnamefont {Pokharel}},
  \bibinfo {author} {\bibfnamefont {A.~F.}\ \bibnamefont {May}}, \bibinfo
  {author} {\bibfnamefont {J.~A.~M.}\ \bibnamefont {Paddison}}, \bibinfo
  {author} {\bibfnamefont {C.}~\bibnamefont {Pasco}}, \bibinfo {author}
  {\bibfnamefont {Y.}~\bibnamefont {Liu}}, \bibinfo {author} {\bibfnamefont
  {K.~M.}\ \bibnamefont {Taddei}}, \bibinfo {author} {\bibfnamefont
  {S.}~\bibnamefont {Calder}}, \bibinfo {author} {\bibfnamefont {D.~G.}\
  \bibnamefont {Mandrus}}, \bibinfo {author} {\bibfnamefont {M.~B.}\
  \bibnamefont {Stone}},\ and\ \bibinfo {author} {\bibfnamefont {A.~D.}\
  \bibnamefont {Christianson}},\ }\bibfield  {title} {\bibinfo {title}
  {{Line-Graph Approach to Spiral Spin Liquids}},\ }\href
  {https://doi.org/10.1103/PhysRevLett.129.237202} {\bibfield  {journal}
  {\bibinfo  {journal} {Phys. Rev. Lett.}\ }\textbf {\bibinfo {volume} {129}},\
  \bibinfo {pages} {237202} (\bibinfo {year} {2022}{\natexlab{b}})}\BibitemShut
  {NoStop}%
\bibitem [{\citenamefont {Yan}\ and\ \citenamefont {Reuther}(2022)}]{Yan2022}%
  \BibitemOpen
  \bibfield  {author} {\bibinfo {author} {\bibfnamefont {H.}~\bibnamefont
  {Yan}}\ and\ \bibinfo {author} {\bibfnamefont {J.}~\bibnamefont {Reuther}},\
  }\bibfield  {title} {\bibinfo {title} {Low-energy structure of spiral spin
  liquids},\ }\href {https://doi.org/10.1103/PhysRevResearch.4.023175}
  {\bibfield  {journal} {\bibinfo  {journal} {Phys. Rev. Research}\ }\textbf
  {\bibinfo {volume} {4}},\ \bibinfo {pages} {023175} (\bibinfo {year}
  {2022})}\BibitemShut {NoStop}%
\bibitem [{\citenamefont {Mulder}\ \emph {et~al.}(2010)\citenamefont {Mulder},
  \citenamefont {Ganesh}, \citenamefont {Capriotti},\ and\ \citenamefont
  {Paramekanti}}]{Mulder2010}%
  \BibitemOpen
  \bibfield  {author} {\bibinfo {author} {\bibfnamefont {A.}~\bibnamefont
  {Mulder}}, \bibinfo {author} {\bibfnamefont {R.}~\bibnamefont {Ganesh}},
  \bibinfo {author} {\bibfnamefont {L.}~\bibnamefont {Capriotti}},\ and\
  \bibinfo {author} {\bibfnamefont {A.}~\bibnamefont {Paramekanti}},\
  }\bibfield  {title} {\bibinfo {title} {{Spiral order by disorder and lattice
  nematic order in a frustrated Heisenberg antiferromagnet on the honeycomb
  lattice}},\ }\href {https://doi.org/10.1103/PhysRevB.81.214419} {\bibfield
  {journal} {\bibinfo  {journal} {Phys. Rev. B}\ }\textbf {\bibinfo {volume}
  {81}},\ \bibinfo {pages} {214419} (\bibinfo {year} {2010})}\BibitemShut
  {NoStop}%
\bibitem [{\citenamefont {Okumura}\ \emph {et~al.}(2010)\citenamefont
  {Okumura}, \citenamefont {Kawamura}, \citenamefont {Okubo},\ and\
  \citenamefont {Motome}}]{Okumura2010}%
  \BibitemOpen
  \bibfield  {author} {\bibinfo {author} {\bibfnamefont {S.}~\bibnamefont
  {Okumura}}, \bibinfo {author} {\bibfnamefont {H.}~\bibnamefont {Kawamura}},
  \bibinfo {author} {\bibfnamefont {T.}~\bibnamefont {Okubo}},\ and\ \bibinfo
  {author} {\bibfnamefont {Y.}~\bibnamefont {Motome}},\ }\bibfield  {title}
  {\bibinfo {title} {{Novel Spin-Liquid States in the Frustrated Heisenberg
  Antiferromagnet on the Honeycomb Lattice}},\ }\href
  {https://doi.org/10.1143/JPSJ.79.114705} {\bibfield  {journal} {\bibinfo
  {journal} {J. Phys. Soc. Jpn.}\ }\textbf {\bibinfo {volume} {79}},\ \bibinfo
  {pages} {114705} (\bibinfo {year} {2010})}\BibitemShut {NoStop}%
\bibitem [{\citenamefont {Lee}\ and\ \citenamefont {Balents}(2008)}]{Lee2008}%
  \BibitemOpen
  \bibfield  {author} {\bibinfo {author} {\bibfnamefont {S.~B.}\ \bibnamefont
  {Lee}}\ and\ \bibinfo {author} {\bibfnamefont {L.}~\bibnamefont {Balents}},\
  }\bibfield  {title} {\bibinfo {title} {{Theory of the ordered phase in A-site
  antiferromagnetic spinels}},\ }\href
  {https://doi.org/10.1103/PhysRevB.78.144417} {\bibfield  {journal} {\bibinfo
  {journal} {Phys. Rev. B}\ }\textbf {\bibinfo {volume} {78}},\ \bibinfo
  {pages} {144417} (\bibinfo {year} {2008})}\BibitemShut {NoStop}%
\bibitem [{\citenamefont {Ferrari}\ and\ \citenamefont
  {Becca}(2020)}]{Ferrari2020}%
  \BibitemOpen
  \bibfield  {author} {\bibinfo {author} {\bibfnamefont {F.}~\bibnamefont
  {Ferrari}}\ and\ \bibinfo {author} {\bibfnamefont {F.}~\bibnamefont
  {Becca}},\ }\bibfield  {title} {\bibinfo {title} {{Dynamical properties of
  N\'eel and valence-bond phases in the $J_1$-$J_2$ model on the honeycomb
  lattice}},\ }\href
  {https://iopscience.iop.org/article/10.1088/1361-648X/ab7f6e/meta?casa_token=sV1HeAfGXZkAAAAA:2k8kWXDZVixb44UlcNdOP9giQ_c9M7Jz2flQDBJ2hGshl8ptFc-RyoxWgOPWKlloHgttS3TsmzuUjTFX2oTQax42qHo}
  {\bibfield  {journal} {\bibinfo  {journal} {J. Phys.: Condens. Matter}\
  }\textbf {\bibinfo {volume} {32}},\ \bibinfo {pages} {274003} (\bibinfo
  {year} {2020})}\BibitemShut {NoStop}%
\bibitem [{\citenamefont {Buessen}\ \emph {et~al.}(2018)\citenamefont
  {Buessen}, \citenamefont {Hering}, \citenamefont {Reuther},\ and\
  \citenamefont {Trebst}}]{Buessen2018}%
  \BibitemOpen
  \bibfield  {author} {\bibinfo {author} {\bibfnamefont {F.~L.}\ \bibnamefont
  {Buessen}}, \bibinfo {author} {\bibfnamefont {M.}~\bibnamefont {Hering}},
  \bibinfo {author} {\bibfnamefont {J.}~\bibnamefont {Reuther}},\ and\ \bibinfo
  {author} {\bibfnamefont {S.}~\bibnamefont {Trebst}},\ }\bibfield  {title}
  {\bibinfo {title} {{Quantum Spin Liquids in Frustrated Spin-1 Diamond
  Antiferromagnets}},\ }\href {https://doi.org/10.1103/PhysRevLett.120.057201}
  {\bibfield  {journal} {\bibinfo  {journal} {Phys. Rev. Lett.}\ }\textbf
  {\bibinfo {volume} {120}},\ \bibinfo {pages} {057201} (\bibinfo {year}
  {2018})}\BibitemShut {NoStop}%
\bibitem [{\citenamefont {Gao}\ \emph {et~al.}(2020)\citenamefont {Gao},
  \citenamefont {Rosales}, \citenamefont {G{\'o}mez~Albarrac{\'\i}n},
  \citenamefont {Tsurkan}, \citenamefont {Kaur}, \citenamefont {Fennell},
  \citenamefont {Steffens}, \citenamefont {Boehm}, \citenamefont {{\v
  C}erm{\'a}k}, \citenamefont {Schneidewind}, \citenamefont {Ressouche},
  \citenamefont {Cabra}, \citenamefont {R{\"u}egg},\ and\ \citenamefont
  {Zaharko}}]{Gao2020}%
  \BibitemOpen
  \bibfield  {author} {\bibinfo {author} {\bibfnamefont {S.}~\bibnamefont
  {Gao}}, \bibinfo {author} {\bibfnamefont {H.~D.}\ \bibnamefont {Rosales}},
  \bibinfo {author} {\bibfnamefont {F.~A.}\ \bibnamefont
  {G{\'o}mez~Albarrac{\'\i}n}}, \bibinfo {author} {\bibfnamefont
  {V.}~\bibnamefont {Tsurkan}}, \bibinfo {author} {\bibfnamefont
  {G.}~\bibnamefont {Kaur}}, \bibinfo {author} {\bibfnamefont {T.}~\bibnamefont
  {Fennell}}, \bibinfo {author} {\bibfnamefont {P.}~\bibnamefont {Steffens}},
  \bibinfo {author} {\bibfnamefont {M.}~\bibnamefont {Boehm}}, \bibinfo
  {author} {\bibfnamefont {P.}~\bibnamefont {{\v C}erm{\'a}k}}, \bibinfo
  {author} {\bibfnamefont {A.}~\bibnamefont {Schneidewind}}, \bibinfo {author}
  {\bibfnamefont {E.}~\bibnamefont {Ressouche}}, \bibinfo {author}
  {\bibfnamefont {D.~C.}\ \bibnamefont {Cabra}}, \bibinfo {author}
  {\bibfnamefont {C.}~\bibnamefont {R{\"u}egg}},\ and\ \bibinfo {author}
  {\bibfnamefont {O.}~\bibnamefont {Zaharko}},\ }\bibfield  {title} {\bibinfo
  {title} {{Fractional antiferromagnetic skyrmion lattice induced by
  anisotropic couplings}},\ }\href
  {https://www.nature.com/articles/s41586-020-2716-8} {\bibfield  {journal}
  {\bibinfo  {journal} {Nature}\ }\textbf {\bibinfo {volume} {586}},\ \bibinfo
  {pages} {37} (\bibinfo {year} {2020})}\BibitemShut {NoStop}%
\bibitem [{\citenamefont {Shimokawa}\ and\ \citenamefont
  {Kawamura}(2019)}]{Shimokawa2019a}%
  \BibitemOpen
  \bibfield  {author} {\bibinfo {author} {\bibfnamefont {T.}~\bibnamefont
  {Shimokawa}}\ and\ \bibinfo {author} {\bibfnamefont {H.}~\bibnamefont
  {Kawamura}},\ }\bibfield  {title} {\bibinfo {title} {{Ripple State in the
  Frustrated Honeycomb-Lattice Antiferromagnet}},\ }\href
  {https://doi.org/10.1103/PhysRevLett.123.057202} {\bibfield  {journal}
  {\bibinfo  {journal} {Phys. Rev. Lett.}\ }\textbf {\bibinfo {volume} {123}},\
  \bibinfo {pages} {057202} (\bibinfo {year} {2019})}\BibitemShut {NoStop}%
\bibitem [{\citenamefont {Shimokawa}\ \emph {et~al.}(2019)\citenamefont
  {Shimokawa}, \citenamefont {Okubo},\ and\ \citenamefont
  {Kawamura}}]{Shimokawa2019b}%
  \BibitemOpen
  \bibfield  {author} {\bibinfo {author} {\bibfnamefont {T.}~\bibnamefont
  {Shimokawa}}, \bibinfo {author} {\bibfnamefont {T.}~\bibnamefont {Okubo}},\
  and\ \bibinfo {author} {\bibfnamefont {H.}~\bibnamefont {Kawamura}},\
  }\bibfield  {title} {\bibinfo {title} {{Multiple-q states of the $J_1$-$J_2$
  classical honeycomb-lattice Heisenberg antiferromagnet under a magnetic
  field}},\ }\href {https://doi.org/10.1103/PhysRevB.100.224404} {\bibfield
  {journal} {\bibinfo  {journal} {Phys. Rev. B}\ }\textbf {\bibinfo {volume}
  {100}},\ \bibinfo {pages} {224404} (\bibinfo {year} {2019})}\BibitemShut
  {NoStop}%
\bibitem [{\citenamefont {Savary}\ \emph {et~al.}(2011)\citenamefont {Savary},
  \citenamefont {Gull}, \citenamefont {Trebst}, \citenamefont {Alicea},
  \citenamefont {Bergman},\ and\ \citenamefont {Balents}}]{Savary2011}%
  \BibitemOpen
  \bibfield  {author} {\bibinfo {author} {\bibfnamefont {L.}~\bibnamefont
  {Savary}}, \bibinfo {author} {\bibfnamefont {E.}~\bibnamefont {Gull}},
  \bibinfo {author} {\bibfnamefont {S.}~\bibnamefont {Trebst}}, \bibinfo
  {author} {\bibfnamefont {J.}~\bibnamefont {Alicea}}, \bibinfo {author}
  {\bibfnamefont {D.}~\bibnamefont {Bergman}},\ and\ \bibinfo {author}
  {\bibfnamefont {L.}~\bibnamefont {Balents}},\ }\bibfield  {title} {\bibinfo
  {title} {{Impurity effects in highly frustrated diamond-lattice
  antiferromagnets}},\ }\href {https://doi.org/10.1103/PhysRevB.84.064438}
  {\bibfield  {journal} {\bibinfo  {journal} {Phys. Rev. B}\ }\textbf {\bibinfo
  {volume} {84}},\ \bibinfo {pages} {064438} (\bibinfo {year}
  {2011})}\BibitemShut {NoStop}%
\bibitem [{\citenamefont {Maryasin}\ and\ \citenamefont
  {Zhitomirsky}(2014)}]{Maryasin2014}%
  \BibitemOpen
  \bibfield  {author} {\bibinfo {author} {\bibfnamefont {V.~S.}\ \bibnamefont
  {Maryasin}}\ and\ \bibinfo {author} {\bibfnamefont {M.~E.}\ \bibnamefont
  {Zhitomirsky}},\ }\bibfield  {title} {\bibinfo {title} {{Order from
  structural disorder in the $XY$ pyrochlore antiferromagnet
  Er$_2$Ti$_2$O$_7$}},\ }\href {https://doi.org/10.1103/PhysRevB.90.094412}
  {\bibfield  {journal} {\bibinfo  {journal} {Phys. Rev. B}\ }\textbf {\bibinfo
  {volume} {90}},\ \bibinfo {pages} {094412} (\bibinfo {year}
  {2014})}\BibitemShut {NoStop}%
\bibitem [{\citenamefont {Ross}\ \emph {et~al.}(2016)\citenamefont {Ross},
  \citenamefont {Krizan}, \citenamefont {Rodriguez-Rivera}, \citenamefont
  {Cava},\ and\ \citenamefont {Broholm}}]{Ross2016}%
  \BibitemOpen
  \bibfield  {author} {\bibinfo {author} {\bibfnamefont {K.~A.}\ \bibnamefont
  {Ross}}, \bibinfo {author} {\bibfnamefont {J.~W.}\ \bibnamefont {Krizan}},
  \bibinfo {author} {\bibfnamefont {J.~A.}\ \bibnamefont {Rodriguez-Rivera}},
  \bibinfo {author} {\bibfnamefont {R.~J.}\ \bibnamefont {Cava}},\ and\
  \bibinfo {author} {\bibfnamefont {C.~L.}\ \bibnamefont {Broholm}},\
  }\bibfield  {title} {\bibinfo {title} {{Static and dynamic $XY$-like
  short-range order in a frustrated magnet with exchange disorder}},\ }\href
  {https://doi.org/10.1103/PhysRevB.93.014433} {\bibfield  {journal} {\bibinfo
  {journal} {Phys. Rev. B}\ }\textbf {\bibinfo {volume} {93}},\ \bibinfo
  {pages} {014433} (\bibinfo {year} {2016})}\BibitemShut {NoStop}%
\bibitem [{\citenamefont {Sarkar}\ \emph {et~al.}(2017)\citenamefont {Sarkar},
  \citenamefont {Krizan}, \citenamefont {Br\"uckner}, \citenamefont {Andrade},
  \citenamefont {Rachel}, \citenamefont {Vojta}, \citenamefont {Cava},\ and\
  \citenamefont {Klauss}}]{Sarkar2017}%
  \BibitemOpen
  \bibfield  {author} {\bibinfo {author} {\bibfnamefont {R.}~\bibnamefont
  {Sarkar}}, \bibinfo {author} {\bibfnamefont {J.~W.}\ \bibnamefont {Krizan}},
  \bibinfo {author} {\bibfnamefont {F.}~\bibnamefont {Br\"uckner}}, \bibinfo
  {author} {\bibfnamefont {E.~C.}\ \bibnamefont {Andrade}}, \bibinfo {author}
  {\bibfnamefont {S.}~\bibnamefont {Rachel}}, \bibinfo {author} {\bibfnamefont
  {M.}~\bibnamefont {Vojta}}, \bibinfo {author} {\bibfnamefont {R.~J.}\
  \bibnamefont {Cava}},\ and\ \bibinfo {author} {\bibfnamefont {H.-H.}\
  \bibnamefont {Klauss}},\ }\bibfield  {title} {\bibinfo {title} {{Spin
  freezing in the disordered pyrochlore magnet NaCaCo$_2$F$_7$: NMR studies and
  Monte Carlo simulations}},\ }\href
  {https://doi.org/10.1103/PhysRevB.96.235117} {\bibfield  {journal} {\bibinfo
  {journal} {Phys. Rev. B}\ }\textbf {\bibinfo {volume} {96}},\ \bibinfo
  {pages} {235117} (\bibinfo {year} {2017})}\BibitemShut {NoStop}%
\bibitem [{\citenamefont {Andrade}\ \emph {et~al.}(2018)\citenamefont
  {Andrade}, \citenamefont {Hoyos}, \citenamefont {Rachel},\ and\ \citenamefont
  {Vojta}}]{Andrade2018}%
  \BibitemOpen
  \bibfield  {author} {\bibinfo {author} {\bibfnamefont {E.~C.}\ \bibnamefont
  {Andrade}}, \bibinfo {author} {\bibfnamefont {J.~A.}\ \bibnamefont {Hoyos}},
  \bibinfo {author} {\bibfnamefont {S.}~\bibnamefont {Rachel}},\ and\ \bibinfo
  {author} {\bibfnamefont {M.}~\bibnamefont {Vojta}},\ }\bibfield  {title}
  {\bibinfo {title} {Cluster-glass phase in pyrochlore $xy$ antiferromagnets
  with quenched disorder},\ }\href
  {https://doi.org/10.1103/PhysRevLett.120.097204} {\bibfield  {journal}
  {\bibinfo  {journal} {Phys. Rev. Lett.}\ }\textbf {\bibinfo {volume} {120}},\
  \bibinfo {pages} {097204} (\bibinfo {year} {2018})}\BibitemShut {NoStop}%
\bibitem [{\citenamefont {Utesov}\ \emph {et~al.}(2015)\citenamefont {Utesov},
  \citenamefont {Sizanov},\ and\ \citenamefont {Syromyatnikov}}]{Utesov2015}%
  \BibitemOpen
  \bibfield  {author} {\bibinfo {author} {\bibfnamefont {O.~I.}\ \bibnamefont
  {Utesov}}, \bibinfo {author} {\bibfnamefont {A.~V.}\ \bibnamefont
  {Sizanov}},\ and\ \bibinfo {author} {\bibfnamefont {A.~V.}\ \bibnamefont
  {Syromyatnikov}},\ }\bibfield  {title} {\bibinfo {title} {{Spiral magnets
  with Dzyaloshinskii-Moriya interaction containing defect bonds}},\ }\href
  {https://doi.org/10.1103/PhysRevB.92.125110} {\bibfield  {journal} {\bibinfo
  {journal} {Phys. Rev. B}\ }\textbf {\bibinfo {volume} {92}},\ \bibinfo
  {pages} {125110} (\bibinfo {year} {2015})}\BibitemShut {NoStop}%
\bibitem [{\citenamefont {Dey}\ \emph {et~al.}(2020)\citenamefont {Dey},
  \citenamefont {Andrade},\ and\ \citenamefont {Vojta}}]{Dey2020}%
  \BibitemOpen
  \bibfield  {author} {\bibinfo {author} {\bibfnamefont {S.}~\bibnamefont
  {Dey}}, \bibinfo {author} {\bibfnamefont {E.~C.}\ \bibnamefont {Andrade}},\
  and\ \bibinfo {author} {\bibfnamefont {M.}~\bibnamefont {Vojta}},\ }\bibfield
   {title} {\bibinfo {title} {Destruction of long-range order in noncollinear
  two-dimensional antiferromagnets by random-bond disorder},\ }\href
  {https://doi.org/10.1103/PhysRevB.101.020411} {\bibfield  {journal} {\bibinfo
   {journal} {Phys. Rev. B}\ }\textbf {\bibinfo {volume} {101}},\ \bibinfo
  {pages} {020411(R)} (\bibinfo {year} {2020})}\BibitemShut {NoStop}%
\bibitem [{\citenamefont {Vojta}(2019)}]{Vojta2019}%
  \BibitemOpen
  \bibfield  {author} {\bibinfo {author} {\bibfnamefont {T.}~\bibnamefont
  {Vojta}},\ }\bibfield  {title} {\bibinfo {title} {{Disorder in Quantum
  Many-Body Systems}},\ }\href
  {https://doi.org/10.1146/annurev-conmatphys-031218-013433} {\bibfield
  {journal} {\bibinfo  {journal} {Annu. Rev. Condens. Matter Phys.}\ }\textbf
  {\bibinfo {volume} {10}},\ \bibinfo {pages} {233} (\bibinfo {year}
  {2019})}\BibitemShut {NoStop}%
\bibitem [{\citenamefont {Hohenberg}(1967)}]{Hohenberg1967}%
  \BibitemOpen
  \bibfield  {author} {\bibinfo {author} {\bibfnamefont {P.~C.}\ \bibnamefont
  {Hohenberg}},\ }\bibfield  {title} {\bibinfo {title} {{Existence of
  Long-Range Order in One and Two Dimensions}},\ }\href
  {https://doi.org/10.1103/PhysRev.158.383} {\bibfield  {journal} {\bibinfo
  {journal} {Phys. Rev.}\ }\textbf {\bibinfo {volume} {158}},\ \bibinfo {pages}
  {383} (\bibinfo {year} {1967})}\BibitemShut {NoStop}%
\bibitem [{\citenamefont {Mermin}\ and\ \citenamefont
  {Wagner}(1966)}]{Mermin1966}%
  \BibitemOpen
  \bibfield  {author} {\bibinfo {author} {\bibfnamefont {N.~D.}\ \bibnamefont
  {Mermin}}\ and\ \bibinfo {author} {\bibfnamefont {H.}~\bibnamefont
  {Wagner}},\ }\bibfield  {title} {\bibinfo {title} {{Absence of Ferromagnetism
  or Antiferromagnetism in One- or Two-Dimensional Isotropic Heisenberg
  Models}},\ }\href {https://doi.org/10.1103/PhysRevLett.17.1133} {\bibfield
  {journal} {\bibinfo  {journal} {Phys. Rev. Lett.}\ }\textbf {\bibinfo
  {volume} {17}},\ \bibinfo {pages} {1133} (\bibinfo {year}
  {1966})}\BibitemShut {NoStop}%
\bibitem [{\citenamefont {Imry}\ and\ \citenamefont {Ma}(1975)}]{Imry1975}%
  \BibitemOpen
  \bibfield  {author} {\bibinfo {author} {\bibfnamefont {Y.}~\bibnamefont
  {Imry}}\ and\ \bibinfo {author} {\bibfnamefont {S.-k.}\ \bibnamefont {Ma}},\
  }\bibfield  {title} {\bibinfo {title} {Random-field instability of the
  ordered state of continuous symmetry},\ }\href
  {https://doi.org/10.1103/PhysRevLett.35.1399} {\bibfield  {journal} {\bibinfo
   {journal} {Phys. Rev. Lett.}\ }\textbf {\bibinfo {volume} {35}},\ \bibinfo
  {pages} {1399} (\bibinfo {year} {1975})}\BibitemShut {NoStop}%
\bibitem [{\citenamefont {Aizenman}\ and\ \citenamefont
  {Wehr}(1989)}]{Aizenman1989}%
  \BibitemOpen
  \bibfield  {author} {\bibinfo {author} {\bibfnamefont {M.}~\bibnamefont
  {Aizenman}}\ and\ \bibinfo {author} {\bibfnamefont {J.}~\bibnamefont
  {Wehr}},\ }\bibfield  {title} {\bibinfo {title} {{Rounding of first-order
  phase transitions in systems with quenched disorder}},\ }\href
  {https://doi.org/10.1103/PhysRevLett.62.2503} {\bibfield  {journal} {\bibinfo
   {journal} {Phys. Rev. Lett.}\ }\textbf {\bibinfo {volume} {62}},\ \bibinfo
  {pages} {2503} (\bibinfo {year} {1989})}\BibitemShut {NoStop}%
\bibitem [{\citenamefont {Maiorano}\ and\ \citenamefont
  {Parisi}(2018)}]{Maiorano2018}%
  \BibitemOpen
  \bibfield  {author} {\bibinfo {author} {\bibfnamefont {A.}~\bibnamefont
  {Maiorano}}\ and\ \bibinfo {author} {\bibfnamefont {G.}~\bibnamefont
  {Parisi}},\ }\bibfield  {title} {\bibinfo {title} {{Support for the value 5/2
  for the spin glass lower critical dimension at zero magnetic field}},\ }\href
  {https://doi.org/10.1073/pnas.1720832115} {\bibfield  {journal} {\bibinfo
  {journal} {PNAS}\ }\textbf {\bibinfo {volume} {115}},\ \bibinfo {pages}
  {5129} (\bibinfo {year} {2018})}\BibitemShut {NoStop}%
\bibitem [{\citenamefont {Lee}\ and\ \citenamefont {Young}(2007)}]{Lee2007}%
  \BibitemOpen
  \bibfield  {author} {\bibinfo {author} {\bibfnamefont {L.~W.}\ \bibnamefont
  {Lee}}\ and\ \bibinfo {author} {\bibfnamefont {A.~P.}\ \bibnamefont
  {Young}},\ }\bibfield  {title} {\bibinfo {title} {{Large-scale Monte Carlo
  simulations of the isotropic three-dimensional Heisenberg spin glass}},\
  }\href {https://doi.org/10.1103/PhysRevB.76.024405} {\bibfield  {journal}
  {\bibinfo  {journal} {Phys. Rev. B}\ }\textbf {\bibinfo {volume} {76}},\
  \bibinfo {pages} {024405} (\bibinfo {year} {2007})}\BibitemShut {NoStop}%
\bibitem [{\citenamefont {Kawamura}(2010)}]{Kawamura2011}%
  \BibitemOpen
  \bibfield  {author} {\bibinfo {author} {\bibfnamefont {H.}~\bibnamefont
  {Kawamura}},\ }\bibfield  {title} {\bibinfo {title} {{Chirality Scenario of
  the Spin-Glass Ordering}},\ }\href {https://doi.org/10.1143/JPSJ.79.011007}
  {\bibfield  {journal} {\bibinfo  {journal} {J. Phys. Soc. Jpn/}\ }\textbf
  {\bibinfo {volume} {79}},\ \bibinfo {pages} {011007} (\bibinfo {year}
  {2010})}\BibitemShut {NoStop}%
\bibitem [{\citenamefont {Ogawa}\ \emph {et~al.}(2020)\citenamefont {Ogawa},
  \citenamefont {Uematsu},\ and\ \citenamefont {Kawamura}}]{Ogawa2020}%
  \BibitemOpen
  \bibfield  {author} {\bibinfo {author} {\bibfnamefont {T.}~\bibnamefont
  {Ogawa}}, \bibinfo {author} {\bibfnamefont {K.}~\bibnamefont {Uematsu}},\
  and\ \bibinfo {author} {\bibfnamefont {H.}~\bibnamefont {Kawamura}},\
  }\bibfield  {title} {\bibinfo {title} {{Monte Carlo studies of the
  spin-chirality decoupling in the three-dimensional Heisenberg spin glass}},\
  }\href {https://doi.org/10.1103/PhysRevB.101.014434} {\bibfield  {journal}
  {\bibinfo  {journal} {Phys. Rev. B}\ }\textbf {\bibinfo {volume} {101}},\
  \bibinfo {pages} {014434} (\bibinfo {year} {2020})}\BibitemShut {NoStop}%
\bibitem [{\citenamefont {Luttinger}\ and\ \citenamefont
  {Tisza}(1946)}]{Luttinger1946}%
  \BibitemOpen
  \bibfield  {author} {\bibinfo {author} {\bibfnamefont {J.~M.}\ \bibnamefont
  {Luttinger}}\ and\ \bibinfo {author} {\bibfnamefont {L.}~\bibnamefont
  {Tisza}},\ }\bibfield  {title} {\bibinfo {title} {Theory of dipole
  interaction in crystals},\ }\href {https://doi.org/10.1103/PhysRev.70.954}
  {\bibfield  {journal} {\bibinfo  {journal} {Phys. Rev.}\ }\textbf {\bibinfo
  {volume} {70}},\ \bibinfo {pages} {954} (\bibinfo {year} {1946})}\BibitemShut
  {NoStop}%
\bibitem [{\citenamefont {Lyons}\ and\ \citenamefont
  {Kaplan}(1960)}]{Lyons1960}%
  \BibitemOpen
  \bibfield  {author} {\bibinfo {author} {\bibfnamefont {D.~H.}\ \bibnamefont
  {Lyons}}\ and\ \bibinfo {author} {\bibfnamefont {T.~A.}\ \bibnamefont
  {Kaplan}},\ }\bibfield  {title} {\bibinfo {title} {Method for determining
  ground-state spin configurations},\ }\href
  {https://doi.org/10.1103/PhysRev.120.1580} {\bibfield  {journal} {\bibinfo
  {journal} {Phys. Rev.}\ }\textbf {\bibinfo {volume} {120}},\ \bibinfo {pages}
  {1580} (\bibinfo {year} {1960})}\BibitemShut {NoStop}%
\bibitem [{\citenamefont {Attig}\ and\ \citenamefont
  {Trebst}(2017)}]{Attig2017}%
  \BibitemOpen
  \bibfield  {author} {\bibinfo {author} {\bibfnamefont {J.}~\bibnamefont
  {Attig}}\ and\ \bibinfo {author} {\bibfnamefont {S.}~\bibnamefont {Trebst}},\
  }\bibfield  {title} {\bibinfo {title} {Classical spin spirals in frustrated
  magnets from free-fermion band topology},\ }\href
  {https://doi.org/10.1103/PhysRevB.96.085145} {\bibfield  {journal} {\bibinfo
  {journal} {Phys. Rev. B}\ }\textbf {\bibinfo {volume} {96}},\ \bibinfo
  {pages} {085145} (\bibinfo {year} {2017})}\BibitemShut {NoStop}%
\bibitem [{\citenamefont {Katsura}\ \emph {et~al.}(1986)\citenamefont
  {Katsura}, \citenamefont {Ide},\ and\ \citenamefont {Morita}}]{Katsura1986}%
  \BibitemOpen
  \bibfield  {author} {\bibinfo {author} {\bibfnamefont {S.}~\bibnamefont
  {Katsura}}, \bibinfo {author} {\bibfnamefont {T.}~\bibnamefont {Ide}},\ and\
  \bibinfo {author} {\bibfnamefont {T.}~\bibnamefont {Morita}},\ }\bibfield
  {title} {\bibinfo {title} {The ground states of the classical heisenberg and
  planar models on the triangular and plane hexagonal lattices},\ }\href
  {https://doi.org/10.1007/BF01127717} {\bibfield  {journal} {\bibinfo
  {journal} {J. Stat. Phys.}\ }\textbf {\bibinfo {volume} {42}},\ \bibinfo
  {pages} {381} (\bibinfo {year} {1986})}\BibitemShut {NoStop}%
\bibitem [{\citenamefont {Fouet}\ \emph {et~al.}(2001)\citenamefont {Fouet},
  \citenamefont {Sindzingre},\ and\ \citenamefont {Lhuillier}}]{Fouet2001}%
  \BibitemOpen
  \bibfield  {author} {\bibinfo {author} {\bibfnamefont {J.~B.}\ \bibnamefont
  {Fouet}}, \bibinfo {author} {\bibfnamefont {P.}~\bibnamefont {Sindzingre}},\
  and\ \bibinfo {author} {\bibfnamefont {C.}~\bibnamefont {Lhuillier}},\
  }\bibfield  {title} {\bibinfo {title} {{An investigation of the quantum
  $J_1$-$J_2$-$J_3$ model on the honeycomb lattice}},\ }\href
  {https://link.springer.com/article/10.1007/s100510170273#citeas} {\bibfield
  {journal} {\bibinfo  {journal} {Eur. Phys. J. B}\ }\textbf {\bibinfo {volume}
  {20}},\ \bibinfo {pages} {241} (\bibinfo {year} {2001})}\BibitemShut
  {NoStop}%
\bibitem [{foo()}]{footnoteJrat0p5}%
  \BibitemOpen
  \href@noop {} {}\bibinfo {note} {In fact, one can rigorously show that the
  number $F$ of continuous degrees of freedom scales with the number of sites
  $N$ according to $F = \sqrt{6N}$ in the thermodynamic limit.}\BibitemShut
  {Stop}%
\bibitem [{\citenamefont {Villain}\ \emph {et~al.}(1980)\citenamefont
  {Villain}, \citenamefont {Bidaux}, \citenamefont {Carton},\ and\
  \citenamefont {Conte}}]{Villain1980}%
  \BibitemOpen
  \bibfield  {author} {\bibinfo {author} {\bibfnamefont {J.}~\bibnamefont
  {Villain}}, \bibinfo {author} {\bibfnamefont {R.}~\bibnamefont {Bidaux}},
  \bibinfo {author} {\bibfnamefont {J.-P.}\ \bibnamefont {Carton}},\ and\
  \bibinfo {author} {\bibfnamefont {R.}~\bibnamefont {Conte}},\ }\bibfield
  {title} {\bibinfo {title} {{Order as an effect of disorder}},\ }\href
  {https://doi.org/10.1051/jphys:0198000410110126300} {\bibfield  {journal}
  {\bibinfo  {journal} {Journal de Physique}\ }\textbf {\bibinfo {volume}
  {41}},\ \bibinfo {pages} {1263} (\bibinfo {year} {1980})}\BibitemShut
  {NoStop}%
\bibitem [{\citenamefont {Henley}(1987)}]{Henley1987}%
  \BibitemOpen
  \bibfield  {author} {\bibinfo {author} {\bibfnamefont {C.~L.}\ \bibnamefont
  {Henley}},\ }\bibfield  {title} {\bibinfo {title} {{Ordering by disorder:
  Ground-state selection in fcc vector antiferromagnets}},\ }\href
  {https://doi.org/10.1063/1.338570} {\bibfield  {journal} {\bibinfo  {journal}
  {J. Appl. Phys.}\ }\textbf {\bibinfo {volume} {61}},\ \bibinfo {pages} {3962}
  (\bibinfo {year} {1987})}\BibitemShut {NoStop}%
\bibitem [{\citenamefont {Henley}(1989)}]{Henley1989}%
  \BibitemOpen
  \bibfield  {author} {\bibinfo {author} {\bibfnamefont {C.~L.}\ \bibnamefont
  {Henley}},\ }\bibfield  {title} {\bibinfo {title} {{Ordering due to disorder
  in a frustrated vector antiferromagnet}},\ }\href
  {https://doi.org/10.1103/PhysRevLett.62.2056} {\bibfield  {journal} {\bibinfo
   {journal} {Phys. Rev. Lett.}\ }\textbf {\bibinfo {volume} {62}},\ \bibinfo
  {pages} {2056} (\bibinfo {year} {1989})}\BibitemShut {NoStop}%
\bibitem [{\citenamefont {Wollny}\ \emph {et~al.}(2011)\citenamefont {Wollny},
  \citenamefont {Fritz},\ and\ \citenamefont {Vojta}}]{Wollny2011}%
  \BibitemOpen
  \bibfield  {author} {\bibinfo {author} {\bibfnamefont {A.}~\bibnamefont
  {Wollny}}, \bibinfo {author} {\bibfnamefont {L.}~\bibnamefont {Fritz}},\ and\
  \bibinfo {author} {\bibfnamefont {M.}~\bibnamefont {Vojta}},\ }\bibfield
  {title} {\bibinfo {title} {Fractional impurity moments in two-dimensional
  noncollinear magnets},\ }\href
  {https://doi.org/10.1103/PhysRevLett.107.137204} {\bibfield  {journal}
  {\bibinfo  {journal} {Phys. Rev. Lett.}\ }\textbf {\bibinfo {volume} {107}},\
  \bibinfo {pages} {137204} (\bibinfo {year} {2011})}\BibitemShut {NoStop}%
\bibitem [{\citenamefont {Friedel}(1958)}]{Friedel1958}%
  \BibitemOpen
  \bibfield  {author} {\bibinfo {author} {\bibfnamefont {J.}~\bibnamefont
  {Friedel}},\ }\bibfield  {title} {\bibinfo {title} {{Metallic alloys}},\
  }\href {https://doi.org/https://doi.org/10.1007/BF02751483} {\bibfield
  {journal} {\bibinfo  {journal} {Nuovo Cim}\ }\textbf {\bibinfo {volume}
  {7}},\ \bibinfo {pages} {287} (\bibinfo {year} {1958})}\BibitemShut {NoStop}%
\bibitem [{\citenamefont {Fetter}\ and\ \citenamefont
  {Walecka}(2003)}]{FetterWalecka}%
  \BibitemOpen
  \bibfield  {author} {\bibinfo {author} {\bibfnamefont {A.~L.}\ \bibnamefont
  {Fetter}}\ and\ \bibinfo {author} {\bibfnamefont {J.~D.}\ \bibnamefont
  {Walecka}},\ }\href@noop {} {\emph {\bibinfo {title} {{Quantum Theory of
  Many-Particle Systems}}}}\ (\bibinfo  {publisher} {Dover Publications, Inc.,
  Mineola, New York},\ \bibinfo {year} {2003})\BibitemShut {NoStop}%
\bibitem [{\citenamefont {Lounis}\ \emph {et~al.}(2011)\citenamefont {Lounis},
  \citenamefont {Zahn}, \citenamefont {Weismann}, \citenamefont {Wenderoth},
  \citenamefont {Ulbrich}, \citenamefont {Mertig}, \citenamefont {Dederichs},\
  and\ \citenamefont {Bl\"ugel}}]{Lounis2011}%
  \BibitemOpen
  \bibfield  {author} {\bibinfo {author} {\bibfnamefont {S.}~\bibnamefont
  {Lounis}}, \bibinfo {author} {\bibfnamefont {P.}~\bibnamefont {Zahn}},
  \bibinfo {author} {\bibfnamefont {A.}~\bibnamefont {Weismann}}, \bibinfo
  {author} {\bibfnamefont {M.}~\bibnamefont {Wenderoth}}, \bibinfo {author}
  {\bibfnamefont {R.~G.}\ \bibnamefont {Ulbrich}}, \bibinfo {author}
  {\bibfnamefont {I.}~\bibnamefont {Mertig}}, \bibinfo {author} {\bibfnamefont
  {P.~H.}\ \bibnamefont {Dederichs}},\ and\ \bibinfo {author} {\bibfnamefont
  {S.}~\bibnamefont {Bl\"ugel}},\ }\bibfield  {title} {\bibinfo {title}
  {{Theory of real space imaging of Fermi surface parts}},\ }\href
  {https://doi.org/10.1103/PhysRevB.83.035427} {\bibfield  {journal} {\bibinfo
  {journal} {Phys. Rev. B}\ }\textbf {\bibinfo {volume} {83}},\ \bibinfo
  {pages} {035427} (\bibinfo {year} {2011})}\BibitemShut {NoStop}%
\bibitem [{\citenamefont {Lau}\ and\ \citenamefont {Kohn}(1978)}]{Lau1978}%
  \BibitemOpen
  \bibfield  {author} {\bibinfo {author} {\bibfnamefont {K.}~\bibnamefont
  {Lau}}\ and\ \bibinfo {author} {\bibfnamefont {W.}~\bibnamefont {Kohn}},\
  }\bibfield  {title} {\bibinfo {title} {{Indirect long-range oscillatory
  interaction between adsorbed atoms}},\ }\href
  {https://doi.org/https://doi.org/10.1016/0039-6028(78)90053-5} {\bibfield
  {journal} {\bibinfo  {journal} {Surf. Sci.}\ }\textbf {\bibinfo {volume}
  {75}},\ \bibinfo {pages} {69} (\bibinfo {year} {1978})}\BibitemShut {NoStop}%
\bibitem [{\citenamefont {Cheianov}\ and\ \citenamefont
  {Fal'ko}(2006)}]{Cheianov2006}%
  \BibitemOpen
  \bibfield  {author} {\bibinfo {author} {\bibfnamefont {V.~V.}\ \bibnamefont
  {Cheianov}}\ and\ \bibinfo {author} {\bibfnamefont {V.~I.}\ \bibnamefont
  {Fal'ko}},\ }\bibfield  {title} {\bibinfo {title} {Friedel oscillations,
  impurity scattering, and temperature dependence of resistivity in graphene},\
  }\href {https://doi.org/10.1103/PhysRevLett.97.226801} {\bibfield  {journal}
  {\bibinfo  {journal} {Phys. Rev. Lett.}\ }\textbf {\bibinfo {volume} {97}},\
  \bibinfo {pages} {226801} (\bibinfo {year} {2006})}\BibitemShut {NoStop}%
\bibitem [{\citenamefont {Fischer}\ and\ \citenamefont
  {Hertz}(1991)}]{Fischer1991}%
  \BibitemOpen
  \bibfield  {author} {\bibinfo {author} {\bibfnamefont {K.~H.}\ \bibnamefont
  {Fischer}}\ and\ \bibinfo {author} {\bibfnamefont {J.~A.}\ \bibnamefont
  {Hertz}},\ }\href@noop {} {\emph {\bibinfo {title} {{Spin Glasses}}}}\
  (\bibinfo  {publisher} {Cambridge University Press, Cambridge},\ \bibinfo
  {year} {1991})\BibitemShut {NoStop}%
\bibitem [{\citenamefont {Chandra}\ \emph {et~al.}(1990)\citenamefont
  {Chandra}, \citenamefont {Coleman},\ and\ \citenamefont
  {Larkin}}]{Chandra1990}%
  \BibitemOpen
  \bibfield  {author} {\bibinfo {author} {\bibfnamefont {P.}~\bibnamefont
  {Chandra}}, \bibinfo {author} {\bibfnamefont {P.}~\bibnamefont {Coleman}},\
  and\ \bibinfo {author} {\bibfnamefont {A.~I.}\ \bibnamefont {Larkin}},\
  }\bibfield  {title} {\bibinfo {title} {{Ising transition in frustrated
  Heisenberg models}},\ }\href {https://doi.org/10.1103/PhysRevLett.64.88}
  {\bibfield  {journal} {\bibinfo  {journal} {Phys. Rev. Lett.}\ }\textbf
  {\bibinfo {volume} {64}},\ \bibinfo {pages} {88} (\bibinfo {year}
  {1990})}\BibitemShut {NoStop}%
\bibitem [{\citenamefont {Weber}\ \emph {et~al.}(2003)\citenamefont {Weber},
  \citenamefont {Capriotti}, \citenamefont {Misguich}, \citenamefont {Becca},
  \citenamefont {Elhajal},\ and\ \citenamefont {Mila}}]{Weber2003}%
  \BibitemOpen
  \bibfield  {author} {\bibinfo {author} {\bibfnamefont {C.}~\bibnamefont
  {Weber}}, \bibinfo {author} {\bibfnamefont {L.}~\bibnamefont {Capriotti}},
  \bibinfo {author} {\bibfnamefont {G.}~\bibnamefont {Misguich}}, \bibinfo
  {author} {\bibfnamefont {F.}~\bibnamefont {Becca}}, \bibinfo {author}
  {\bibfnamefont {M.}~\bibnamefont {Elhajal}},\ and\ \bibinfo {author}
  {\bibfnamefont {F.}~\bibnamefont {Mila}},\ }\bibfield  {title} {\bibinfo
  {title} {{Ising Transition Driven by Frustration in a 2D Classical Model with
  Continuous Symmetry}},\ }\href
  {https://doi.org/10.1103/PhysRevLett.91.177202} {\bibfield  {journal}
  {\bibinfo  {journal} {Phys. Rev. Lett.}\ }\textbf {\bibinfo {volume} {91}},\
  \bibinfo {pages} {177202} (\bibinfo {year} {2003})}\BibitemShut {NoStop}%
\bibitem [{\citenamefont {Baxter}(1973)}]{Baxter1973}%
  \BibitemOpen
  \bibfield  {author} {\bibinfo {author} {\bibfnamefont {R.~J.}\ \bibnamefont
  {Baxter}},\ }\bibfield  {title} {\bibinfo {title} {{Potts model at the
  critical temperature}},\ }\href
  {https://iopscience.iop.org/article/10.1088/0022-3719/6/23/005/meta}
  {\bibfield  {journal} {\bibinfo  {journal} {J. Phys. C: Solid State Phys.}\
  }\textbf {\bibinfo {volume} {6}},\ \bibinfo {pages} {L445} (\bibinfo {year}
  {1973})}\BibitemShut {NoStop}%
\bibitem [{\citenamefont {Miranda}\ \emph {et~al.}(2021)\citenamefont
  {Miranda}, \citenamefont {Almeida}, \citenamefont {Andrade},\ and\
  \citenamefont {Hoyos}}]{Miranda2021}%
  \BibitemOpen
  \bibfield  {author} {\bibinfo {author} {\bibfnamefont {M.~M.~J.}\
  \bibnamefont {Miranda}}, \bibinfo {author} {\bibfnamefont {I.~C.}\
  \bibnamefont {Almeida}}, \bibinfo {author} {\bibfnamefont {E.~C.}\
  \bibnamefont {Andrade}},\ and\ \bibinfo {author} {\bibfnamefont {J.~A.}\
  \bibnamefont {Hoyos}},\ }\bibfield  {title} {\bibinfo {title} {{Phase diagram
  of a frustrated Heisenberg model: From disorder to order and back again}},\
  }\href {https://doi.org/10.1103/PhysRevB.104.054201} {\bibfield  {journal}
  {\bibinfo  {journal} {Phys. Rev. B}\ }\textbf {\bibinfo {volume} {104}},\
  \bibinfo {pages} {054201} (\bibinfo {year} {2021})}\BibitemShut {NoStop}%
\bibitem [{\citenamefont {Blankschtein}\ \emph {et~al.}(1984)\citenamefont
  {Blankschtein}, \citenamefont {Shapir},\ and\ \citenamefont
  {Aharony}}]{Blankschtein1984}%
  \BibitemOpen
  \bibfield  {author} {\bibinfo {author} {\bibfnamefont {D.}~\bibnamefont
  {Blankschtein}}, \bibinfo {author} {\bibfnamefont {Y.}~\bibnamefont
  {Shapir}},\ and\ \bibinfo {author} {\bibfnamefont {A.}~\bibnamefont
  {Aharony}},\ }\bibfield  {title} {\bibinfo {title} {{Potts models in random
  fields}},\ }\href {https://doi.org/10.1103/PhysRevB.29.1263} {\bibfield
  {journal} {\bibinfo  {journal} {Phys. Rev. B}\ }\textbf {\bibinfo {volume}
  {29}},\ \bibinfo {pages} {1263} (\bibinfo {year} {1984})}\BibitemShut
  {NoStop}%
\bibitem [{\citenamefont {Eichhorn}\ and\ \citenamefont
  {Binder}(1996)}]{Eichhorn1996}%
  \BibitemOpen
  \bibfield  {author} {\bibinfo {author} {\bibfnamefont {K.}~\bibnamefont
  {Eichhorn}}\ and\ \bibinfo {author} {\bibfnamefont {K.}~\bibnamefont
  {Binder}},\ }\bibfield  {title} {\bibinfo {title} {{Monte Carlo investigation
  of the three-dimensional random-field three-state Potts model}},\ }\href
  {https://iopscience.iop.org/article/10.1088/0953-8984/8/28/005} {\bibfield
  {journal} {\bibinfo  {journal} {J. Phys.: Condens. Matter}\ }\textbf
  {\bibinfo {volume} {8}},\ \bibinfo {pages} {5209} (\bibinfo {year}
  {1996})}\BibitemShut {NoStop}%
\bibitem [{\citenamefont {Campos}\ \emph {et~al.}(2006)\citenamefont {Campos},
  \citenamefont {Cotallo-Aban}, \citenamefont {Martin-Mayor}, \citenamefont
  {Perez-Gaviro},\ and\ \citenamefont {Tarancon}}]{Campos2006}%
  \BibitemOpen
  \bibfield  {author} {\bibinfo {author} {\bibfnamefont {I.}~\bibnamefont
  {Campos}}, \bibinfo {author} {\bibfnamefont {M.}~\bibnamefont
  {Cotallo-Aban}}, \bibinfo {author} {\bibfnamefont {V.}~\bibnamefont
  {Martin-Mayor}}, \bibinfo {author} {\bibfnamefont {S.}~\bibnamefont
  {Perez-Gaviro}},\ and\ \bibinfo {author} {\bibfnamefont {A.}~\bibnamefont
  {Tarancon}},\ }\bibfield  {title} {\bibinfo {title} {{Spin-Glass Transition
  of the Three-Dimensional Heisenberg Spin Glass}},\ }\href
  {https://doi.org/10.1103/PhysRevLett.97.217204} {\bibfield  {journal}
  {\bibinfo  {journal} {Phys. Rev. Lett.}\ }\textbf {\bibinfo {volume} {97}},\
  \bibinfo {pages} {217204} (\bibinfo {year} {2006})}\BibitemShut {NoStop}%
\bibitem [{\citenamefont {Fernandez}\ \emph {et~al.}(2009)\citenamefont
  {Fernandez}, \citenamefont {Martin-Mayor}, \citenamefont {Perez-Gaviro},
  \citenamefont {Tarancon},\ and\ \citenamefont {Young}}]{Fernandez2009}%
  \BibitemOpen
  \bibfield  {author} {\bibinfo {author} {\bibfnamefont {L.~A.}\ \bibnamefont
  {Fernandez}}, \bibinfo {author} {\bibfnamefont {V.}~\bibnamefont
  {Martin-Mayor}}, \bibinfo {author} {\bibfnamefont {S.}~\bibnamefont
  {Perez-Gaviro}}, \bibinfo {author} {\bibfnamefont {A.}~\bibnamefont
  {Tarancon}},\ and\ \bibinfo {author} {\bibfnamefont {A.~P.}\ \bibnamefont
  {Young}},\ }\bibfield  {title} {\bibinfo {title} {{Phase transition in the
  three dimensional Heisenberg spin glass: Finite-size scaling analysis}},\
  }\href {https://doi.org/10.1103/PhysRevB.80.024422} {\bibfield  {journal}
  {\bibinfo  {journal} {Phys. Rev. B}\ }\textbf {\bibinfo {volume} {80}},\
  \bibinfo {pages} {024422} (\bibinfo {year} {2009})}\BibitemShut {NoStop}%
\bibitem [{\citenamefont {Cole}\ \emph {et~al.}(2023)\citenamefont {Cole},
  \citenamefont {Streeter}, \citenamefont {Fumega}, \citenamefont {Yao},
  \citenamefont {Wang}, \citenamefont {Feng}, \citenamefont {Cao},
  \citenamefont {Lado}, \citenamefont {Nagler},\ and\ \citenamefont
  {Tafti}}]{Cole2023}%
  \BibitemOpen
  \bibfield  {author} {\bibinfo {author} {\bibfnamefont {A.}~\bibnamefont
  {Cole}}, \bibinfo {author} {\bibfnamefont {A.}~\bibnamefont {Streeter}},
  \bibinfo {author} {\bibfnamefont {A.~O.}\ \bibnamefont {Fumega}}, \bibinfo
  {author} {\bibfnamefont {X.}~\bibnamefont {Yao}}, \bibinfo {author}
  {\bibfnamefont {Z.-C.}\ \bibnamefont {Wang}}, \bibinfo {author}
  {\bibfnamefont {E.}~\bibnamefont {Feng}}, \bibinfo {author} {\bibfnamefont
  {H.}~\bibnamefont {Cao}}, \bibinfo {author} {\bibfnamefont {J.~L.}\
  \bibnamefont {Lado}}, \bibinfo {author} {\bibfnamefont {S.~E.}\ \bibnamefont
  {Nagler}},\ and\ \bibinfo {author} {\bibfnamefont {F.}~\bibnamefont
  {Tafti}},\ }\bibfield  {title} {\bibinfo {title} {{Extreme sensitivity of the
  magnetic ground state to halide composition in
  ${\mathrm{FeCl}}_{3\ensuremath{-}x}{\mathrm{Br}}_{x}$}},\ }\href
  {https://doi.org/10.1103/PhysRevMaterials.7.064401} {\bibfield  {journal}
  {\bibinfo  {journal} {Phys. Rev. Mater.}\ }\textbf {\bibinfo {volume} {7}},\
  \bibinfo {pages} {064401} (\bibinfo {year} {2023})}\BibitemShut {NoStop}%
\bibitem [{\citenamefont {Haraguchi}\ \emph {et~al.}(2019)\citenamefont
  {Haraguchi}, \citenamefont {Nawa}, \citenamefont {Michioka}, \citenamefont
  {Ueda}, \citenamefont {Matsuo}, \citenamefont {Kindo}, \citenamefont
  {Avdeev}, \citenamefont {Sato},\ and\ \citenamefont
  {Yoshimura}}]{Haraguchi2019}%
  \BibitemOpen
  \bibfield  {author} {\bibinfo {author} {\bibfnamefont {Y.}~\bibnamefont
  {Haraguchi}}, \bibinfo {author} {\bibfnamefont {K.}~\bibnamefont {Nawa}},
  \bibinfo {author} {\bibfnamefont {C.}~\bibnamefont {Michioka}}, \bibinfo
  {author} {\bibfnamefont {H.}~\bibnamefont {Ueda}}, \bibinfo {author}
  {\bibfnamefont {A.}~\bibnamefont {Matsuo}}, \bibinfo {author} {\bibfnamefont
  {K.}~\bibnamefont {Kindo}}, \bibinfo {author} {\bibfnamefont
  {M.}~\bibnamefont {Avdeev}}, \bibinfo {author} {\bibfnamefont {T.~J.}\
  \bibnamefont {Sato}},\ and\ \bibinfo {author} {\bibfnamefont
  {K.}~\bibnamefont {Yoshimura}},\ }\bibfield  {title} {\bibinfo {title}
  {{Frustrated magnetism in the $J_1$-$J_2$ honeycomb lattice compounds
  MgMnO$_3$ and ZnMnO$_3$ synthesized via a metathesis reaction}},\ }\href
  {https://doi.org/10.1103/PhysRevMaterials.3.124406} {\bibfield  {journal}
  {\bibinfo  {journal} {Phys. Rev. Mater.}\ }\textbf {\bibinfo {volume} {3}},\
  \bibinfo {pages} {124406} (\bibinfo {year} {2019})}\BibitemShut {NoStop}%
\bibitem [{\citenamefont {Blaizot}\ and\ \citenamefont
  {Ripka}(1986)}]{Blaizot1986}%
  \BibitemOpen
  \bibfield  {author} {\bibinfo {author} {\bibfnamefont {J.}~\bibnamefont
  {Blaizot}}\ and\ \bibinfo {author} {\bibfnamefont {G.}~\bibnamefont
  {Ripka}},\ }\href@noop {} {\emph {\bibinfo {title} {{Quantum Theory of Finite
  Systems}}}}\ (\bibinfo  {publisher} {MIT Press, Cambridge, MA, USA},\
  \bibinfo {year} {1986})\BibitemShut {NoStop}%
\bibitem [{\citenamefont {Sachdev}(2011)}]{Sachdev}%
  \BibitemOpen
  \bibfield  {author} {\bibinfo {author} {\bibfnamefont {S.}~\bibnamefont
  {Sachdev}},\ }\href@noop {} {\emph {\bibinfo {title} {{Quantum Phase
  Transitions}}}}\ (\bibinfo  {publisher} {Cambridge University Press, New
  York},\ \bibinfo {year} {2011})\BibitemShut {NoStop}%
\bibitem [{\citenamefont {Bender}\ and\ \citenamefont
  {Orszag}(1999)}]{Bender1999}%
  \BibitemOpen
  \bibfield  {author} {\bibinfo {author} {\bibfnamefont {C.~M.}\ \bibnamefont
  {Bender}}\ and\ \bibinfo {author} {\bibfnamefont {S.~A.}\ \bibnamefont
  {Orszag}},\ }\href@noop {} {\emph {\bibinfo {title} {{Advanced Mathematical
  Methods for Scientists and Engineers I: Asymptotics Methods and Perturbation
  Theory}}}}\ (\bibinfo  {publisher} {Springer New York},\ \bibinfo {year}
  {1999})\BibitemShut {NoStop}%
\bibitem [{\citenamefont {Press}\ \emph {et~al.}(1996)\citenamefont {Press},
  \citenamefont {Teukolsky}, \citenamefont {Vetterling},\ and\ \citenamefont
  {Flannery}}]{NumericalRecipes}%
  \BibitemOpen
  \bibfield  {author} {\bibinfo {author} {\bibfnamefont {W.~H.}\ \bibnamefont
  {Press}}, \bibinfo {author} {\bibfnamefont {S.~A.}\ \bibnamefont
  {Teukolsky}}, \bibinfo {author} {\bibfnamefont {W.~T.}\ \bibnamefont
  {Vetterling}},\ and\ \bibinfo {author} {\bibfnamefont {B.~P.}\ \bibnamefont
  {Flannery}},\ }\href@noop {} {\emph {\bibinfo {title} {{Numerical Recipes in
  Fortran 90: The Art of Parallel Scientific Computing}}}}\ (\bibinfo
  {publisher} {Cambridge University Press, New York},\ \bibinfo {year}
  {1996})\BibitemShut {NoStop}%
\end{thebibliography}%

\end{document}